# New Possible Properties of Atomic Nuclei Investigated by Non Linear Methods, Fractal and Recurrence Quantification Analysis.

Elio Conte(*)
(*) Department of Pharmacology and Human Physiology – Tires – Center for Innovative Technologies for Signal Detection and Processing, University of Bari , Italy;
International Center for Studies on Radioactivity, Bari, Italy.
Andrei Yu. Khrennikov (+)
(+) International Center for Mathematical Modeling in Physics and Cognitive Sciences, University of Växjo, S-35195 Sweden
and
Joseph P. Zbilut (°)
(°) Department of Molecular Biophysics and Physiology, Rush University,1653 W. Congress, Chicago, IL60612, USA.

**Abstract**: For the first time we apply the methodologies of non linear analysis to investigate atomic matter. The sense is that we use such methods in analysis of Atomic Weights and of Mass Number of atomic nuclei. Using AutoCorrelation Function and Mutual Information we establish the presence of non linear effects in the mechanism of increasing mass of atomic nuclei considered as function of the $Z$ atomic number. We also operate reconstruction in phase space and we obtain values for Lyapunov spectrum and $D_2$ - correlation dimension. We find that such mechanism of increasing mass is divergent, possibly chaotic. Non integer values of $D_2$ are found. According to previous studies of V. Paar et al. [ 5 ] we also investigate the possible existence of a Power Law for atomic nuclei and , using also the technique of the variogram, we arrive to conclude that a fractal regime could superintend to the mechanism of increasing mass for nuclei. Finally , using Hurst exponent, evidence is obtained that the mechanism of increasing mass in atomic nuclei is fractional Brownian regime with long range correlations. The most interesting results are obtained by using the Recurrence Quantification Analysis (RQA). We estimate % Recurrences, % Determinism, Entropy and Max Line one time in an embedded phase space with dimension D=2 and the other time in embedding dimension D=1. New recurrences, psudoperiodicities, self-resemblance and class of self-similarities are identified with values of determinism showing oscillating values indicating the presence of more or less stability during the process of increasing mass of atomic nuclei All the data were analyzed using shuffled data for comparison.
In brief, new regimes of regularities are identified for atomic nuclei that deserve to be deepened by future researches. In particular an accurate analysis of binding energy values by non linear methods is further required.

## Introduction

It is well known that the mass represents one of the most basic properties of an atomic nucleus.
It is also a complex and non trivial quantity whose basic properties still must be investigated deeply and properly understood.
The celebrated Einstein's mass law is known

$$m = \frac{E}{c^2} \qquad (1)$$

On this basis some different contributions of energy are stored inside a nucleus, and contribute to its mass.
During nucleus formation in its ground state, a certain amount of energy B will be released in the process so that

$$Mc^2 = \sum_j m_j c^2 - B\,. \qquad (2)$$

There are different sources of such energy B. It contributes the strong attractive interaction of nucleons . However, despite the immense amount of data about nuclear properties, the basic understanding of the nuclear strong interaction, as example, still lacks. We have a basic model of meson exchange that of course works at a qualitative level but it does not provide a satisfactory approach to the description of such basic interaction. Still it contributes Coulomb repulsion between protons, and in addition we have also surface effects and still many other contributions that in a phenomenological picture are tentatively taken into account invoking some models as example the liquid drop elaboration as von Weizsacker [1].

It is known some other nuclear mass models may be considered and, despite the numerous parameters that are contained in these different models and the intrinsic conceptual differences adopted in their formulation, some common features arise from these calculations. All such models, [2], give similar results for the known masses, their calculations yield a typical accuracy that results about $5\times10^{-4}$ for a medium-heavy nucleus having binding energy of the order of $1000\,MeV$, but the predictions of such different mass models strongly give a net divergence when applied to unknown regions.
One consequence of such two indications seems rather evident. According to [2], there is the possibility that a basic underlying mechanism oversees the process of mass formation of atomic nuclei, and it is not presently incorporated and considered in the present nuclear models of the traditional nuclear physics.
In fact, some astonishing results are not lacking as far as this problem is concerned.
Owing to the presence of Pauli's exclusion principle, when nucleons are put together to form a bound state , there are not at rest and thus their kinetic energy also contributes to $B$ given in (2) and thus to the mass of the nucleus. Still according to [2], a part of this energy, that is to say, that one that varies smoothly with the number of the nucleons, is taken into account in the liquid drop model but the remaining part of this energy fluctuates with the number of nucleons.
The proper nature of such fluctuations should be more investigated.
P. Leboeuf [2] has extensively analyzed this problem and his conclusion is that the motion of the nucleons inside the nucleus has a regular plus a chaotic component. We will not enter into details here [2] but we only remember here that traditionally in nuclear physics dynamical effects in the structure of nuclei have been referred to as shell effects with the pioneer studies of A. Bohr and B.R. Mottelson [3] and V.M. Strutinsky [4]. The experience here derives from atomic physics where the symmetries of the Hamiltonian generates strong degeneracy of the electronic levels and such degeneracy produce oscillations in the electronic binding energy. Shell effects should be due to deviations of the single particle levels with respect to their average properties.
According to the different approaches that have been introduced to reproduce the systematics of the observed nuclear masses that in part are inspired to liquid drop models or Thomas Fermi approximations, the total energy may be expressed as the sum of two contributions:

$$U(N,Z,x)=\overline{U}(N,Z,x)+\hat{U}(N,Z,x) \qquad (3)$$

with $x$ a parameter set defining the shape of the nucleus. $\overline{U}$ is describing the bulk or macroscopic properties of the nucleus and $\hat{U}$ instead describes shell effects. This term could be splitted in two components [2], the first representing the regular component and the second representing instead the chaotic contribution. The same thing we should have for the mass

$$Mc^2=\sum_j m_j c^2 - B$$

with

$$B(E,x)=\overline{B}(E,x)-\hat{B}(E,x) \qquad (4)$$

There is now another important but independent contribution that deserves to be mentioned here.
Rather recently V. Paar et al [5] introduced a power law for description of the line of stability of atomic nuclei, and in particular for the description of atomic weights. They compared the found power law with the semi-empirical formula of the liquid drop model, and showed that the power law corresponds to a reduction of neutron excess in superheavy nuclei with respect to the semi-empirical formula. Some fractal features of the nuclear valley of stability were analyzed and the value of fractal dimension was determined.
It is well known that a power law may be often connected with an underlying fractal geometry of the considered system. If confirmed for atomic nuclei, according to [5], it could be proposed a new approach to the problem of stability of atomic nuclei. In this case the aim should be to identify the basic features in underlying dynamics giving rise to the structure of the atomic nuclei. Of course, it was pointed out the role of fractal geometry in quantum physics and quark dynamics [6] and in particular it was analyzed the self-similarity of paths of the Feynman path integral.

Finally, M. Pitkanen repeatedly outlined that his TGD model predicts that universe is 4-D spin glass and this kind of fractal energy landscape might be present in some geometric degrees of freedom such as shape of nuclear outer surface or, if nuclear string picture is accepted, in the folding dynamics of the nuclear string [7].

Still examining the problem under a different point of view, we must outline here the results that recently were obtained in [8].

These authors found non linear dynamical properties of giant monopole resonances in atomic nuclei investigated within the time-dependent relativistic mean field model.

Finally, in ref [9], the statistics of the radioactive decay of heavy nuclei was the subject of experimental interest. It was considered that, owing to the intrinsic fluctuations of the decay rate, the counting statistics could depart from the simple Poissonian behaviour. Several experiments carried out with alfa and beta sources have found that the counting variance for long counting periods, is higher than the Poissonian value by more than one order of magnitude. This anomalous large variance has been taken as an experimental indication that the power spectrum of the decay rate fluctuations has a contribution that grows as the inverse of the frequency $f$ at low frequencies.

In conclusion, also considered the problem from several and different view points, there are some different arising evidences that it deserves to be analyzed by the methods of non linear dynamics in order to obtain some detailed result.

This was precisely the aim of the present paper. We analyzed the Atomic Weights, $W_a(Z)$ and the Mass Number A(Z) as function of the atomic number Z for stable atomic nuclei and we applied to such data our non linear test methods, Fractal and Recurrence Quantification Analysis. The results are reported and discussed in detail in the following section.

## 2. Preparation of the Experimental Data.

It is known that the trends of nuclear stability may be represented in a well known $N, Z$ chart of nuclides where each nucleus with $Z$ protons and $N$ neutrons has mass Number $A = Z + N$. A line of stability may be realized by taking for each atomic number $Z$, the stable nucleus of the isotope having the largest relative abundance.

The atomic weights of a naturally occurring element are given by averaging the corresponding isotope weights, weighted so to take into account the relative isotopic abundances. In this paper the data for stable nuclei with $Z$ values until $Z = 83$ were considered. The data were taken using the IUPAC 1997, standard atomic weights, at www.webelements.com. $W_a(Z)$ and $A(Z)$ are given in Fig.1.

**Fig.1**

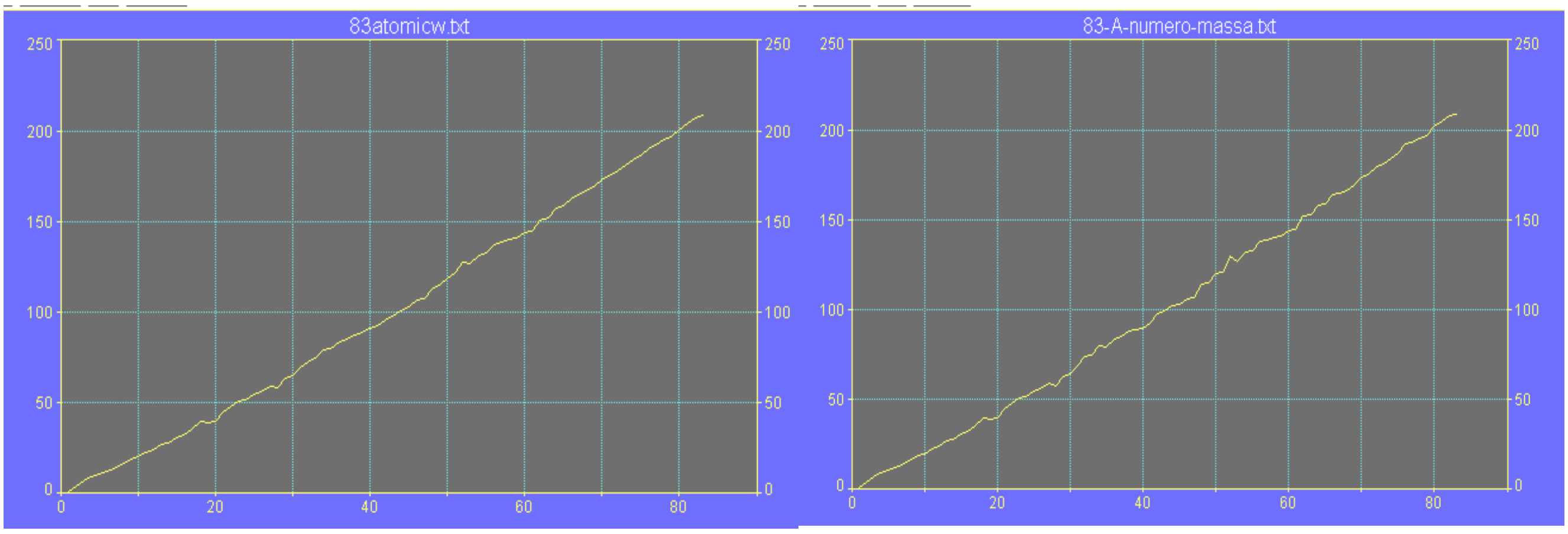

. Atomic Weights: $W_a(Z)$ Mass Number $A(Z)$

**3.Tests by using Mutual Information.**

Autocorrelation Function. The autocorrelation $\rho(\tau)$ is given by the correlation of a time series with itself using $x(t)$ and $x(t+\tau)$ two time series in the correlation formula. For time series it measures as well correlated values of the given time series result under different delays. A choice for the delay time to use when embedding time series data should be the first time the autocorrelation crosses zero. It represents the lowest value of delay for which the values are uncorrelated. The important thing here is that the autocorrelation function is a linear measure and thus it should not provide accurate results in all the situations in which important non linear contributions are expected.

In the present case we examine two series that are $W_a(Z)$ and $A(Z)$. Here we have not a time variable respect to which the delay must be characterized but instead it is the atomic number Z that takes the place of time t. Therefore we will speak here of $Z-shift$ instead of time –lags in our embedding procedure.

In Fig.2 we report the results of our calculations for autocorrelation function (ACF) in the case of Atomic Weights and Mass Number respectively. Both the ACF for $W_a(Z)$ and $A(Z)$ were calculated for $Z-shift$ ranging from 1 to 80. The first value of $\boldsymbol{Z}$ the ACF crosses the zero was obtained for $W_a(Z)$ and $A(Z)$, and it resulted $Z-shift=30$. A typical behaviour was obtained for ACF, in the cases of $W_a(Z)$ and $A(Z)$ respectively, resulting in progressively, positive but decreasing values of ACF until the value $Z-shift=30$, and a subsequent negative half-wave for $Z-shift$ values greater than 30. This seems an interesting result that deserves in some manner a careful interpretation.

**Fig.2**

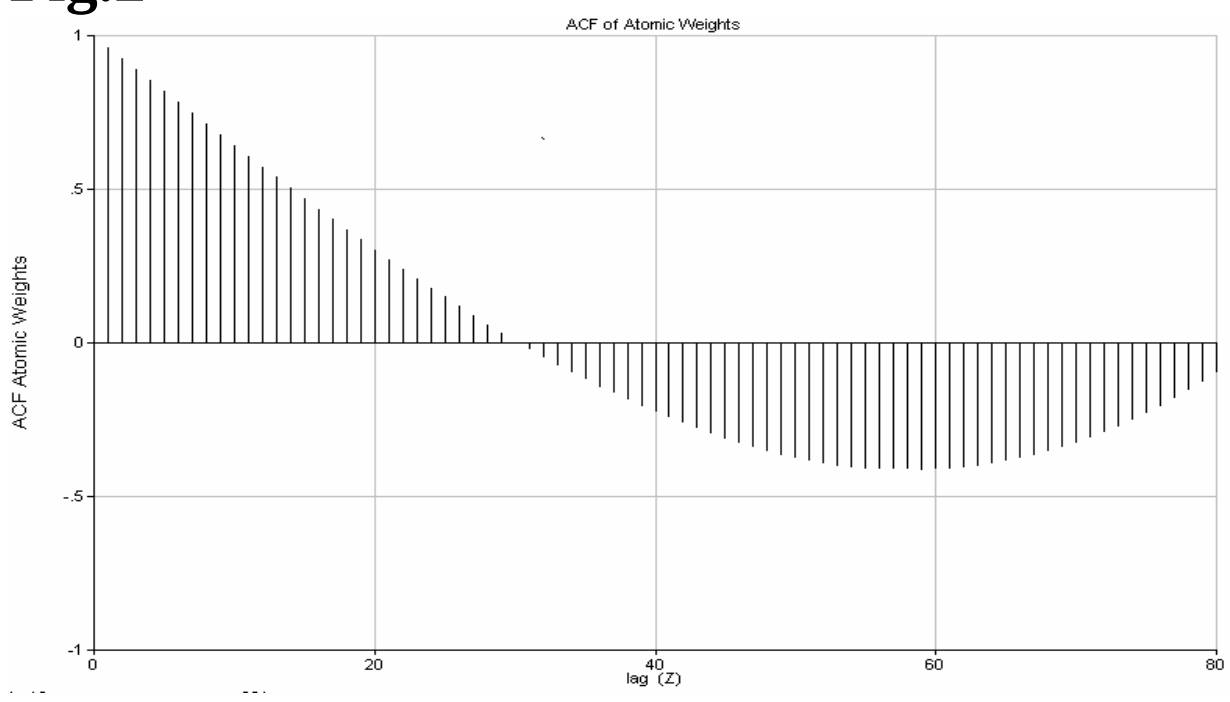

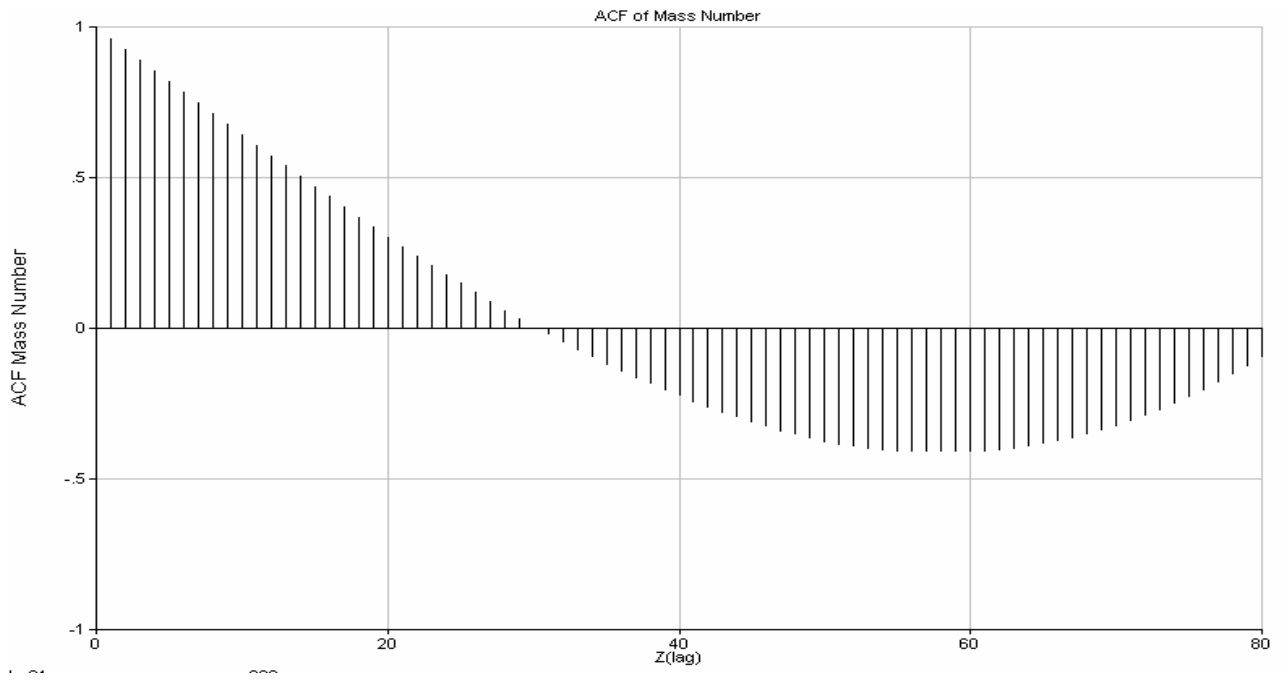

ACF for $Z-shift$ values ranging from 1 to 80 in the case of $W_a(Z)$.

ACF for $Z-shift$ values ranging from 1 to 80 in the case of $A(Z)$.

**The Mutual Information**. It is usually used to determine a useful time delay for attractor reconstruction of a given time series. Generally speaking, we may observe only a variable from a system, $x(t)$, and we wish to reconstruct a higher dimensional attractor. We have to consider $[x(t), x(t+\tau), x(t+2\tau), ......., x(t+n\tau)]$ to produce a $(n+1)$ dimensional representation. Consequently, the problem is to choose a proper value for the delay $\tau$. If the delay is chosen too short, then $x(t)$ is very similar to $x(t+\tau)$. Of course, for a too large delay, then the corresponding coordinates result essentially independent and no information can be gained. The method of Mutual Information [10] involves the idea that a good choice for $\tau$ is one that, given $x(t)$ provides new information with measurement $x(t+\tau)$. In other terms, given a measurement of $x(t)$, how many bits on the average can be predicted about $x(t+\tau)$? In the general case, as $\tau$ is increased, Mutual Information decreases and then usually rises again. The first minimum in Mutual Information is used to select a proper $\tau$. The important thing is here that the Mutual Information function takes non linear correlations into account.

Before to consider the results that we have obtained, it is important to take into account that they change in some manner our traditional manner to approach the discussion on atomic weights and mass numbers of atomic nuclei. In fact, we do not consider here values obtained for a single atomic weight or for a single mass number. Instead, using M.I., we evaluate M.I. values computed for pairs of Atomic Weights, i.e. $W_a(Z)$ and $W_a(Z+Z-shift)$, for any possible $Z$ and for each considered $Z-shift$. The same thing happens for pairs of atomic nuclei with Mass Numbers $A(Z)$ and $A(Z+Z-shift)$.
In Fig.3 we give our results for analysis of $W_a(Z)$. The calculated $Z-shift$ resulted $Z=3$.In Fig.4 we give instead the results for $A(Z)$. In this case the calculated $Z-shift$ resulted to be $Z=2$. To complete our results, in Fig.5 we give also the results of M.I computed for $N(Z)$, being this time $N$ the number of neutrons considered as $N(Z)$. Finally, Fig. 6 compares Mutual Information of $W_a(Z)$, $A(Z)$, $N(Z)$.

**Fig. 3: Mutual Information**

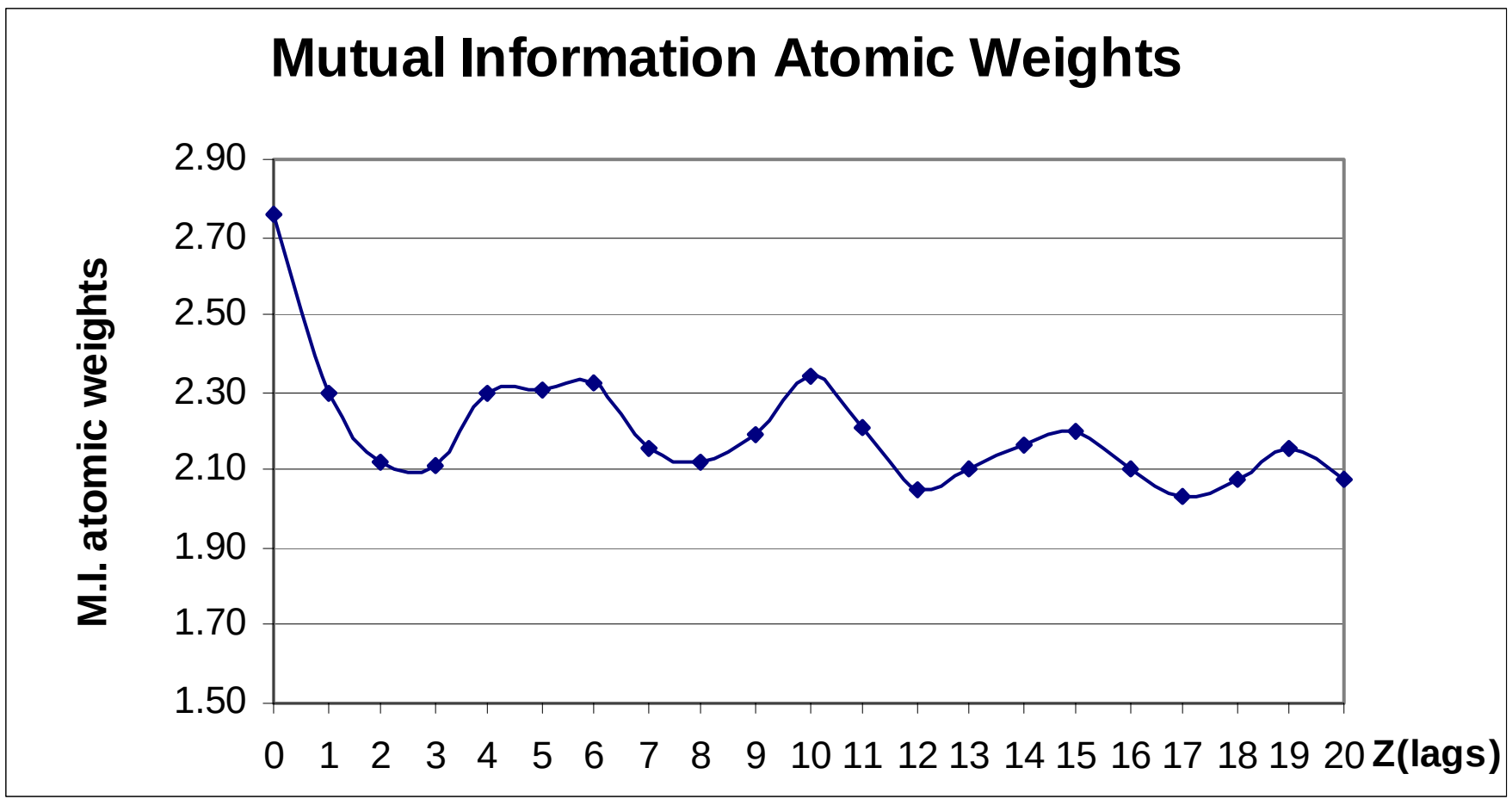

| Z-shift | M. I. |
|---|---|
| 0 | 2.75572 |
| 1 | 2.29477 |
| 2 | 2.12415 |
| 3 | 2.11128 |
| 4 | 2.29507 |
| 5 | 2.30601 |
| 6 | 2.32549 |
| 7 | 2.15915 |
| 8 | 2.11995 |
| 9 | 2.19359 |
| 10 | 2.34003 |
| 11 | 2.21110 |
| 12 | 2.05087 |
| 13 | 2.09827 |
| 14 | 2.16302 |
| 15 | 2.19951 |
| 16 | 2.09898 |
| 17 | 2.03026 |
| 18 | 2.07486 |
| 19 | 2.15473 |
| 20 | 2.07256 |

**Fig. 4: Mutual Information**

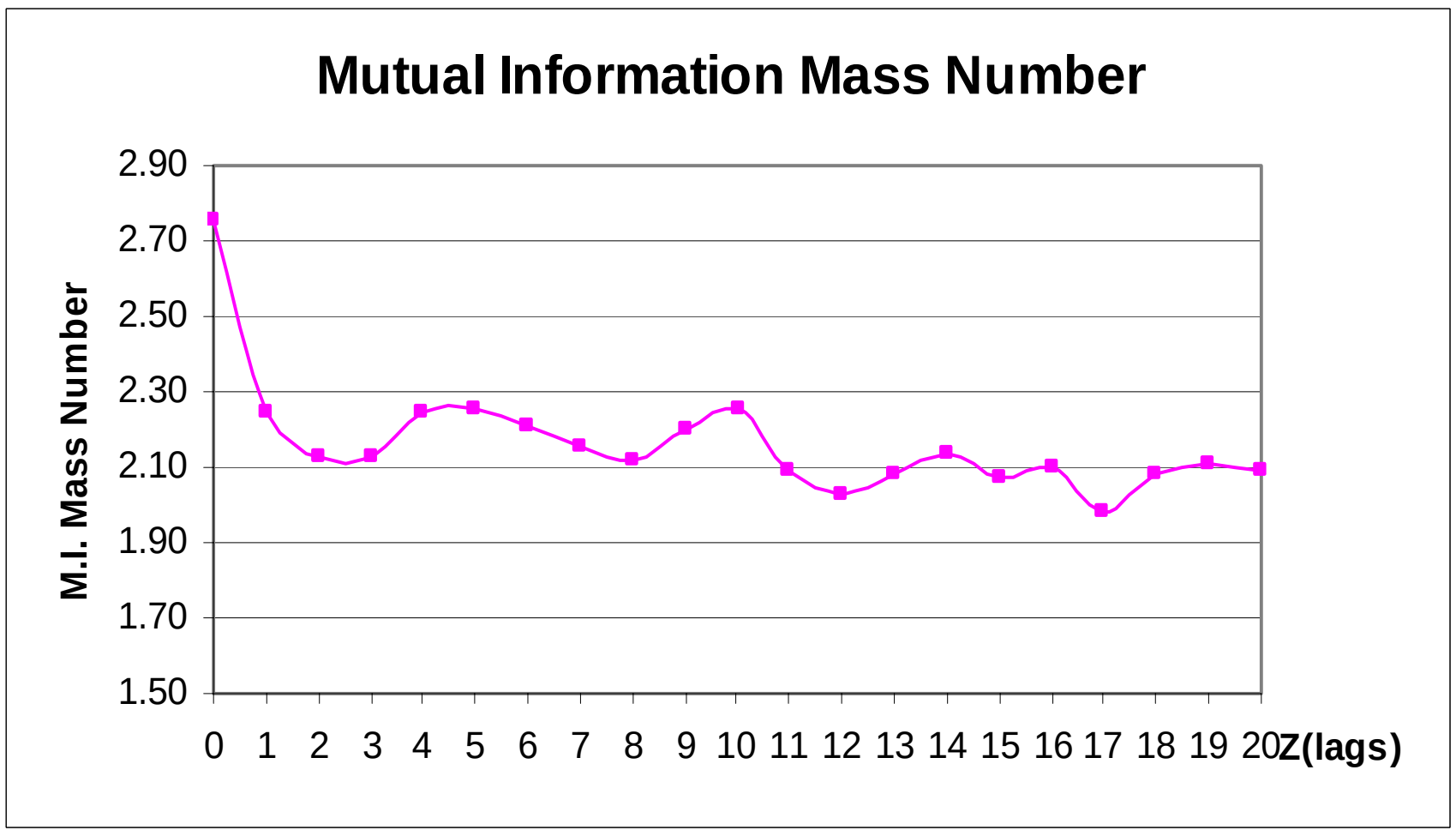

| Z-shift | M. I. |
|---|---|
| 0 | 2.75025 |
| 1 | 2.24680 |
| 2 | 2.12359 |
| 3 | 2.12379 |
| 4 | 2.24418 |
| 5 | 2.25447 |
| 6 | 2.21247 |
| 7 | 2.15885 |
| 8 | 2.11560 |
| 9 | 2.19567 |
| 10 | 2.25399 |
| 11 | 2.09334 |
| 12 | 2.02933 |
| 13 | 2.08003 |
| 14 | 2.13427 |
| 15 | 2.07279 |
| 16 | 2.09985 |
| 17 | 1.98171 |
| 18 | 2.08486 |
| 19 | 2.11095 |
| 20 | 2.09009 |

**Fig. 5: Mutual Information**

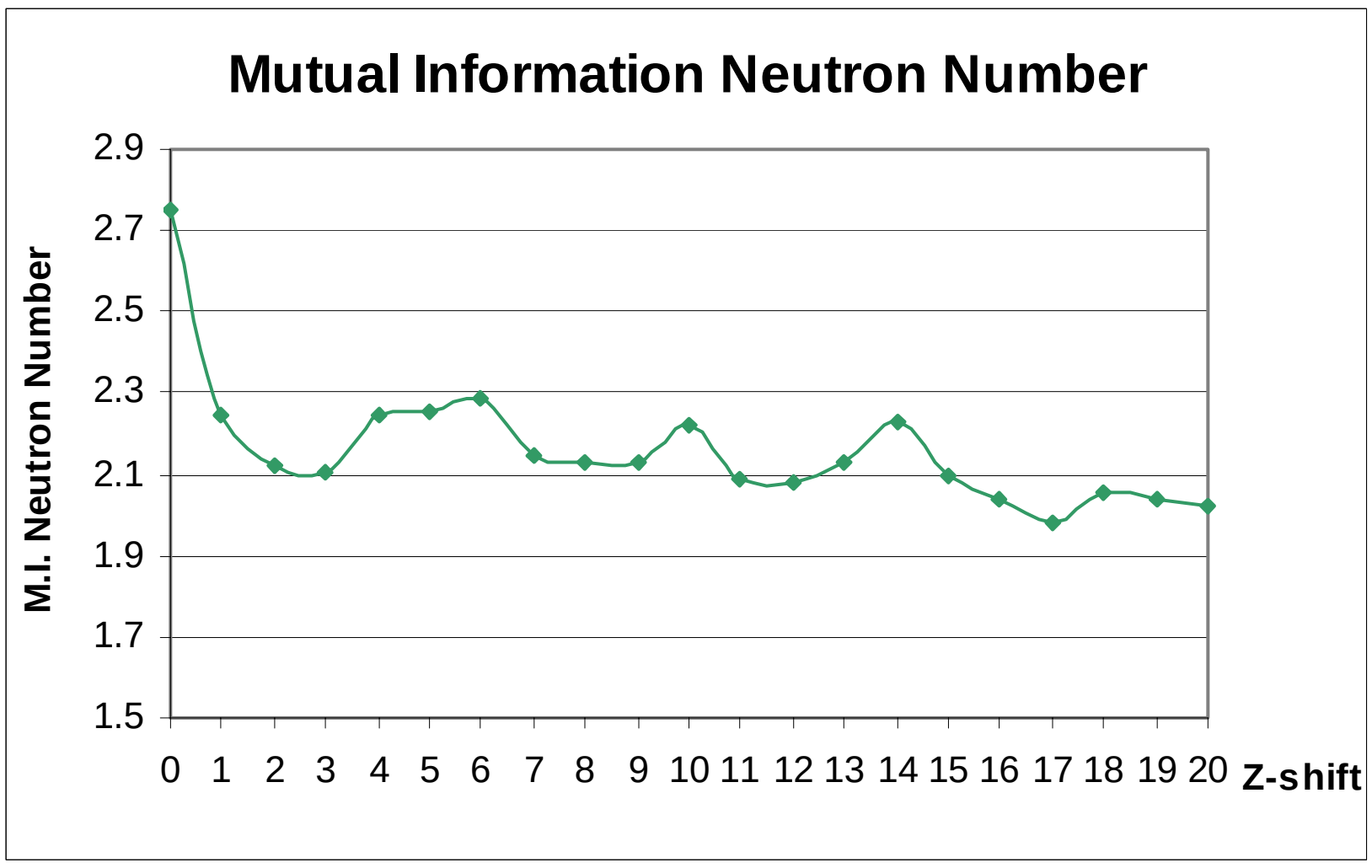

| Z-shift | M.I. |
|---|---|
| 0 | 2.746829 |
| 1 | 2.247466 |
| 2 | 2.117802 |
| 3 | 2.108233 |
| 4 | 2.248289 |
| 5 | 2.251926 |
| 6 | 2.288034 |
| 7 | 2.144352 |
| 8 | 2.131585 |
| 9 | 2.133403 |
| 10 | 2.219288 |
| 11 | 2.086516 |
| 12 | 2.082935 |
| 13 | 2.129288 |
| 14 | 2.226088 |
| 15 | 2.092718 |
| 16 | 2.034579 |
| 17 | 1.983242 |
| 18 | 2.053253 |
| 19 | 2.038391 |
| 20 | 2.018196 |

**Fig.6 : Mutual Information.**

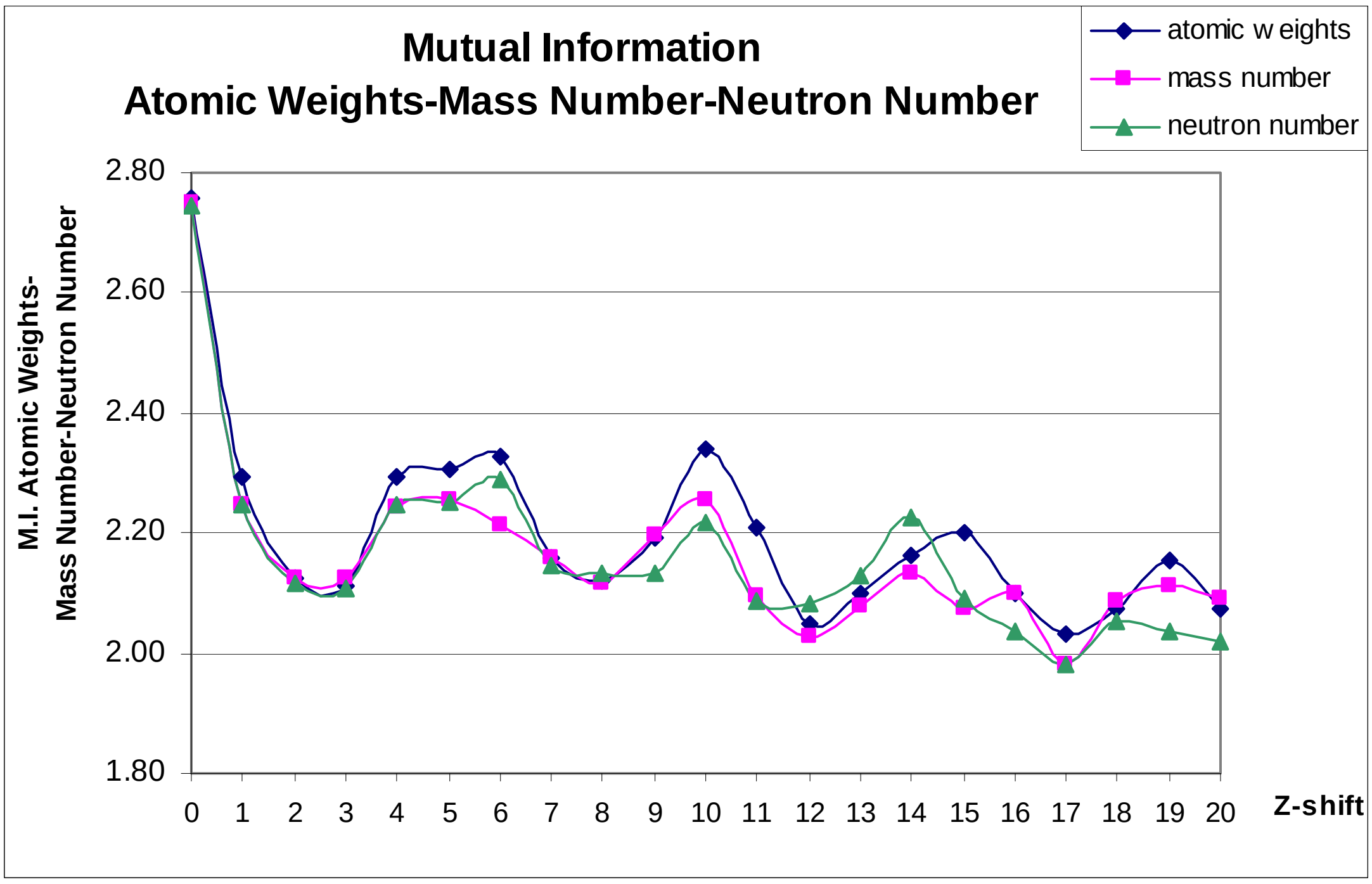

We are now in the condition to reassume some results.

Using autocorrelation function, ACF (Linear Analysis), a $Z-shift$ value of $Z=30$ is obtained for both $W_a(Z)$ and $A(Z)$.

Using Mutual Information (Non Linear Analysis) it is obtained instead $Z-shift=3$ for $W_a(Z)$ and $Z-shift=2$ for $A(Z)$. Also $N(Z)$ gave $Z-shift=3$.

We have a preliminary indication that the mechanism of increasing mass in atomic nuclei is a non linear mechanism. Of course, this could be an important indication in understanding of the basic features of nuclear matter .Therefore it becomes of relevant importance to attempt to confirm such conclusion on the basis of a more deepened control. The test that in such cases one uses in analysis of non linear dynamics of time series data is that one of surrogate data. Here we used shuffled data. The results are given in Fig.7 for $W_a(Z)$ and in Fig.8 for $A(Z)$.

**Fig.7 : Surrogate Data Analysis**

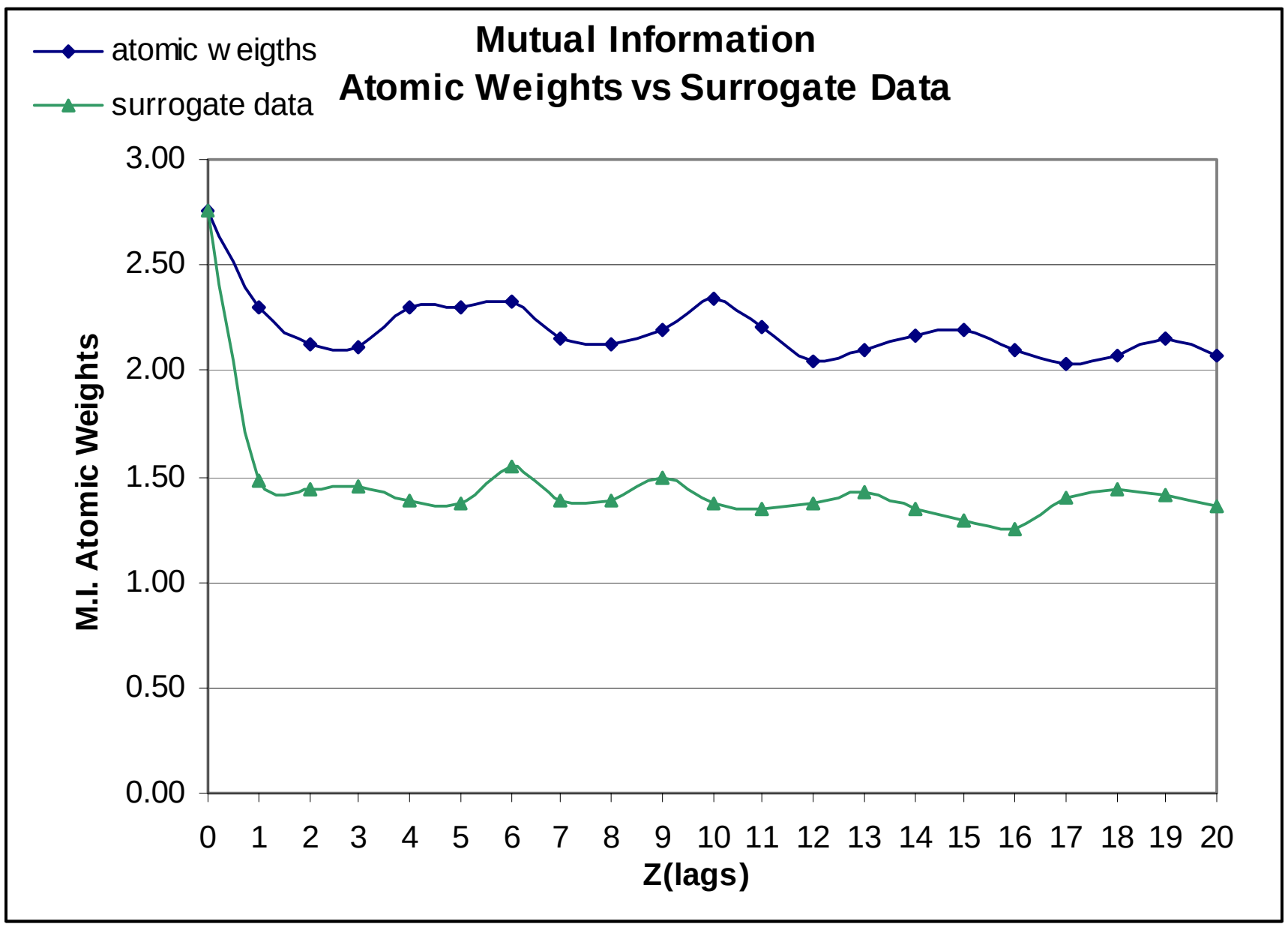

| Z-lags | M.I.-Surrogate Data |
|---|---|
| 0 | 2.76156 |
| 1 | 1.48621 |
| 2 | 1.44105 |
| 3 | 1.44755 |
| 4 | 1.38917 |
| 5 | 1.36923 |
| 6 | 1.54505 |
| 7 | 1.38976 |
| 8 | 1.38138 |
| 9 | 1.48849 |
| 10 | 1.36547 |
| 11 | 1.34689 |
| 12 | 1.37347 |
| 13 | 1.42643 |
| 14 | 1.34143 |
| 15 | 1.29609 |
| 16 | 1.25763 |
| 17 | 1.40470 |
| 18 | 1.43727 |
| 19 | 1.41390 |
| 20 | 1.36288 |

**Fig. 8 : Surrogate Data Analysis**

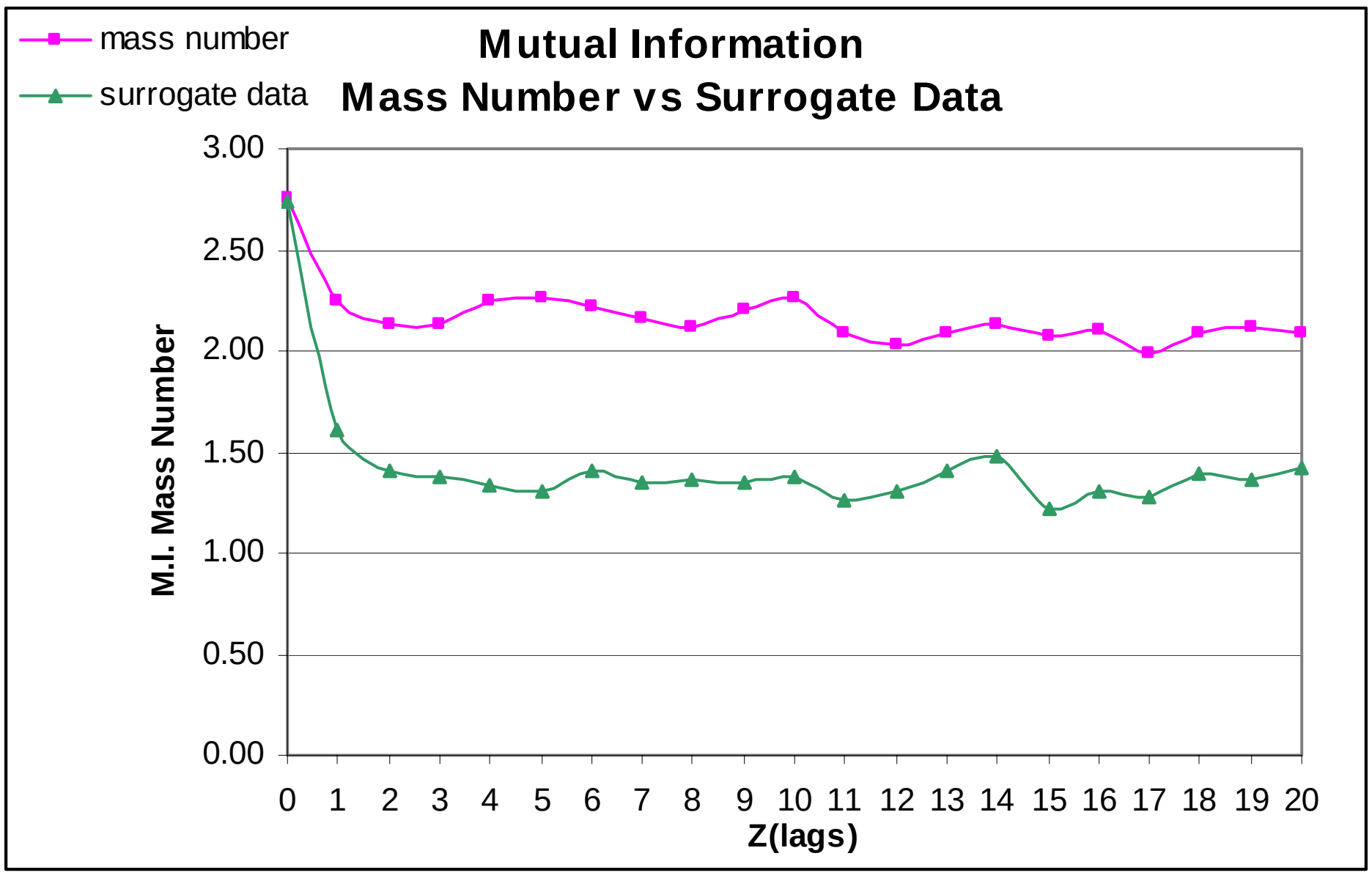

| Z-lags | M.I.-Surrogate Data |
|---|---|
| 0 | 2.73606 |
| 1 | 1.61453 |
| 2 | 1.39896 |
| 3 | 1.37041 |
| 4 | 1.33673 |
| 5 | 1.30566 |
| 6 | 1.41145 |
| 7 | 1.34616 |
| 8 | 1.35618 |
| 9 | 1.34365 |
| 10 | 1.37382 |
| 11 | 1.25994 |
| 12 | 1.30270 |
| 13 | 1.41078 |
| 14 | 1.47545 |
| 15 | 1.21417 |
| 16 | 1.31047 |
| 17 | 1.27582 |
| 18 | 1.38925 |
| 19 | 1.35851 |
| 20 | 1.41464 |

The results obtained by using shuffled data clearly confirm that we are in presence of a on linear mechanism in the process of increasing mass of atomic nuclei.
We also tested statistically the obtained differences between M.I. for original and surrogate data for the case of Atomic Weights as well as for the case of Mass Numbers. In the case of M.I for Atomic Weights vs M.I. Atomic Weights – Surrogate Data , using Unpaired t test we obtained a P value P<0.0001 and the same value was found in the case of M.I. Mass Number vs M.I. Mass Number – Surrogate Data.

In conclusion, by accepting the presence of non linearity, we have reached the first relevant conclusion of the present paper.

Looking now to Figures 3, 4, 5, 6 one may identify now new properties for atomic nuclei. Remember that we are considering each time, pairs of Atomic Weights or pairs of Mass Numbers or still pairs of Neutron Numbers for atomic nuclei with $Z - shifted$ values ranging from 1 to 20. What one should expect in this case is to find a minimum of Mutual Information followed soon after by a rather constant behaviour for M.I. Examination of the results reveals instead that we have some definite maxima and some definite minima at given values of $Z - shift$ that are quite different in the two and three cases that we have examined.

In detail the maxima for Atomic Weights are given at Z-shift values = 6,10,15,19,…. . Minima instead are given at Z-shift values =3,8,12,17,…… .

The maxima for Mass Numbers are given at Z-shift values=5,10,14,16(19) while the minima are given at Z-shift values =2,8,12,15,17. For Neutrons we have Z-shift values = 3,9,12,17 for the minima and 6,10,14,18 for the maxima. In conclusion: Still repeating here that each time we are exploring the M.I value for pairs of atomic weights, or of mass number or of neutrons, shifted in the $Z - value$ by some given values ranging between 1 and 20, we find that some pairs of nuclei show maxima MI values while other pairs of nuclei show minima MI value. Therefore we have new and interesting properties identified in atomic nuclei when analyzed by pairs as in the present methodology. We may call such new identified regularities for atomic nuclei as pseudo periodicities in pairs of atomic nuclei.

For Mass Numbers we may write as example that

$$\Delta A = \Delta N + Z - shift$$

Fixed a value of $Z$, we have consequently $A_1 = Z + N_1$. For an atomic nucleus with muss number $A_2$ and Z-shifted, we will have $A_2 = Z + Z - shift + N_2$.

Consequently, $\Delta A = A_2 - A_1 = \Delta N + Z - shift$ with $\Delta N = N_2 - N_1$. For Z-shift values=5,10,14,16(19), the considered pairs of atomic nuclei will show maxima of M.I.. Instead for Z-shift values =2,8,12,15,17, such M.I. values will reach a minimum value.

Let us go in more detail in the analysis.

First of all we have also to note that the values of MI, calculated for each $Z - shift$ ranging from 1 to 20, result to be quite different in $W_a(Z)$ respect to $A(Z)$, and N(Z). In addition, as previously said, Mutual Information measures how much, given two random variables, and knowing one of these two variables, is reduced our uncertainty about the other. Mutual Information must thus be intended essentially as estimation of mutual dependence of two variables. In our case we find that the pairs of atomic weights, or of mass number or of neutrons in atomic nuclei that are shifted by some definite values of the atomic number $Z$, show strong dependence (maxima values of M.I.) or, respectively, they show very low dependence (minima values of M.I.). We have some new pseudoperiodicities that in some manner recall a new kind of pseudo isotopies. All that seems to be realized in a full regime of non linearity.

**4. Phase Space Reconstruction of $W_a(Z)$ and $A(Z)$.**

We may now attempt to obtain for the first time a phase space reconstruction of Atomic Weights and Mass Number of atomic nuclei.

To reach this objective one must estimate Embedding Dimension using the False Nearest Neighbors Criterion (FFN). We applied it using a $Z - shift$ = 3 for $W_a(Z)$ and a $Z - shift$ =2 for $A(Z)$ as previously found. A false criterion distance was considered to be 4.42 for both the cases of the analysis. The results are reported in Fig.9 for atomic weights, $W_a(Z)$, and in Fig.10 for Mass Number, $A(Z)$.

**Fig. 9 :False Nearest Neighbors for Atomic Weights**

**Fig.10 : False Nearest Neighbors for Mass Number**

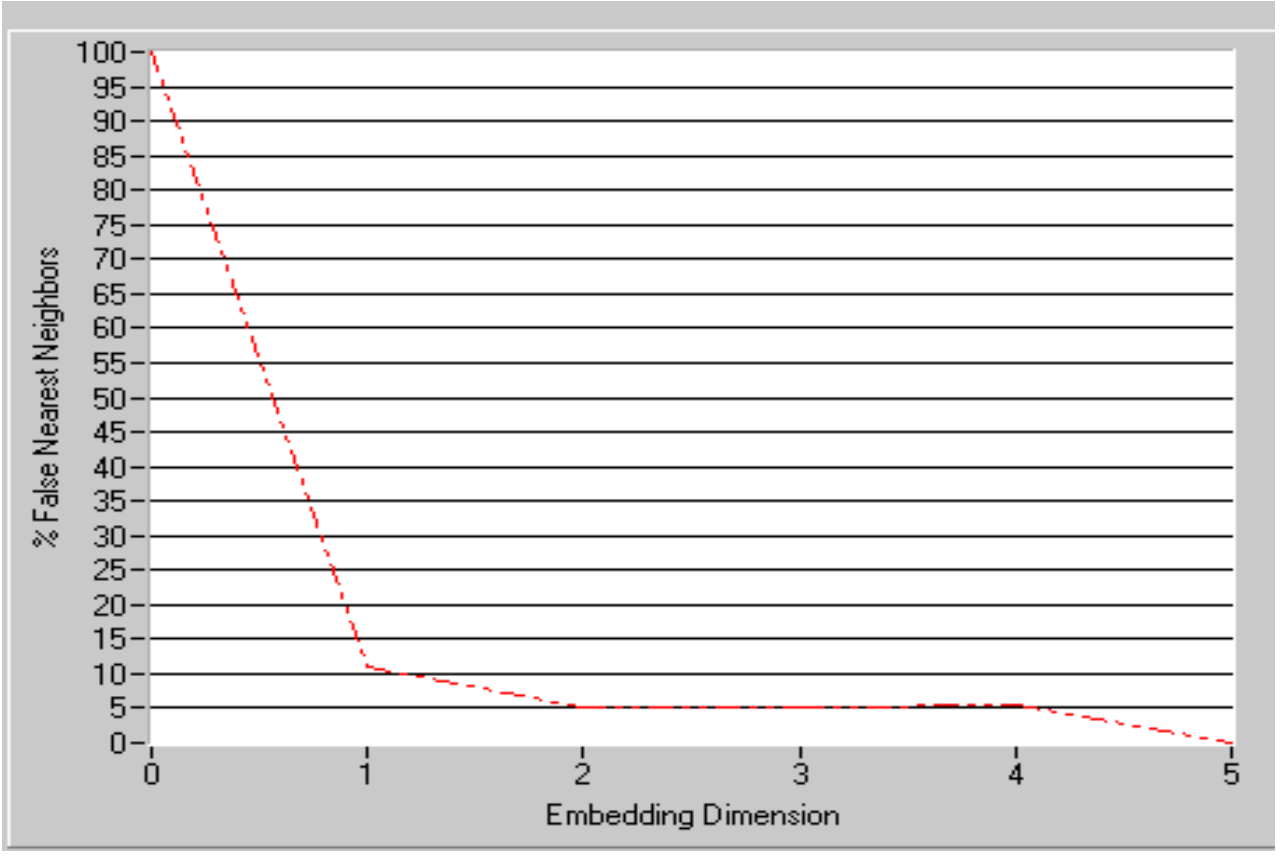

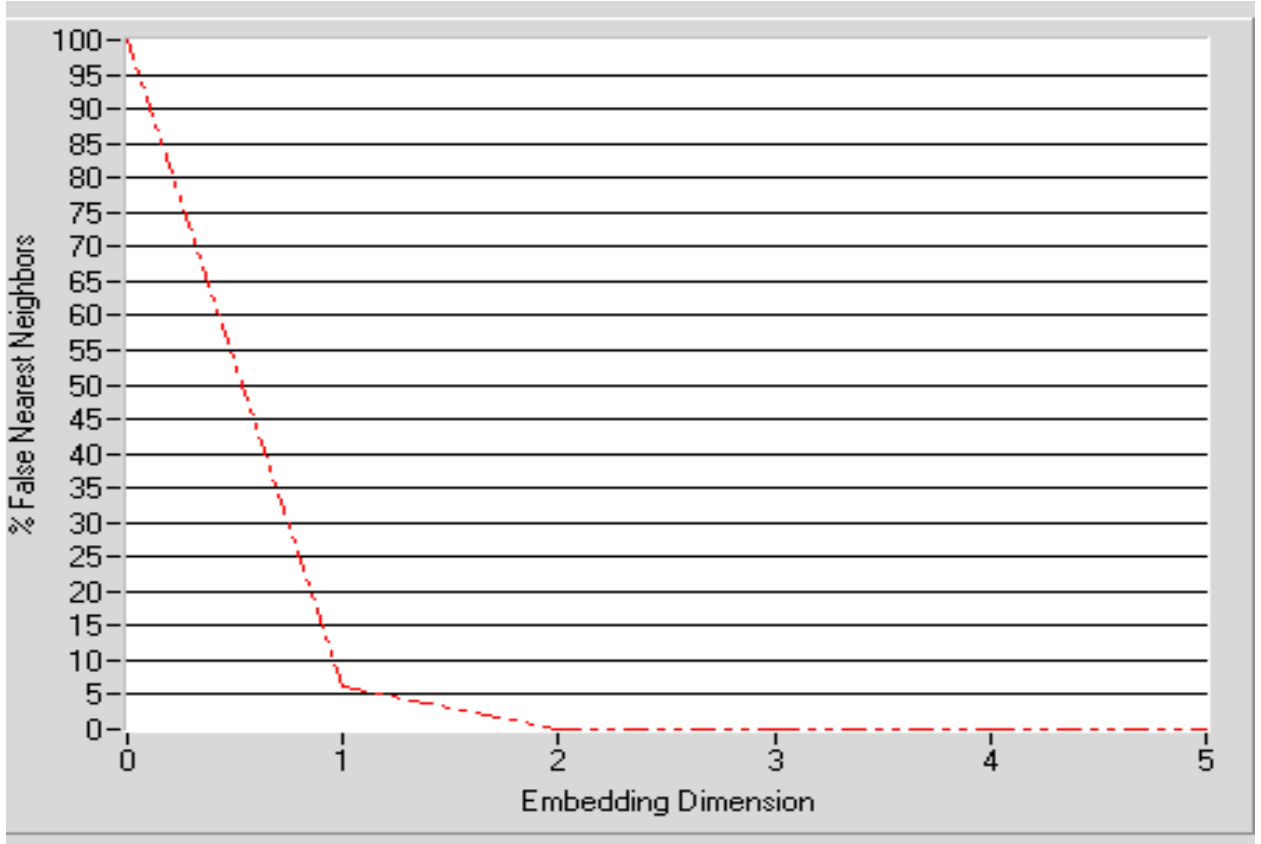

The evaluation of the results given in Figures 9, 10 enables us to conclude that the phase space reconstruction for atomic weights and mass number requires an estimated embedding dimension ,that results to be included between 1 and 2. We may assume to consider $D = 2$. Atomic weights and mass number of atomic nuclei may be approximately represented in a bi dimensional phase space. Consequently, according to the general framework of the theory on non linear dynamics of systems, we may conclude that a very few number of independent variables is required in order to describe the mechanism of increasing mass of atomic nuclei. We may accept to consider that they are two variables that, with greatest prudence, we may accept to identify as being the proton and the neutron numbers, respectively. The phase space description of atomic weights $W_a(Z)$ and of Mass Number $A(Z)$ requires with approximation, the use of such two variables.

Since this result has been obtained in a closed form, we may now attempt to analyze if the two given $W_a(Z)$ and $A(Z)$ exhibit or not properties of divergence.

To this purpose we may calculate Lyapunov spectrum in the embedded phase space. The results that we obtained, are reported in Fig. 11 and in Fig.12 for $W_a(Z)$ and $A(Z)$, respectively.

**Fig.11 : Lyapunov spectrum of atomic weights**

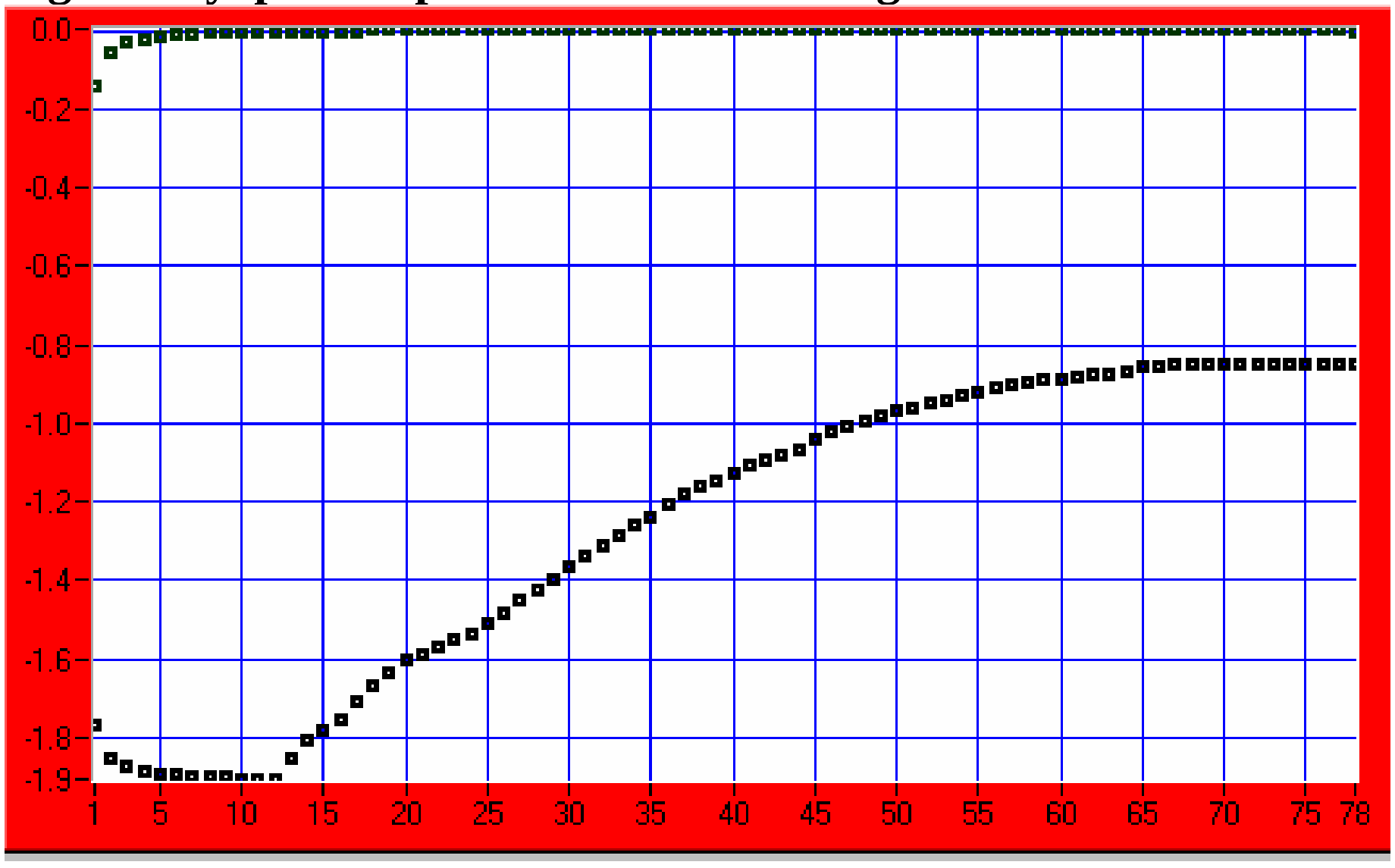

| iteration, | exponents | |
|---|---|---|
| 1 | -0.139132 | -1.767928 |
| 2 | -0.054642 | -1.852418 |
| 3 | -0.032002 | -1.875058 |
| 4 | -0.021497 | -1.885562 |

| iteration, | exponents | |
|---|---|---|
| 5 | -0.015293 | -1.891767 |
| 6 | -0.011169 | -1.895891 |
| 7 | -0.008224 | -1.898835 |
| 8 | -0.006016 | -1.901043 |

| iteration, | exponents | |
|---|---|---|
| 9 | -0.004299 | -1.902761 |
| 10 | -0.002925 | -1.904134 |
| 11 | -0.001801 | -1.905258 |
| 12 | -0.000864 | -1.906195 |
| 13 | -0.000946 | -1.854474 |
| 14 | -0.000219 | -1.810264 |
| 15 | 0.000026 | -1.779827 |
| 16 | 0.000273 | -1.753227 |
| 17 | 0.000425 | -1.705629 |
| 18 | 0.000914 | -1.667297 |
| 19 | 0.001349 | -1.632997 |
| 20 | 0.001738 | -1.602124 |
| 21 | 0.002186 | -1.585933 |
| 22 | 0.002567 | -1.570617 |
| 23 | 0.002785 | -1.552573 |
| 24 | 0.003007 | -1.536054 |
| 25 | 0.003233 | -1.508474 |
| 26 | 0.003433 | -1.483006 |
| 27 | 0.003304 | -1.450959 |
| 28 | 0.003100 | -1.421118 |
| 29 | 0.003155 | -1.394447 |
| 30 | 0.002734 | -1.365179 |
| 31 | 0.002481 | -1.336786 |
| 32 | 0.002367 | -1.309490 |
| 33 | 0.002297 | -1.283733 |
| 34 | 0.002104 | -1.261030 |
| 35 | 0.002071 | -1.235585 |
| 36 | 0.002229 | -1.208038 |
| 37 | 0.002491 | -1.182091 |
| 38 | 0.002558 | -1.161598 |
| 39 | 0.002648 | -1.143582 |
| 40 | 0.002781 | -1.125369 |
| 41 | 0.002912 | -1.108048 |
| 42 | 0.003353 | -1.094097 |
| 43 | 0.003464 | -1.078832 |
| 44 | 0.003659 | -1.064133 |

| iteration, | exponents | |
|---|---|---|
| 45 | 0.003833 | -1.043353 |
| 46 | 0.003724 | -1.022886 |
| 47 | 0.003713 | -1.008188 |
| 48 | 0.003649 | -0.995043 |
| 49 | 0.003576 | -0.982424 |
| 50 | 0.003419 | -0.970124 |
| 51 | 0.003116 | -0.958012 |
| 52 | 0.002781 | -0.948094 |
| 53 | 0.002434 | -0.938526 |
| 54 | 0.002281 | -0.929092 |
| 55 | 0.002133 | -0.920000 |
| 56 | 0.001982 | -0.911222 |
| 57 | 0.001851 | -0.902767 |
| 58 | 0.001885 | -0.897098 |
| 59 | 0.001945 | -0.891648 |
| 60 | 0.002004 | -0.886751 |
| 61 | 0.002097 | -0.882050 |
| 62 | 0.002113 | -0.877724 |
| 63 | 0.002311 | -0.872891 |
| 64 | 0.002398 | -0.865438 |
| 65 | 0.002469 | -0.858201 |
| 66 | 0.002539 | -0.854135 |
| 67 | 0.002586 | -0.850170 |
| 68 | 0.002372 | -0.850370 |
| 69 | 0.002156 | -0.850555 |
| 70 | 0.001942 | -0.850731 |
| 71 | 0.001733 | -0.850900 |
| 72 | 0.001529 | -0.851065 |
| 73 | 0.001330 | -0.851224 |
| 74 | 0.001136 | -0.851379 |
| 75 | 0.000947 | -0.851530 |
| 76 | 0.000763 | -0.851676 |
| 77 | 0.000585 | -0.851819 |
| 78 | 0.000410 | -0.851958 |

**Fig.12 : Lyapunov spectrum of mass number**

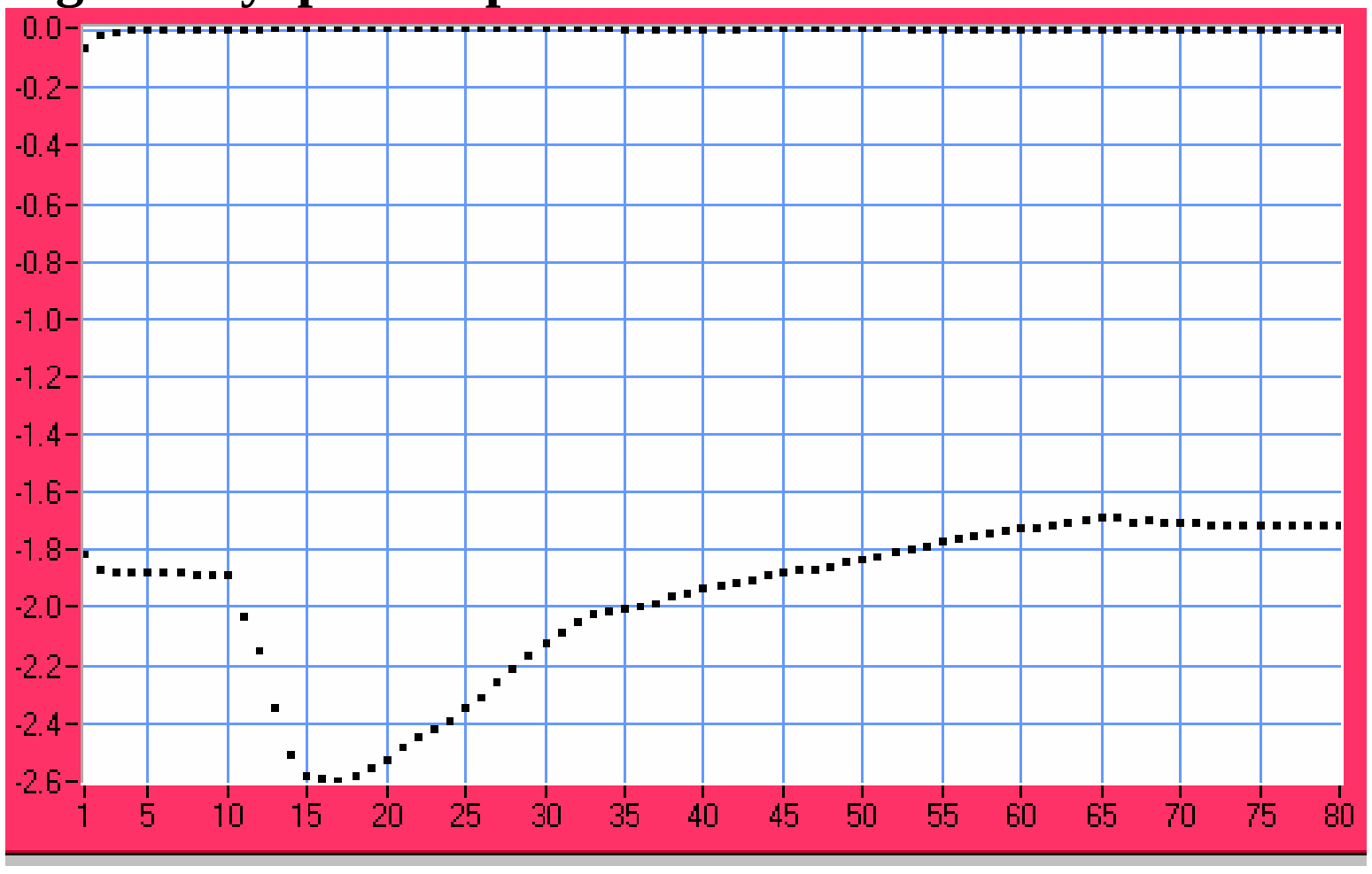

| iteration, | exponents | |
|---|---|---|
| 1 | -0.063750 | -1.815136 |
| 2 | -0.013655 | -1.865231 |
| 3 | -0.004496 | -1.874390 |

| iteration, | exponents | |
|---|---|---|
| 4 | -0.000940 | -1.877946 |
| 5 | 0.001067 | -1.879953 |
| 6 | 0.002390 | -1.881276 |

| iteration, | exponents | |
|---|---|---|
| 7 | 0.003333 | -1.882218 |
| 8 | 0.004039 | -1.882925 |
| 9 | 0.004589 | -1.883475 |
| 10 | 0.005029 | -1.883914 |
| 11 | 0.005907 | -2.030549 |
| 12 | 0.005902 | -2.152007 |
| 13 | 0.007184 | -2.344940 |
| 14 | 0.007900 | -2.508600 |
| 15 | 0.008538 | -2.585360 |
| 16 | 0.008446 | -2.593648 |
| 17 | 0.008353 | -2.600949 |
| 18 | 0.007751 | -2.585976 |
| 19 | 0.008750 | -2.554168 |
| 20 | 0.008150 | -2.531907 |
| 21 | 0.008186 | -2.482481 |
| 22 | 0.007894 | -2.450329 |
| 23 | 0.008806 | -2.420931 |
| 24 | 0.009633 | -2.393973 |
| 25 | 0.009816 | -2.350399 |
| 26 | 0.009954 | -2.310147 |
| 27 | 0.009880 | -2.258690 |
| 28 | 0.009662 | -2.210759 |
| 29 | 0.009018 | -2.163469 |
| 30 | 0.008250 | -2.119600 |
| 31 | 0.007832 | -2.082773 |
| 32 | 0.007498 | -2.052309 |
| 33 | 0.007109 | -2.023616 |
| 34 | 0.006523 | -2.013469 |
| 35 | 0.005691 | -2.003623 |
| 36 | 0.005143 | -1.995782 |
| 37 | 0.004718 | -1.988460 |
| 38 | 0.003966 | -1.963859 |
| 39 | 0.004116 | -1.946858 |
| 40 | 0.004339 | -1.930786 |
| 41 | 0.004649 | -1.922343 |
| 42 | 0.005367 | -1.917894 |
| 43 | 0.007274 | -1.903393 |

| iteration, | exponents | |
|---|---|---|
| 44 | 0.008790 | -1.889480 |
| 45 | 0.010447 | -1.876392 |
| 46 | 0.010388 | -1.871786 |
| 47 | 0.009970 | -1.869670 |
| 48 | 0.009374 | -1.855670 |
| 49 | 0.008541 | -1.841986 |
| 50 | 0.007498 | -1.829730 |
| 51 | 0.006726 | -1.821934 |
| 52 | 0.005940 | -1.808215 |
| 53 | 0.005215 | -1.795046 |
| 54 | 0.004497 | -1.783516 |
| 55 | 0.004030 | -1.771905 |
| 56 | 0.003596 | -1.760725 |
| 57 | 0.002953 | -1.750916 |
| 58 | 0.002916 | -1.743006 |
| 59 | 0.002961 | -1.735446 |
| 60 | 0.003004 | -1.728808 |
| 61 | 0.003032 | -1.722373 |
| 62 | 0.003108 | -1.713120 |
| 63 | 0.003229 | -1.704206 |
| 64 | 0.003291 | -1.693598 |
| 65 | 0.003299 | -1.689223 |
| 66 | 0.003127 | -1.691107 |
| 67 | 0.002824 | -1.702828 |
| 68 | 0.002620 | -1.700615 |
| 69 | 0.002292 | -1.708755 |
| 70 | 0.002307 | -1.709571 |
| 71 | 0.002341 | -1.710383 |
| 72 | 0.002445 | -1.714966 |
| 73 | 0.002536 | -1.719414 |
| 74 | 0.002212 | -1.719409 |
| 75 | 0.001877 | -1.719385 |
| 76 | 0.001547 | -1.719358 |
| 77 | 0.001225 | -1.719332 |
| 78 | 0.000912 | -1.719306 |
| 79 | 0.000606 | -1.719280 |
| 80 | 0.000308 | -1.719255 |

To calculate the Lyapunov spectrum in the case of the Atomic Weights, $W_a(Z)$, we used a number of 22 fitted points in the embedded phase space.
These are the following results for the calculated Lyapunov exponents:
$\lambda_1$ = 0.000410 and $\lambda_2$ = -0.851958. It is seen that we have $\lambda_1 > 0$ and $\lambda_2 < 0$ with $\lambda_1 + \lambda_2 < 0$ as required. In conclusion we are in presence of a divergent system and such divergence may be indicative a pure chaotic regime for Atomic Weights.
In the case of the Mass Number, $A(Z)$, we utilized a number of 17 fitted points in the embedded phase space. These are the results we obtained for the calculated Lyapunov exponents: $\lambda_1 = 0.000308$, $\lambda_2 = -1.719255$ with $\lambda_1 > 0$, $\lambda_2 < 0$ and $\lambda_1 + \lambda_2 < 0$. Also in the case of Mass Number, $A(Z)$, we are in presence of a divergent system and it could be indicative of a pure chaotic regime.
In brief, we have reached the following conclusions:

1) In the process of increasing mass of atomic nuclei we are in presence of a non linear mechanism. Remember that the presence of non linear contributions in the dynamic of a system gives often origin to chaotic regimes.
2) The mechanism of increasing mass in Atomic Weights and in Mass Number for pairs of atomic nuclei also exhibits some pseudo periodicities at some definite $Z - shifts$ of atomic nuclei.

Therefore, we could be in presence of an ordered regime of increasing mass but in the complex of a whole structure that is divergent and possibly chaotic.

3) A phase space reconstruction has been realized for Atomic Weights and Mass Number of atomic nuclei, respectively. In our opinion this is a relevant result that is obtained here. In fact, from the analysis performed by using F.F.N, it does not result in a so clear manner that the reconstructed phase space has dimension D=2. We have F.F.N. values that suspend embedding dimension between 1 and 2. In this case in the analysis it is adopted the greatest value. In conclusion we may accept an embedding dimension D=2, and only in this condition we have that only few, two variables, are required in order to describe the mechanism of increasing mass of atomic nuclei in phase space with respect to atomic weights and mass number . The first variable should be the atomic number, $Z$, and the other variable should be the Neutron Number , $N = A - Z$.
4) The analysis of $W_a(Z)$ and $A(Z)$ reveals a new important features when we analyze such two systems by calculation of Lyapunov spectrum. It results that we are in presence of divergent systems in both case of stable nuclei analyzed by $W_a(Z)$ and $A(Z)$. Such divergent property, linked to the previously found on non linearity, could be indicative that we are in presence of a chaotic regime in the mechanism of the increasing mass of atomic nuclei when seen as function of $Z$.

We may go on by a further step calculating the Correlation Dimension in the reconstructed phase space for both $W_a(Z)$ and $A(Z)$. In Fig.13 we report the results for $W_a(Z)$. In Fig. 14 we give instead the results for $A(Z)$. Finally in Figures 15 and 16 we have the results using surrogate (shuffled) data.

**Fig.13 : Atomic Weights**.

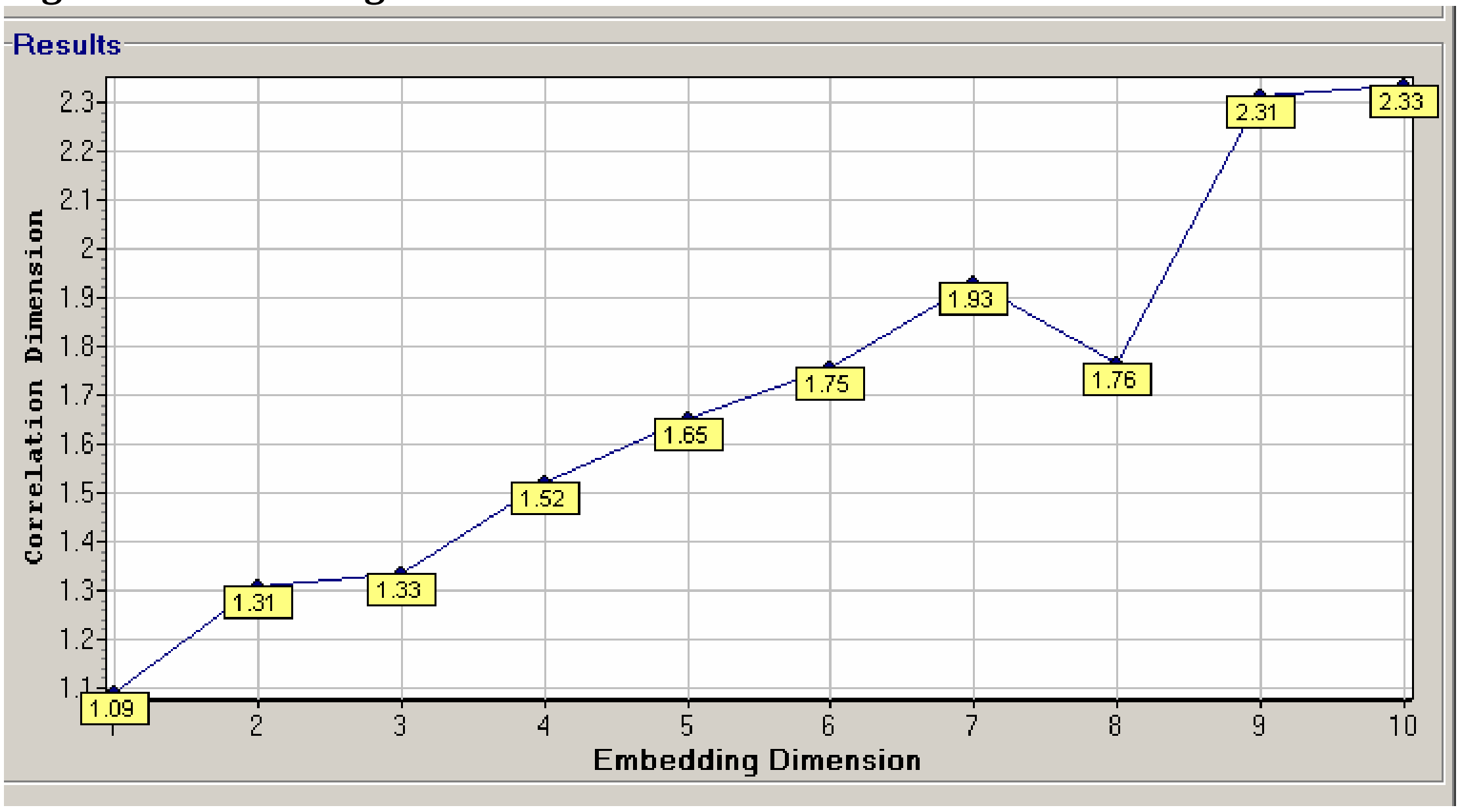

**Fig.14 : Mass Number.**

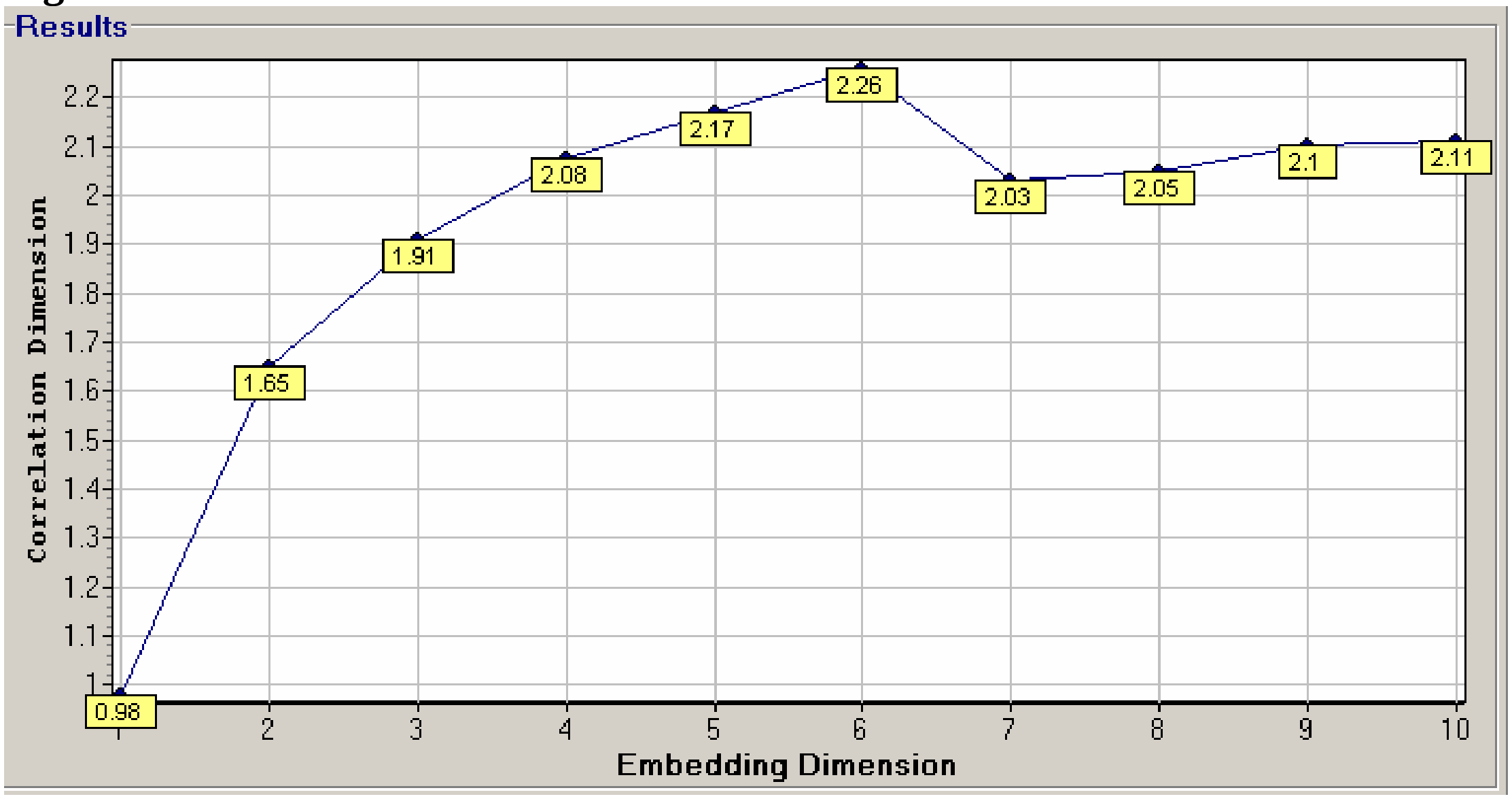

For the atomic weights it is obtained $D_2 = 1.955 \pm 0.296$ as value for Correlation Dimension. For Mass Number it results instead $D_2 = 2.120 \pm 0.084$.

It is important to observe that in both case we obtain non integer values of such topological dimension in phase space reconstruction.

We may now consider the results for surrogate data.

**Fig.15 : Results on Correlation Dimension, Atomic Weights** (surrogate data).

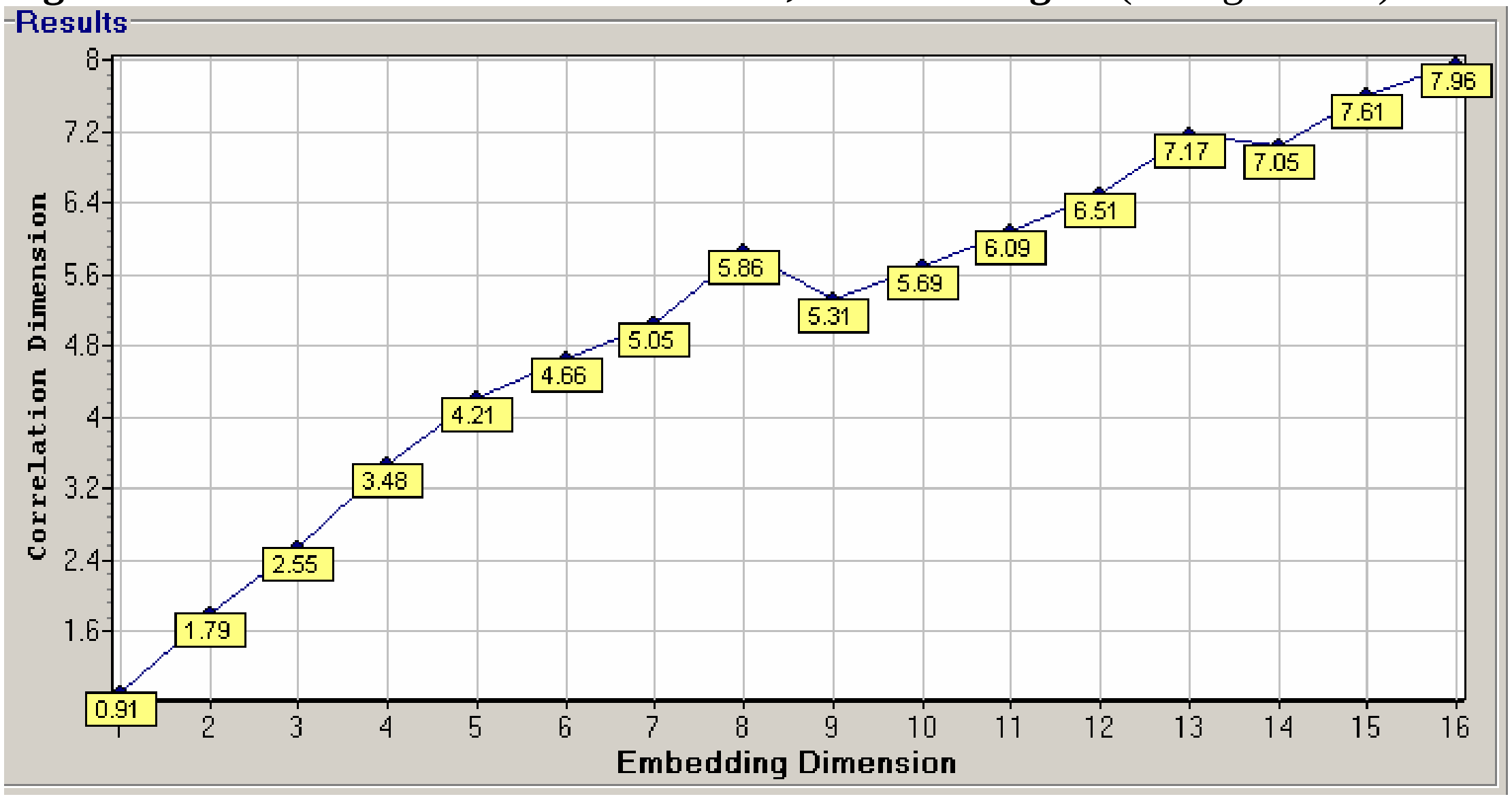

**Fig.16 : Results on Correlation Dimension, Mass Number** (surrogate data).

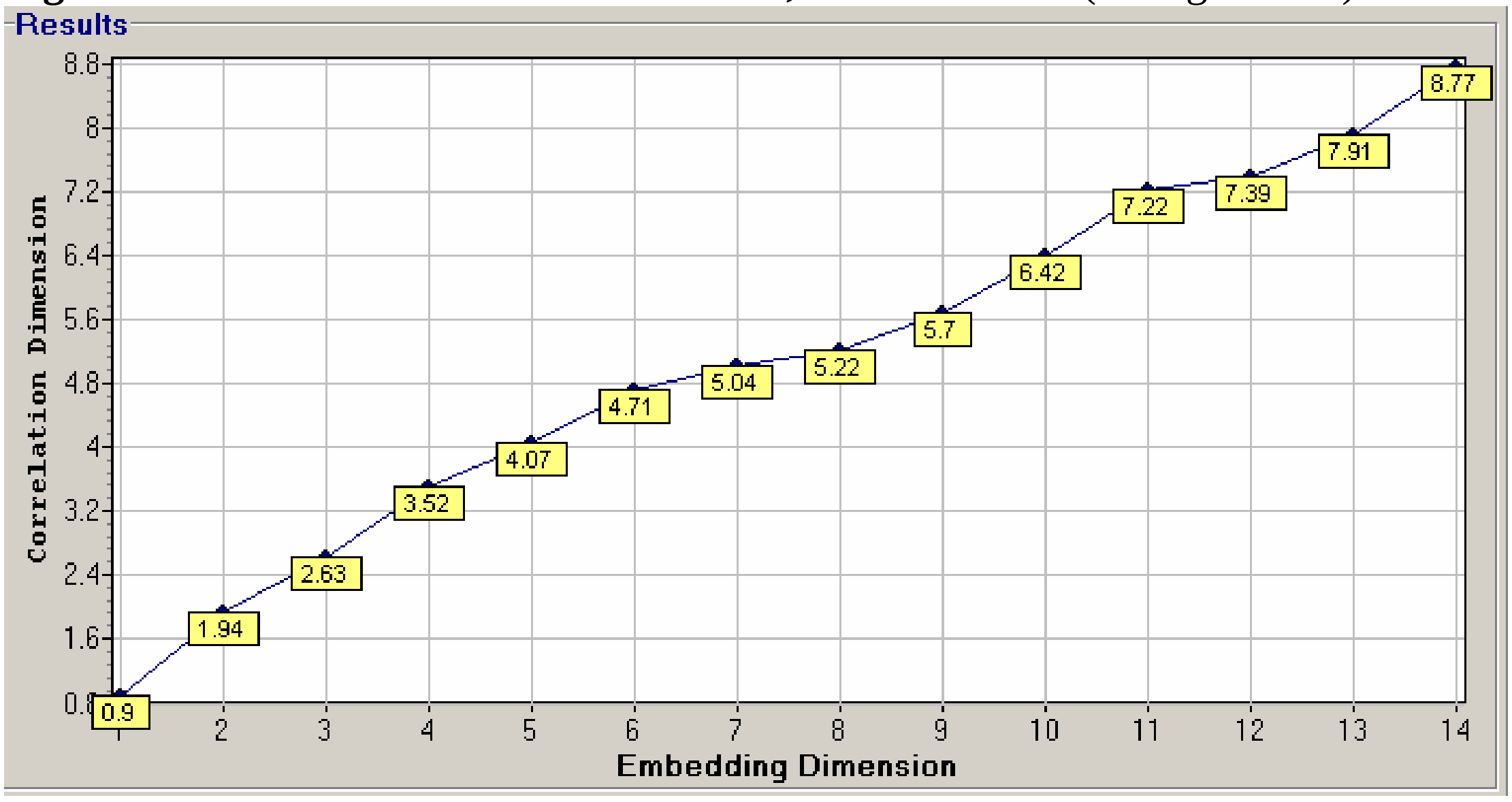

In the case of $W_a(Z)$ we obtain $D_2 = 5.130 \pm 0.624$ while instead in the case of $A(Z)$ we have $D_2 = 5.193 \pm 0.810$

As seen through the results, a net difference is obtained in comparison of original with surrogate data. They may be quantified in the following manner:

$$S = \frac{\left|D_{2,original} - D_{2,surrogate}\right|}{\sigma_{surrogate}} = 3.678(Atomic\ Weights\ Z - lag = 2), 5.088\,(Atomic\ Weights\ Z - lag = 4),$$

$$3.793(\ Mass\ Number\ Z - lag = 2)$$

The null hypothesis may be rejected. In conclusion we have a non integer topological dimension for both $W_a(Z)$ and $A(Z)$.

Therefore it has a sense to attempt to ascertain if we are in presence of a fractal behaviour for both the system of data that we have in examination.

## 5. On a Possible Existing Power Law to Represent Increasing Mass in Atomic Weights and Mass Number of Atomic Nuclei.

As we indicated in the introduction in the present paper, rather recently V. Paar et al [6] introduced a power law for description of the line of stability of atomic nuclei, and in particular for the description of atomic weights. They compared the found power law with the semi-empirical formula of the liquid drop model, and showed that the power law corresponds to a reduction of neutron excess in superheavy nuclei with respect to the semi-empirical formula. Some fractal features of the nuclear valley of stability were analyzed and the value of fractal dimension was determined.

It is well known that a power law may be often connected with an underlying fractal geometry of the considered system. If confirmed for atomic nuclei, according to [5], it could be proposed a new approach to the problem of stability of atomic nuclei. In this case the aim should be to identify the basic features in underlying dynamics giving rise to the structure of the atomic nuclei.

The aim is to perform here such kind of analysis for $W_a(Z)$ and $A(Z)$.

For atomic weights let us introduce the following Power Law :

$$W_a(Z) = aZ^{\beta} \qquad (5)$$

while instead for the Mass Number let us introduce the following power law

$$A(Z) = cZ^{\gamma} \qquad (6)$$

The problem is now to estimate $(a, \beta)$ and $(c, \gamma)$ by a fitting procedure.

We give here the obtained results for the Atomic Weights.

**Curve Fit Report**

Y Variable: C2. X Variable: C1. Model Fit: A*X^B

**Parameter Estimates for All Groups**

| **Groups** | **Count** | **Iter's** | **R2** | **A** | **B** |
|---|---|---|---|---|---|
| All | 83 | 24 | 0.99954 | 1.47335 | 1.12133 |

**Combined Plot Section**

**Fig. 17**

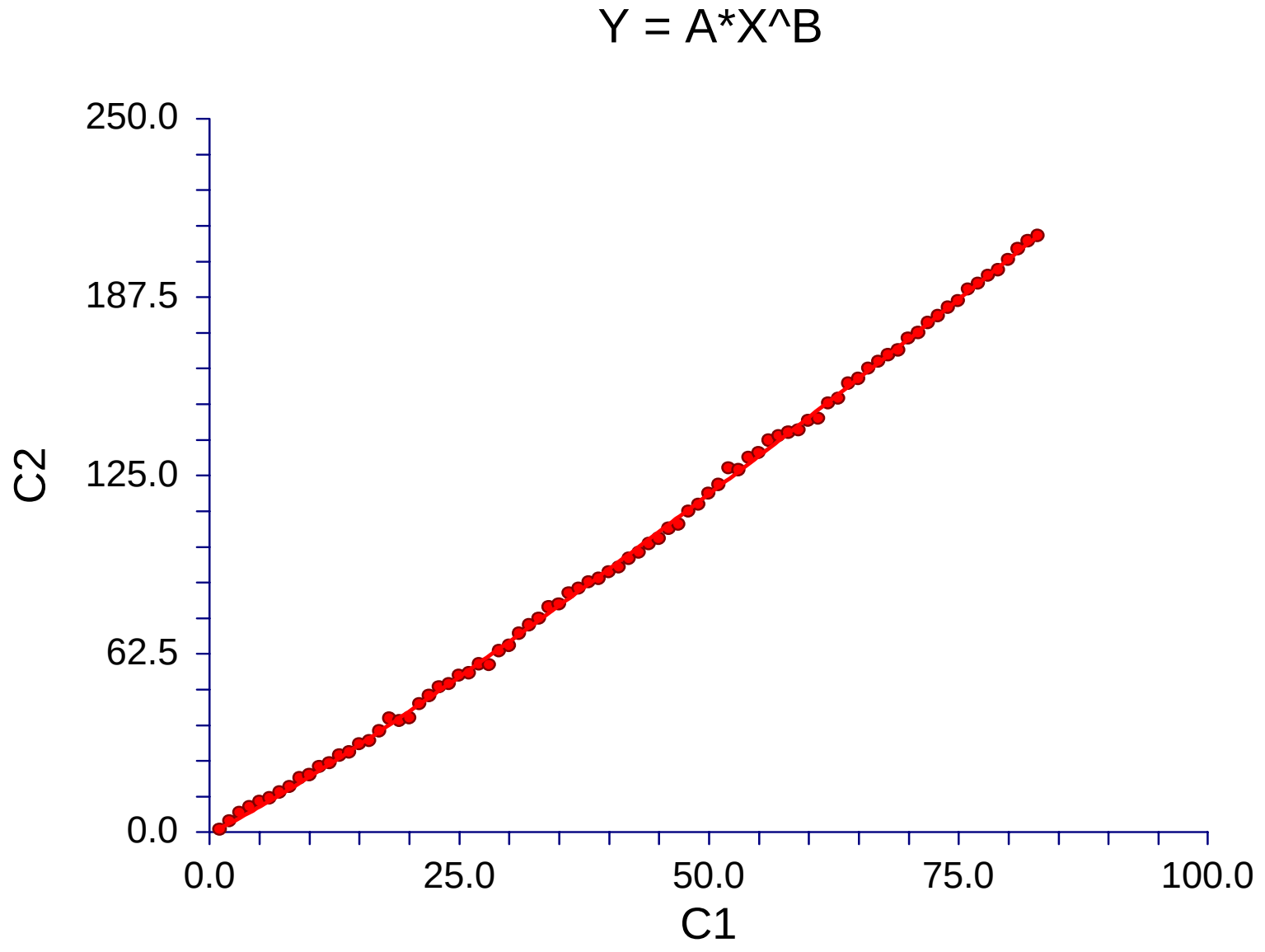

**Model Estimation Section**

| **Parameter Name** | **Parameter Estimate** | **Asymptotic Standard Error** | **Lower 95% C.L.** | **Upper 95% C.L.** |
|---|---|---|---|---|
| A | 1.47335 | 0.02448 | 1.42464 | 1.52206 |
| B | 1.12133 | 0.00399 | 1.11338 | 1.12927 |
| Iterations | 24 | Rows Read | 83 | |
| R-Squared | 0.999538 | Rows Used | 83 | |
| Random Seed | 9839 | Total Count | 83 | |

**Estimated Model**

**Curve Fit Report**

Y Variable: C2. X Variable: C1.

**Plot Section**

**Fig.18**

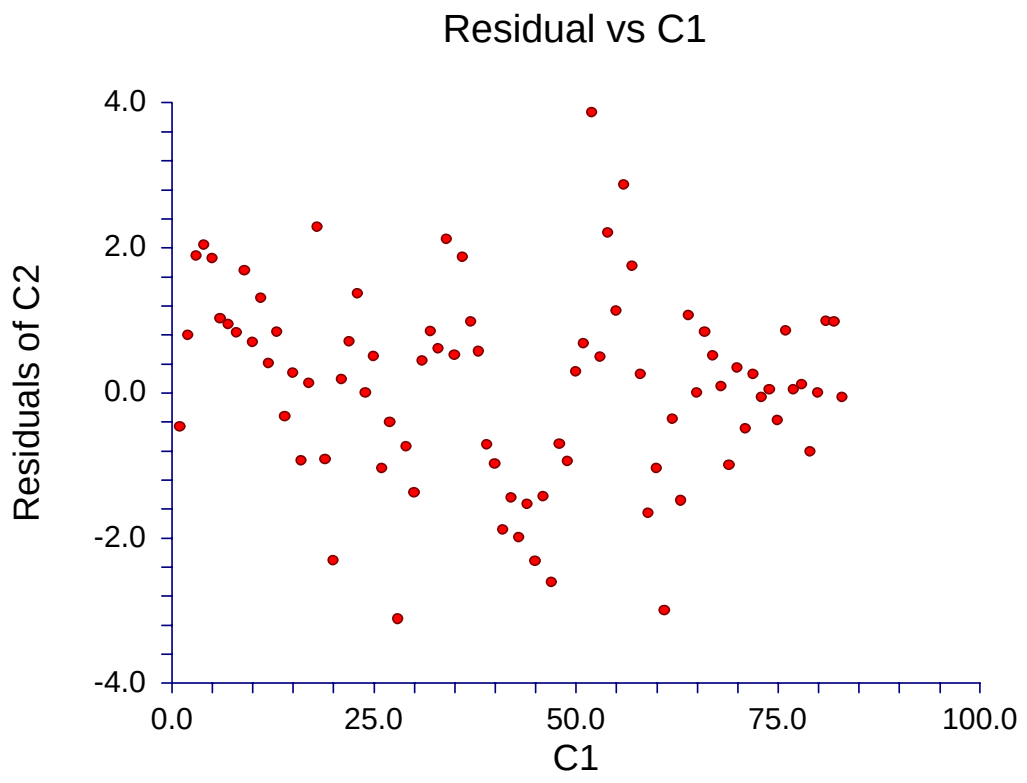

In conclusion, for the (5) we obtained a = 1.47335 and β = 1.12133.
V. Paar et al. [5] obtained instead that a =1.44±0.02 and β=1.120±0.004 and β=1.19±0.01 by using the Box Counting method. There is an excellent agreement.
As it may be seen the obtained values significantly differ from the line. In addition, the obtained values strongly give evidence for a possible fractal regime.
Let us see now the results that we obtained for the (6) concerning Mass Number.

**Curve Fit Report**
Y Variable: C2. X Variable: C1.
Model Fit: A*X^B
**Parameter Estimates for All Groups**

| Groups | Count | Iter's | R2 | A | B |
|---|---|---|---|---|---|
| All | 83 | 23 | 0.99929 | 1.46185 | 1.12389 |

**Combined Plot Section**

**Fig.19**

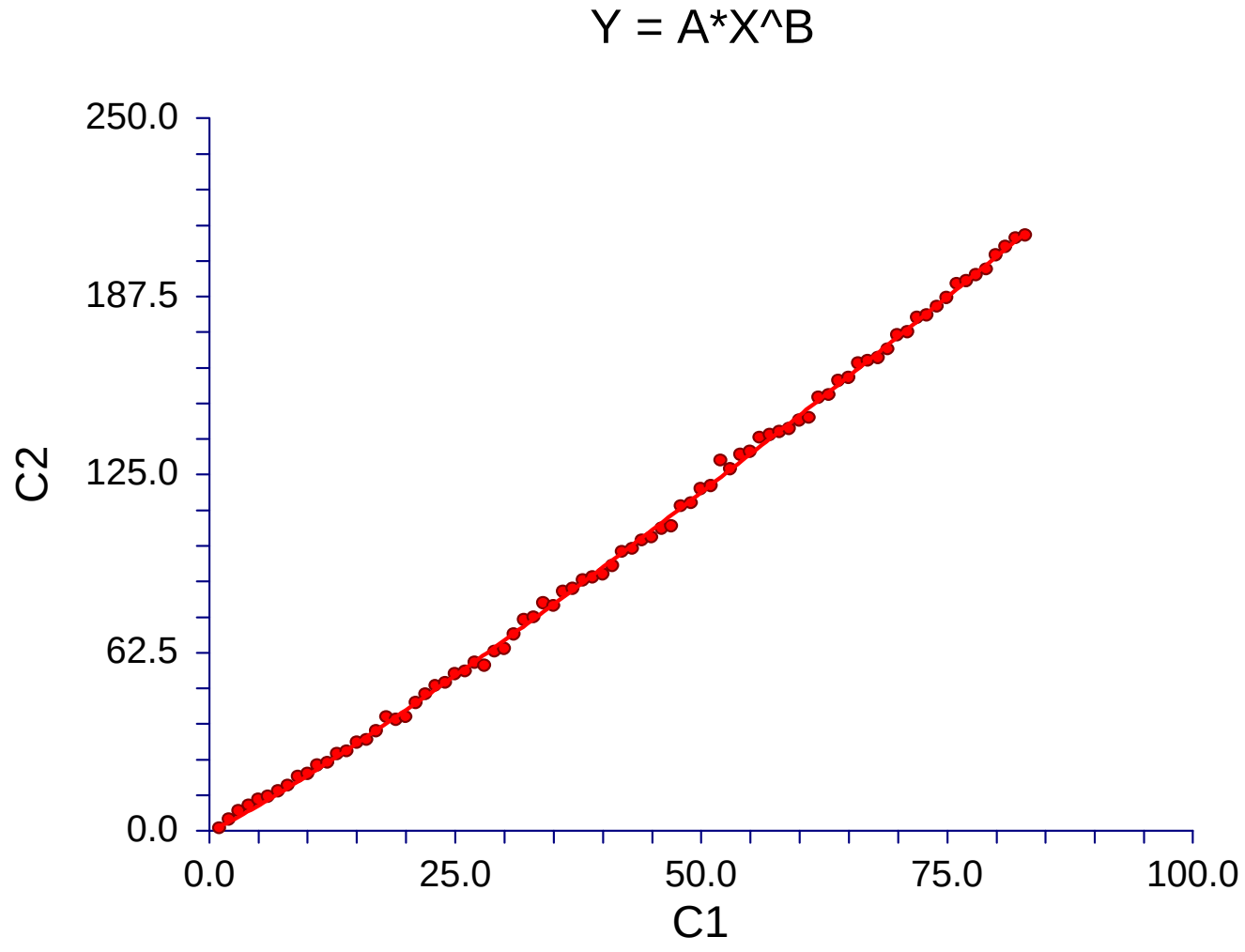

**Model Estimation Section**

| Parameter Name | Parameter Estimate | Asymptotic Standard Error | Lower 95% C.L. | Upper 95% C.L. |
|---|---|---|---|---|
| A | 1.46185 | 0.03010 | 1.40195 | 1.52174 |
| B | 1.12389 | 0.00495 | 1.11404 | 1.13373 |
| Iterations | 23 | Rows Read | 83 | |
| R-Squared | 0.999294 | Rows Used | 83 | |
| Random Seed | 10007 | Total Count | 83 | |

**Estimated Model**

**Curve Fit Report**

Y Variable: C2. X Variable: C1.
**Plot Section**

**Fig.20**

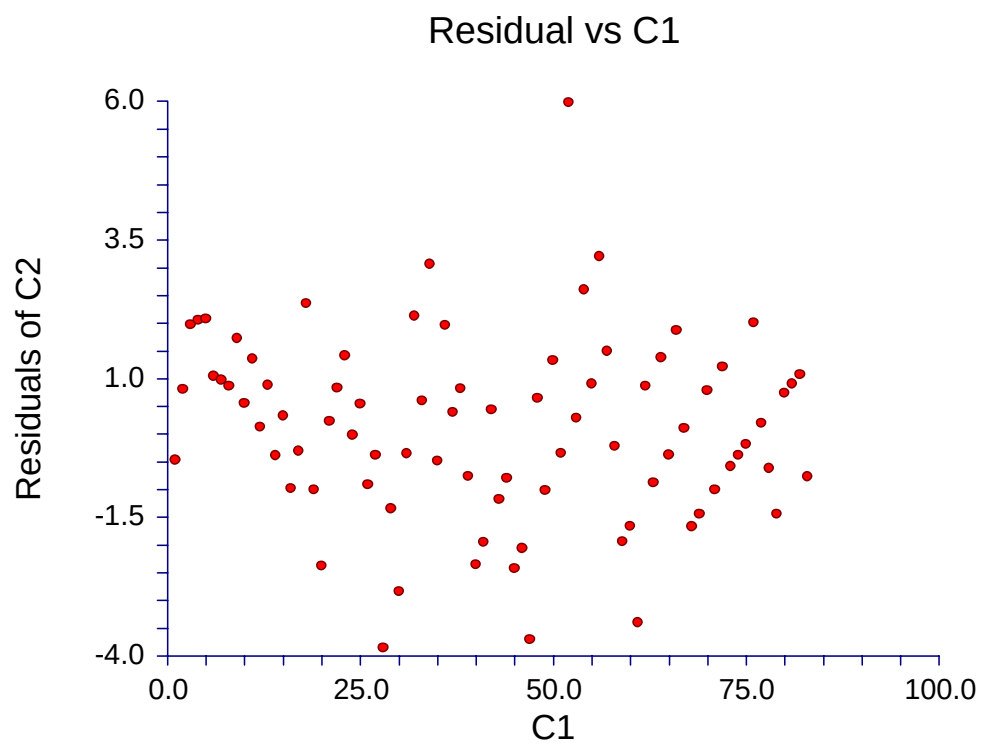

In conclusion, for the (6) we obtained: c=1.46185, $\gamma$=1.12389.
V. Paar et al. obtained [6] c=1.47±0.02 and $\gamma$=1.123±0.005 in excellent agreement.
Again we may conclude for values that significantly differ from line. In addition, the obtained values strongly give evidence for a possible fractal regime.
The possible existence of a fractal regime in the mechanism of increasing mass in atomic weights and mass umber of atomic nuclei changes radically our traditional manner to conceive nuclear matter. Consequently, it becomes of relevant importance to attempt to deepen such result so to reach the highest possible level of certainty on it. A way to deepen such kind of analysis is to follow the way of variogram method. Variograms usually give powerful indications on the variability of the examined data, on their self-similarity and self-affine behaviour. In particular, they enable us to calculate the Generalized Fractal Dimension [ for details see ref.11].
The semivariogram is given in the following manner

$$\gamma(h) = \frac{E(R(x) - R(x+h))^2}{2} \qquad (7)$$

For Atomic Weights it is:

$$\gamma(h) = \frac{E(W_a(Z) - W_a(Z+h))^2}{2} \qquad (8)$$

and for Mass Number it is

$$\gamma(h) = \frac{E(M_n(Z) - M_n(Z+h))^2}{2}$$ . $Z - shift$ is indicated here by $h = 1, 2, .....$ . (9)

Still, in the general case we may write

$$\gamma(h) = \frac{1}{2(N-h)} \sum_{i=1}^{N-h} (R(x_i + h) - R(x_i))^2 \qquad (10)$$

For a self-affine series the semivariogram scales according to

$$\gamma(h) = C\, h^D \qquad (11)$$

being $D$ the Generalized Fractal Dimension. It is linked to $H_a$ by $D = 2H_a$ being $H_a$ the Hausdorff dimension.
We may also estimate the corresponding Probability Density Function that is given in the following manner

$$P(h) = \frac{a}{k^a} h^{a-1} \qquad (12)$$

being $D = a - 1$ and $k$ is a scale parameter.
Let us introduce now the results that we obtained for Atomic Weights.

## RESULTS

**Fig. 21: Variogram of atomic weights**

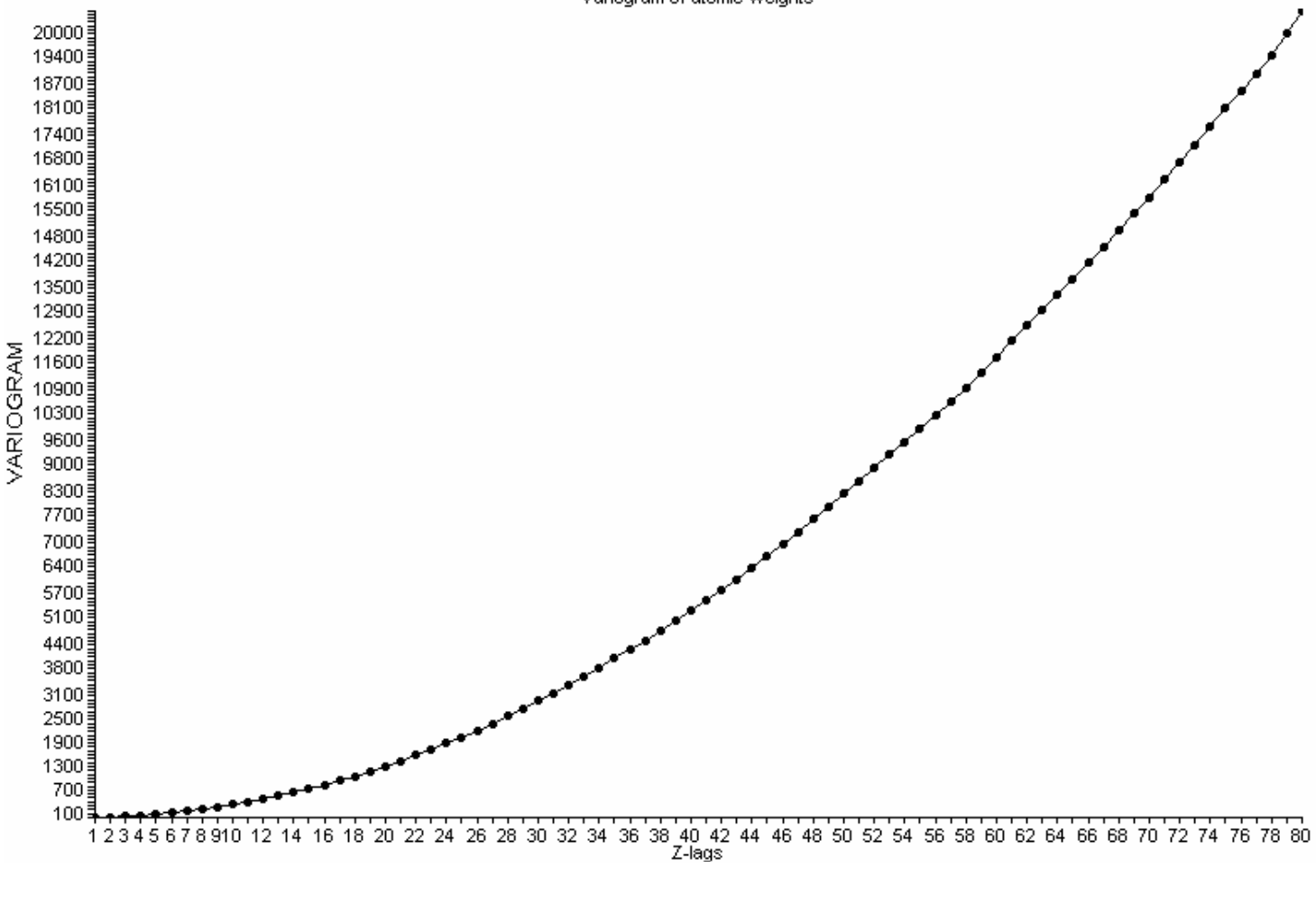

| Z-shift | value | Z-shift | value | Z-shift | value |
|---|---|---|---|---|---|
| 1 | 4.1038233 | 28 | 2583.4415 | 55 | 9936.2738 |
| 2 | 13.890723 | 29 | 2774.3962 | 56 | 10263.722 |
| 3 | 30.73476 | 30 | 2974.6827 | 57 | 10614.884 |
| 4 | 53.281165 | 31 | 3177.0212 | 58 | 10974.746 |
| 5 | 82.706811 | 32 | 3391.3496 | 59 | 11360.41 |
| 6 | 118.39056 | 33 | 3607.1587 | 60 | 11745.812 |
| 7 | 161.00905 | 34 | 3826.8629 | 61 | 12160.26 |
| 8 | 209.83439 | 35 | 4053.0824 | 62 | 12559.606 |
| 9 | 265.1562 | 36 | 4283.8682 | 63 | 12969.082 |
| 10 | 326.91086 | 37 | 4517.609 | 64 | 13342.27 |
| 11 | 395.38036 | 38 | 4760.8117 | 65 | 13738.976 |
| 12 | 470.35989 | 39 | 5010.0734 | 66 | 14163.507 |
| 13 | 551.85195 | 40 | 5268.4938 | 67 | 14573.838 |
| 14 | 640.44287 | 41 | 5533.4834 | 68 | 14979.658 |
| 15 | 735.22246 | 42 | 5805.6797 | 69 | 15415.604 |
| 16 | 837.44107 | 43 | 6083.8624 | 70 | 15830.968 |
| 17 | 946.67776 | 44 | 6370.1678 | 71 | 16277.052 |
| 18 | 1060.8571 | 45 | 6665.5819 | 72 | 16709.305 |
| 19 | 1183.535 | 46 | 6969.6876 | 73 | 17165.872 |
| 20 | 1314.6237 | 47 | 7285.3773 | 74 | 17613.299 |
| 21 | 1448.8492 | 48 | 7605.9824 | 75 | 18096.431 |
| 22 | 1590.1856 | 49 | 7928.8907 | 76 | 18533.473 |
| 23 | 1734.6391 | 50 | 8261.1897 | 77 | 18994.677 |
| 24 | 1888.6115 | 51 | 8592.0056 | 78 | 19457.937 |
| 25 | 2047.654 | 52 | 8917.3226 | 79 | 20009.547 |
| 26 | 2217.1471 | 53 | 9259.8625 | 80 | 20578.44 |
| 27 | 2394.576 | 54 | 9591.9184 | | |

Using the (11) in a Ln-Ln representation, we may now estimate the Generalized Fractal Dimension. We obtained the following results.

**Curve Fit Report**

Y Variable: C2. X Variable: C1.

Model Fit: C2=A+B*(C1) Simple Linear

**Parameter Estimates for All Groups**

| **Groups** | **Count** | **Iter's** | **R2** | **A** | **B** |
|---|---|---|---|---|---|
| All | 81 | 6 | 0.99986 | 1.25542 | 1.98086 |

**Combined Plot Section: Ln-Ln variogram fitting**

**Fig.22**

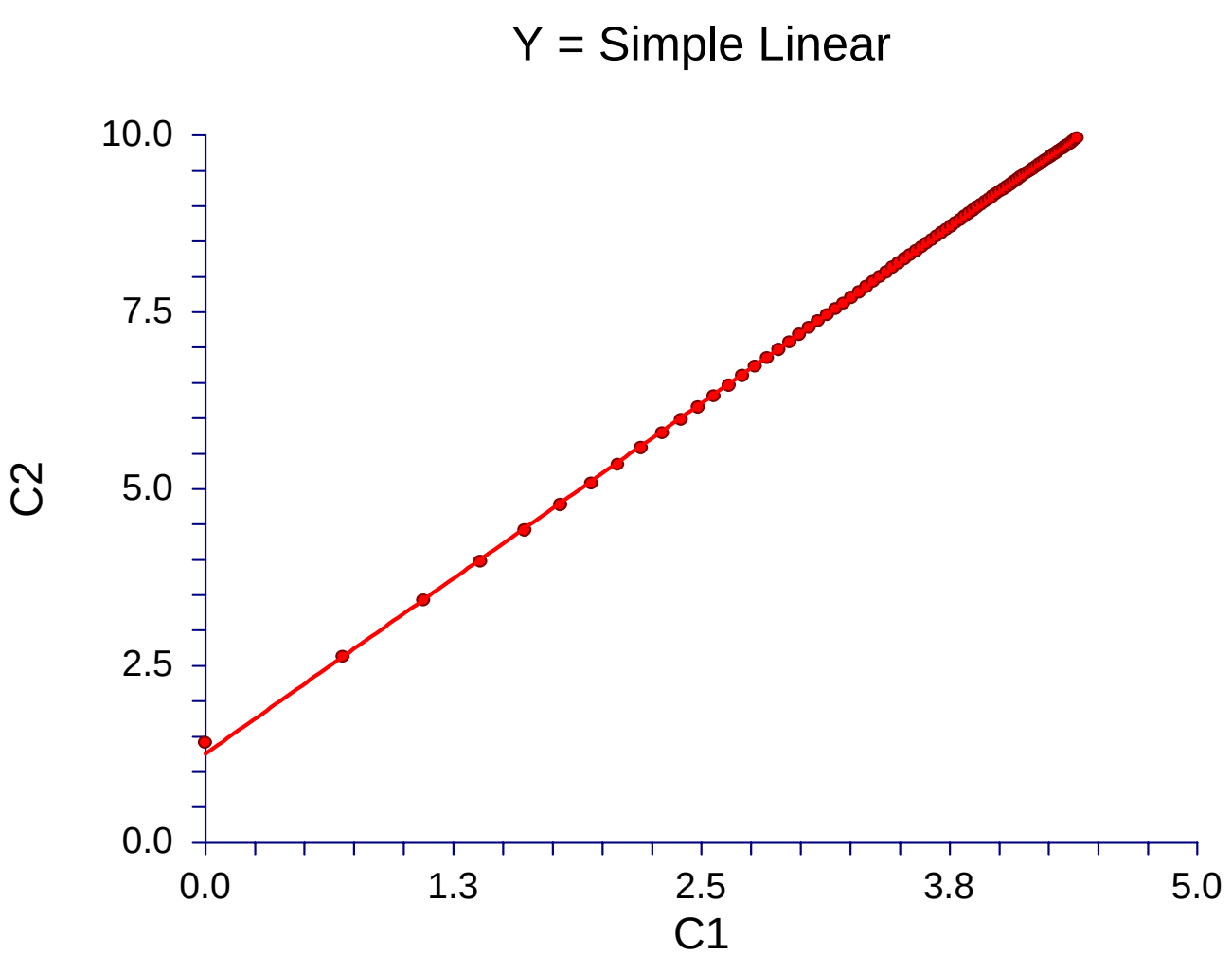

**Model Estimation Section**

| Parameter Name | Parameter Estimate | Asymptotic Standard Error | Lower 95% C.L. | Upper 95% C.L. |
|---|---|---|---|---|
| A | 1.25542 | 0.00939 | 1.23674 | 1.27410 |
| B | 1.98086 | 0.00264 | 1.97560 | 1.98612 |
| Iterations | 6 | Rows Read | 81 | |
| R-Squared | 0.999859 | Rows Used | 81 | |
| Random Seed | 7364 | Total Count | 81 | |

**Estimated Model**
(1.25542232430747)+(1.98086225009967)*(C1)

**Curve Fit Report**
Y Variable: C2.  X Variable: C1.
Model Fit: C2=A+B*(C1)  Simple Linear

**Plot Section**

**Fig.23**

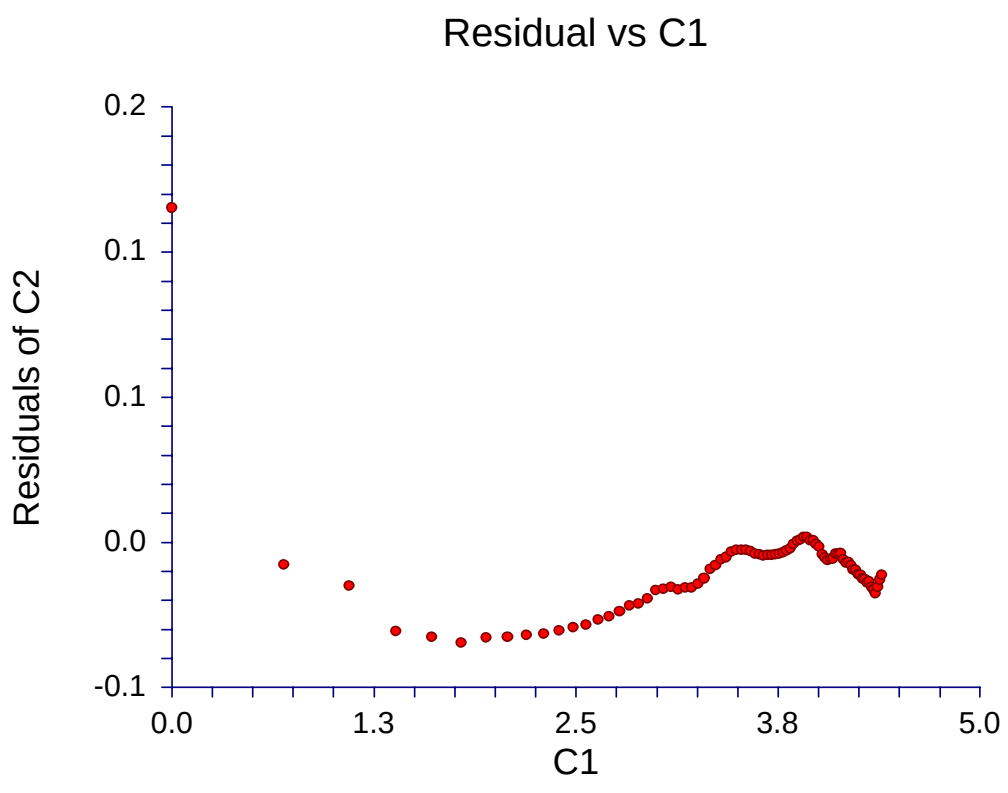

In conclusion, we obtain for Atomic Weights the following results:
Generalized Fractal Dimension $D = 1.98086$
Hausdorff dimension $H_a = 0.99043$.

By using the (12) we may now calculate the Probability Density Function. For atomic weights it results that

$$P(Z) = 5.673\times10^{-6}\, Z^{1.98086} \qquad (13)$$

that is given in Fig.24.

**Fig. 24**

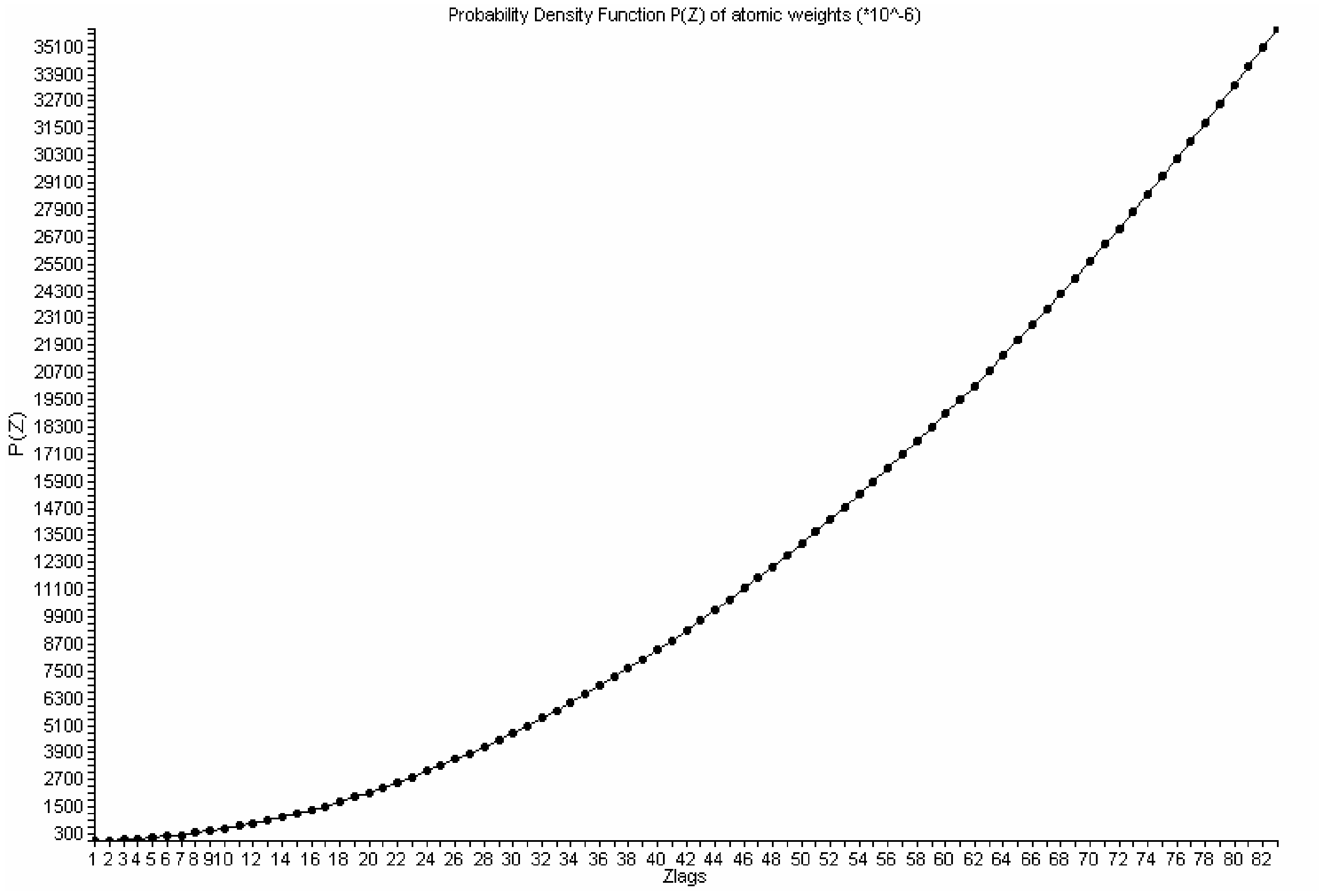

In order to deepen our analysis we may also employ a modified version of standard variogram analysis, using this time a light modification of its usual form in the following way

$$\gamma_{2N}(h) = \frac{1}{2N}\sum_{i=1}^{N-h}(R(x_i) - R(x_i + h))^2 \qquad (14)$$

where we calculate now by $(1/2N)$ instead of $1/2(N-h)$ being $2(N-1)$ the number of degrees of freedom for the whole system taken in consideration.

We have the results in the case of the variogram $\gamma_{2N}$ for atomic weights in Figures 25, 26, 27.

**Fig. 25**

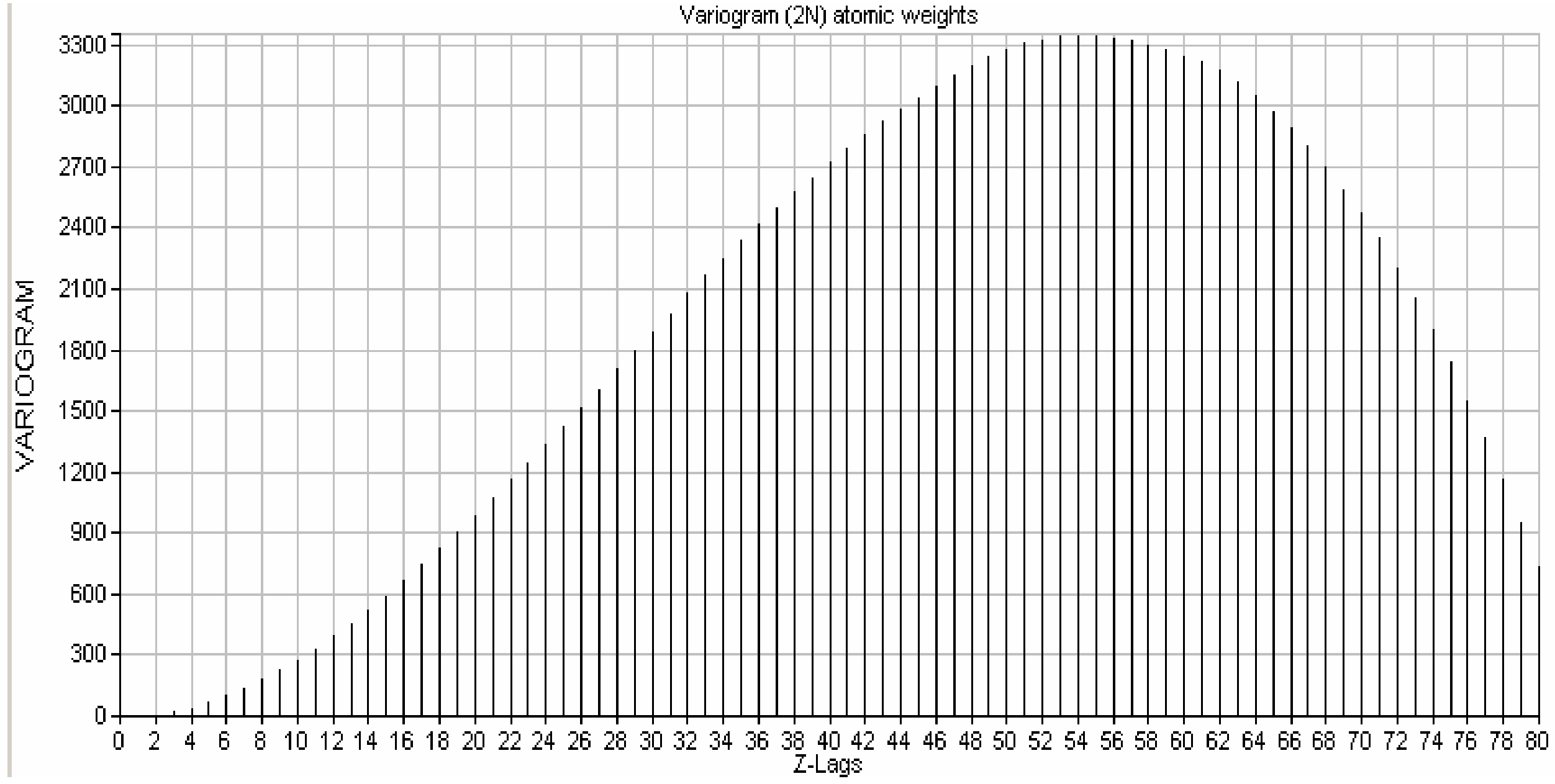

**Fig.26**

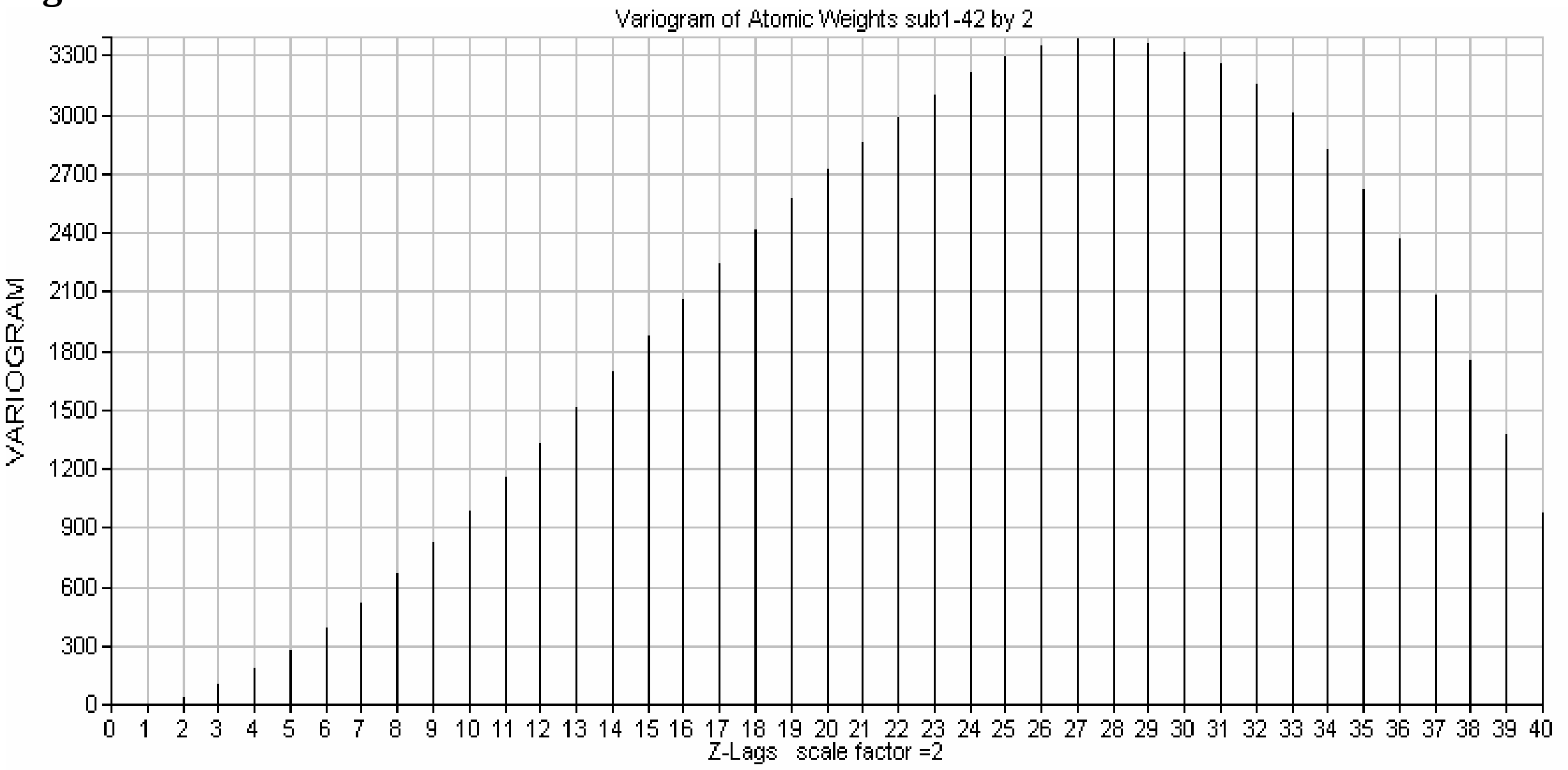

**Fig.27**

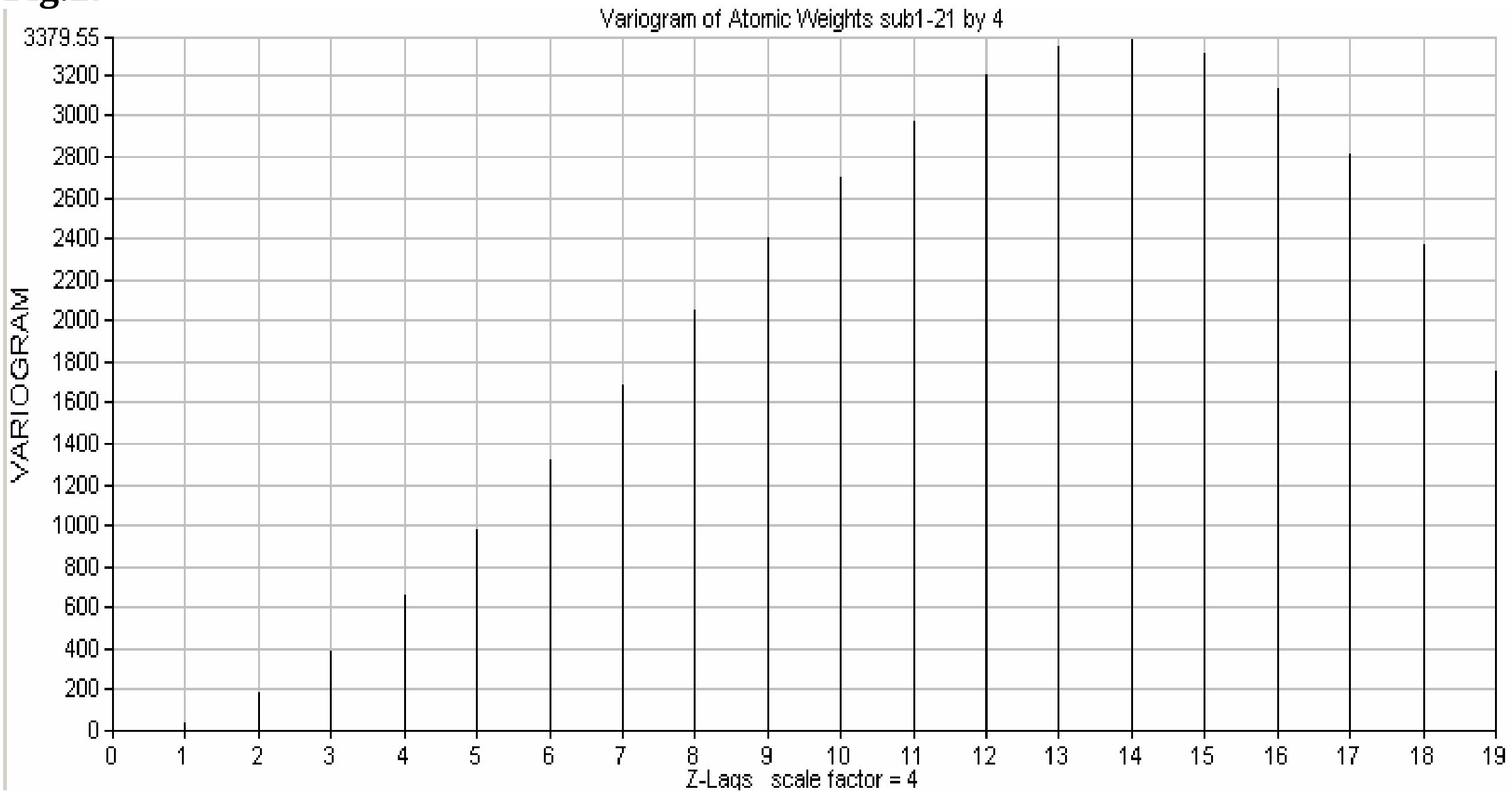

As see, passing from variogram in Fig.25 to variogram in Fig.26 and, finally, in variogram in Fig.27 we have used each time a different factor of scale and, in spite of such different factors of scale, the behaviour of the correspondent variograms, remain unchanged. This result may be taken as further indication that we are in presence of a fractal regime.

In addition, by the $\gamma_{2N}$ variogram, we may now re-calculate the Generalized Fractal Dimension using a Ln-Ln scale.

The results are given in the following scheme.

**Curve Fit Report**

Y Variable: C2. X Variable: C1.

Model Fit: C2=A+B*(C1) Simple Linear

**Parameter Estimates for All Groups**

| **Groups** | **Count** | **Iter's** | **R2** | **A** | **B** |
|---|---|---|---|---|---|
| All | 52 | 6 | 0.99288 | 1.65610 | 1.70683 |

**Combined Plot Section**

**Fig.28**

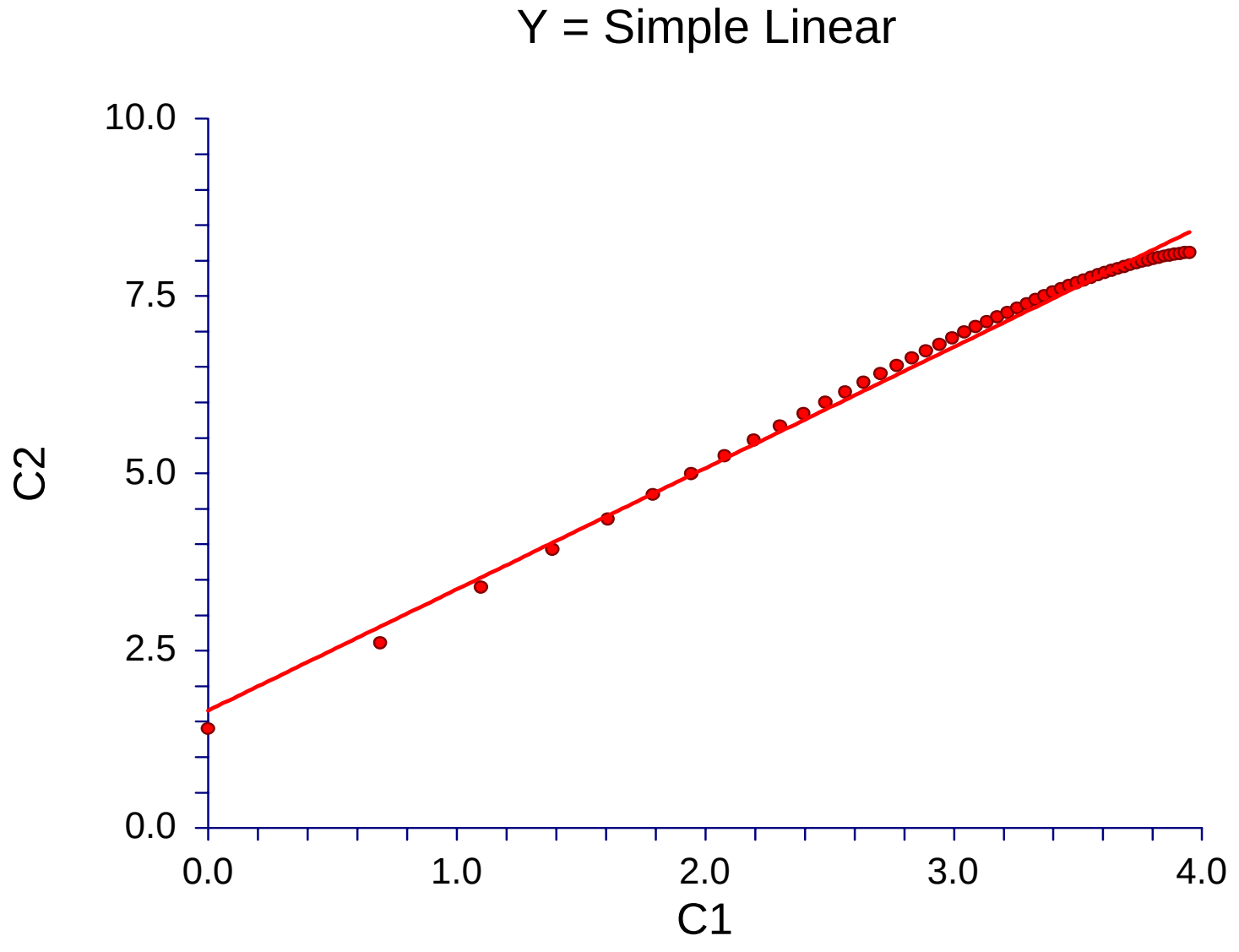

**Model Estimation Section**

| Parameter Name | Parameter Estimate | Asymptotic Standard Error | Lower 95% C.L. | Upper 95% C.L. |
|---|---|---|---|---|
| A | 1.65610 | 0.06408 | 1.52739 | 1.78481 |
| B | 1.70683 | 0.02045 | 1.66576 | 1.74790 |
| Iterations | 6 | Rows Read | 52 | |
| R-Squared | 0.992875 | Rows Used | 52 | |
| Random Seed | 10882 | Total Count | 52 | |

**Estimated Model**

(1.65609637745411)+(1.70683194557246)*(C1)

**Curve Fit Report**

Y Variable: C2. X Variable: C1.

Model Fit: C2=A+B*(C1) Simple Linear

**Plot Section**

**Fig.29**

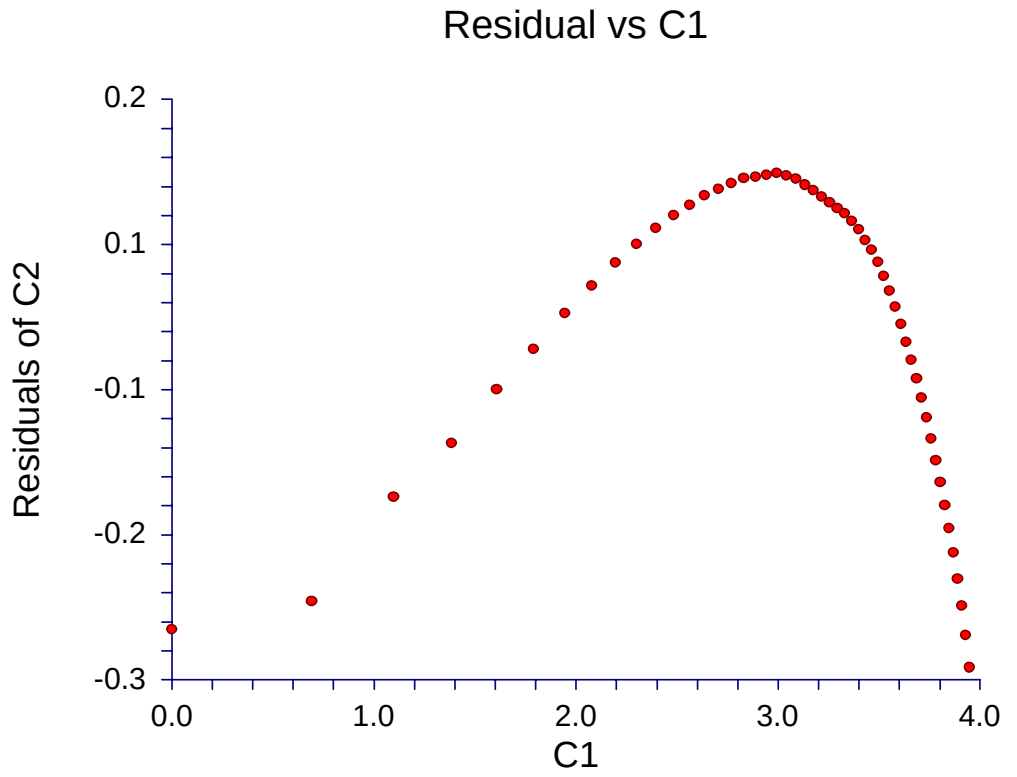

In conclusion, also in this case a non integer value of the Generalized Fractal Dimension is obtained. It results D=1.70683 with Hausdorff dimension $H_a = 0.853415$. Such values result in satisfactory accord with those previously had in the case of the standard variogram.

We may now consider the results that we obtained in the corresponding analysis for Mass Number.

**Fig. 30:Variogram of Mass Number**

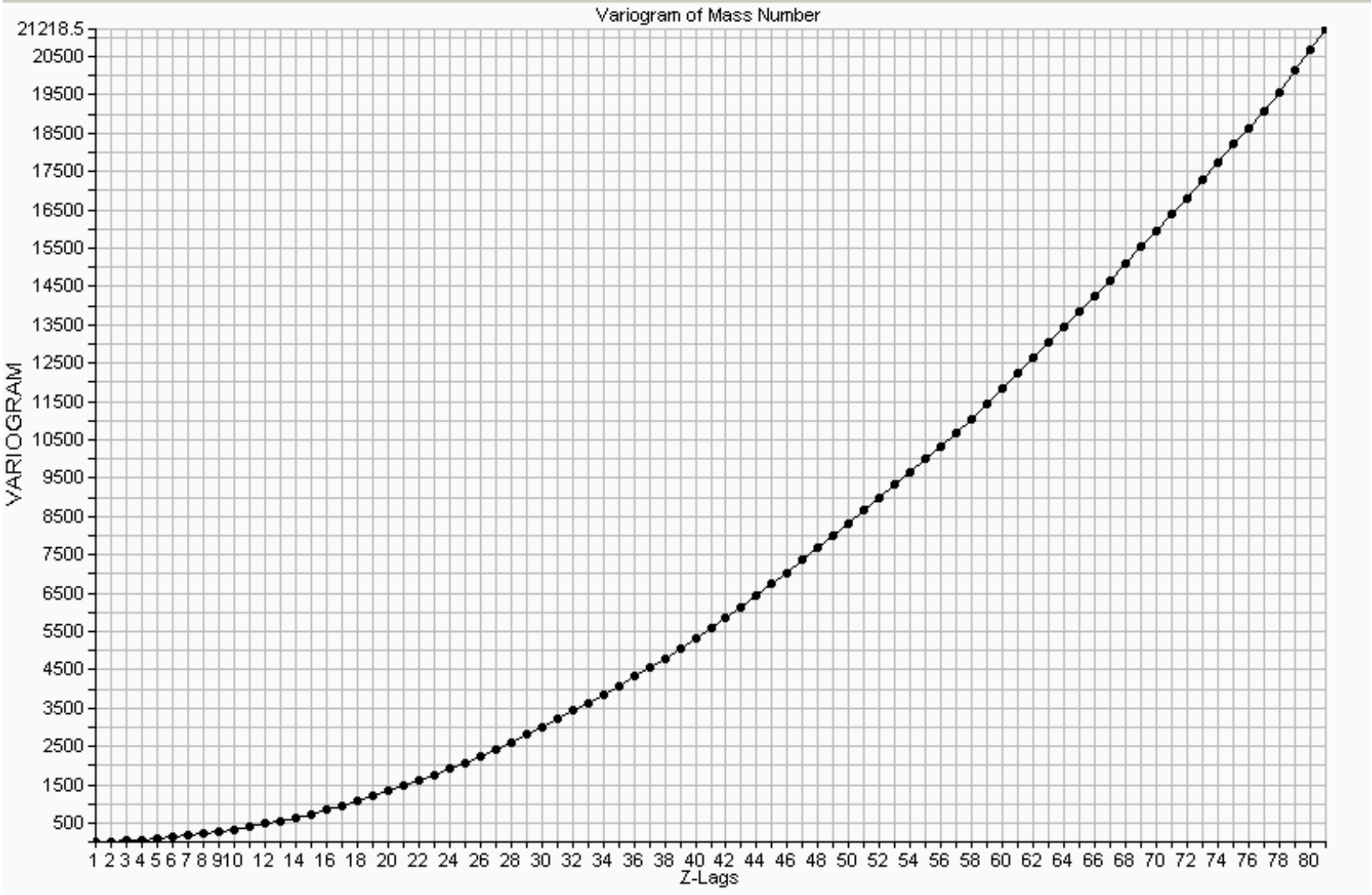

Variogram values:

| Z-lags | Variogram- value |
|---|---|
| 1 | 5.243902 |
| 2 | 14.7284 |
| 3 | 32.275 |
| 4 | 54.56329 |
| 5 | 84.44231 |
| 6 | 119.9221 |
| 7 | 163.0592 |
| 8 | 212.2067 |
| 9 | 268.1824 |
| 10 | 329.637 |
| 11 | 399.0625 |
| 12 | 474.9789 |
| 13 | 557.2214 |
| 14 | 646.1667 |
| 15 | 741.9779 |
| 16 | 843.4851 |
| 17 | 953.9318 |
| 18 | 1068.915 |
| 19 | 1192.695 |
| 20 | 1324.532 |
| 21 | 1460.766 |
| 22 | 1604.451 |
| 23 | 1749.508 |
| 24 | 1903.856 |
| 25 | 2063.957 |
| 26 | 2234.096 |
| 27 | 2412.473 |
| 28 | 2603.482 |
| 29 | 2796.472 |
| 30 | 3000.292 |
| 31 | 3205.394 |
| 32 | 3423.01 |
| 33 | 3639.03 |
| 34 | 3860.969 |
| 35 | 4091.052 |
| 36 | 4326.67 |
| 37 | 4561.033 |
| 38 | 4804.122 |
| 39 | 5055.284 |
| 40 | 5320.965 |
| 41 | 5591.036 |

| Z-lags | Variogram- value |
|---|---|
| 42 | 5867.11 |
| 43 | 6145.663 |
| 44 | 6428.603 |
| 45 | 6726.934 |
| 46 | 7036.689 |
| 47 | 7356.458 |
| 48 | 7675.757 |
| 49 | 7997.926 |
| 50 | 8333.5 |
| 51 | 8671.016 |
| 52 | 8996.565 |
| 53 | 9339.117 |
| 54 | 9664.879 |
| 55 | 10009.8 |
| 56 | 10334.06 |
| 57 | 10688.52 |
| 58 | 11053.98 |
| 59 | 11444.02 |
| 60 | 11836.33 |
| 61 | 12256.52 |
| 62 | 12651.02 |
| 63 | 13058.83 |
| 64 | 13431.24 |
| 65 | 13832.53 |
| 66 | 14249.85 |
| 67 | 14660.66 |
| 68 | 15086.9 |
| 69 | 15531.39 |
| 70 | 15943.19 |
| 71 | 16399.63 |
| 72 | 16814.68 |
| 73 | 17282.35 |
| 74 | 17737.39 |
| 75 | 18219.06 |
| 76 | 18626.5 |
| 77 | 19078.67 |
| 78 | 19562.9 |
| 79 | 20150.38 |
| 80 | 20672.67 |
| 81 | 21218.5 |

We may now give the estimation of the Generalized Fractal Dimension.

**Curve Fit Report (Ln-Ln plot)**
Y Variable: C2.  X Variable: C1.
Model Fit: C2=A+B*(C1)  Simple Linear
**Parameter Estimates for All Groups**

| Groups | Count | Iter's | R2 | A | B |
|---|---|---|---|---|---|
| All | 81 | 4 | 0.99943 | 1.32691 | 1.96370 |

**Combined Plot Section**

**Fig.31**

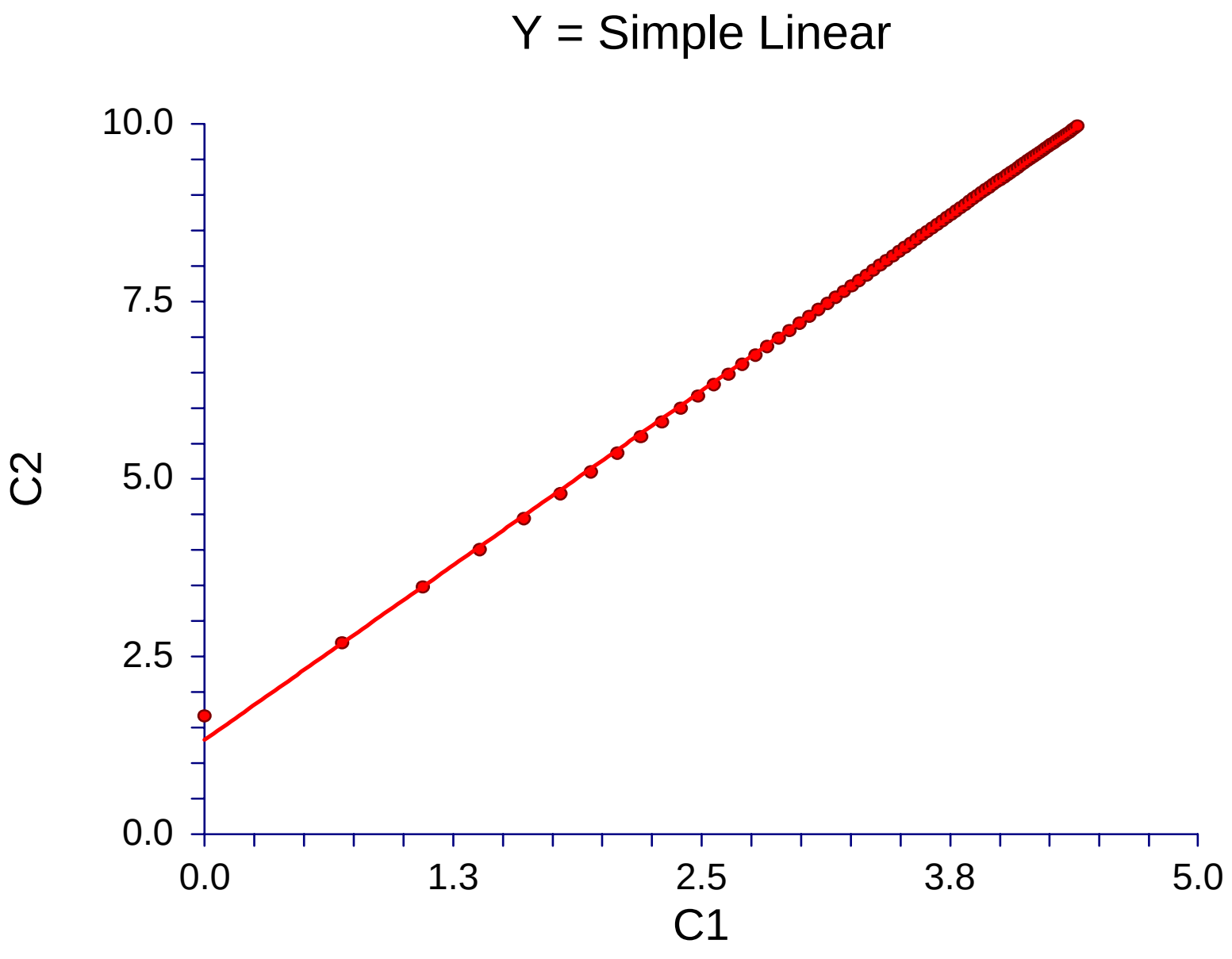

**Model Estimation Section**

| Parameter Name | Parameter Estimate | Asymptotic Standard Error | Lower 95% C.L. | Upper 95% C.L. |
|---|---|---|---|---|
| A | 1.32691 | 0.01877 | 1.28955 | 1.36427 |
| B | 1.96370 | 0.00528 | 1.95318 | 1.97422 |
| Iterations | 4 | Rows Read | 81 | |
| R-Squared | 0.999428 | Rows Used | 81 | |
| Random Seed | 2960 | Total Count | 81 | |

**Estimated Model**
(1.32690928509202)+(1.96369892532774)*(C1)
**Curve Fit Report**
Y Variable: C2.  X Variable: C1.
Model Fit: C2=A+B*(C1)  Simple Linear
**Plot Section**

**Fig.32**

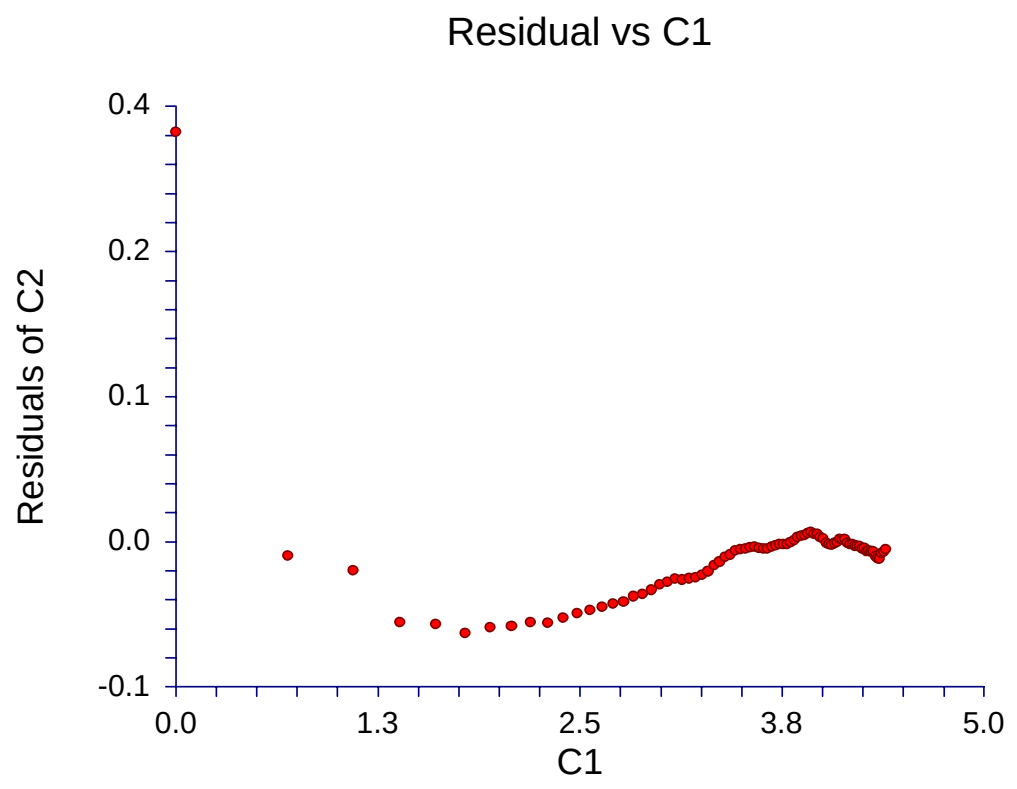

The analysis enables us to give the following results:
Generalized Fractal Dimension D = 1.96370
Hausdorff dimension $H_a$ = 0.98185

We may now calculate the Probability Density Function. It assumes the following form

$$\mathbf{P(Z) = 6.786 \times 10^{-6}} \; Z^{\mathbf{1.9637}}$$

**Fig. 33**

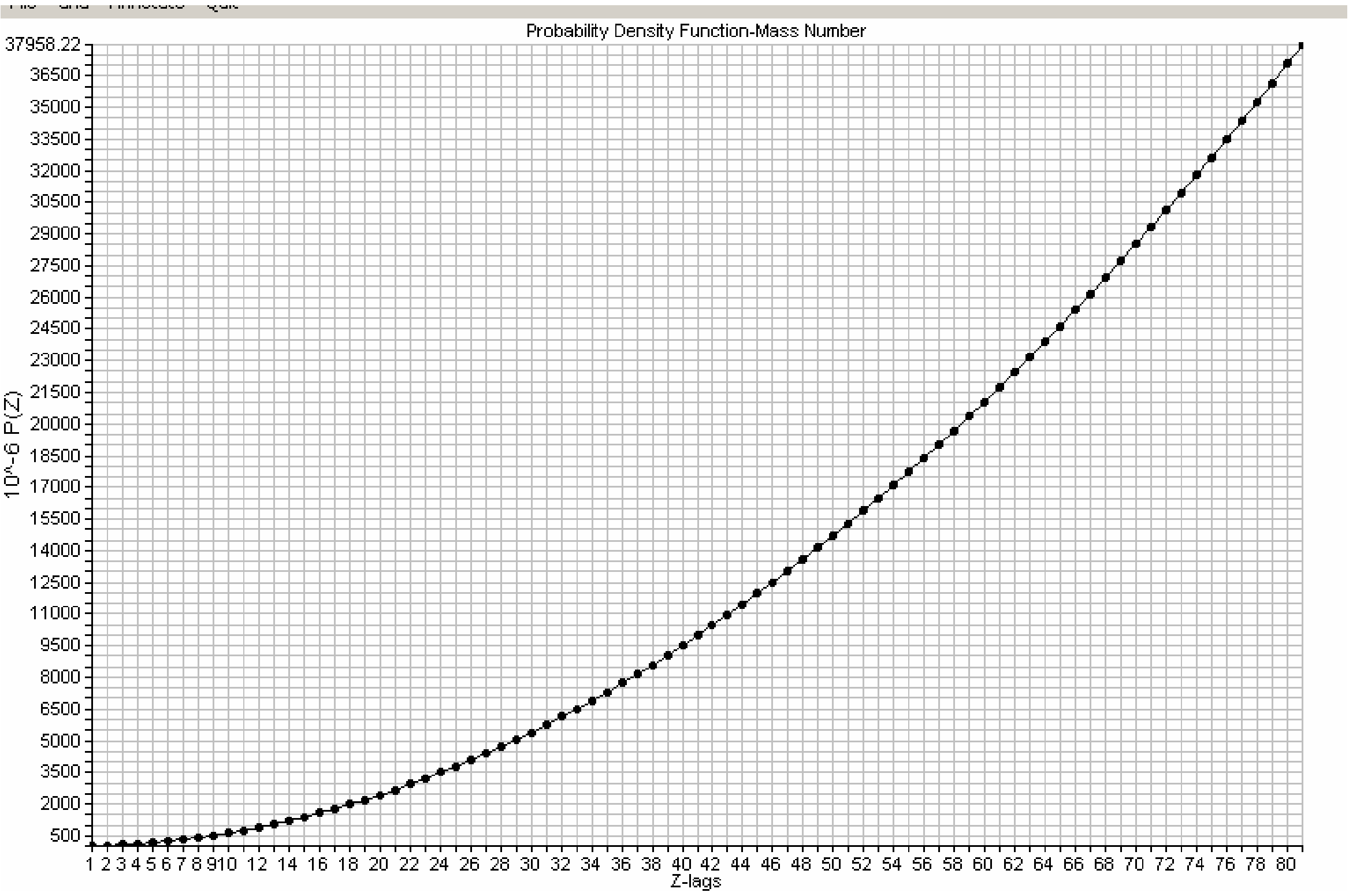

Let us proceed estimating $\gamma_{2N}$ at different scale factors.

**Fig. 34**

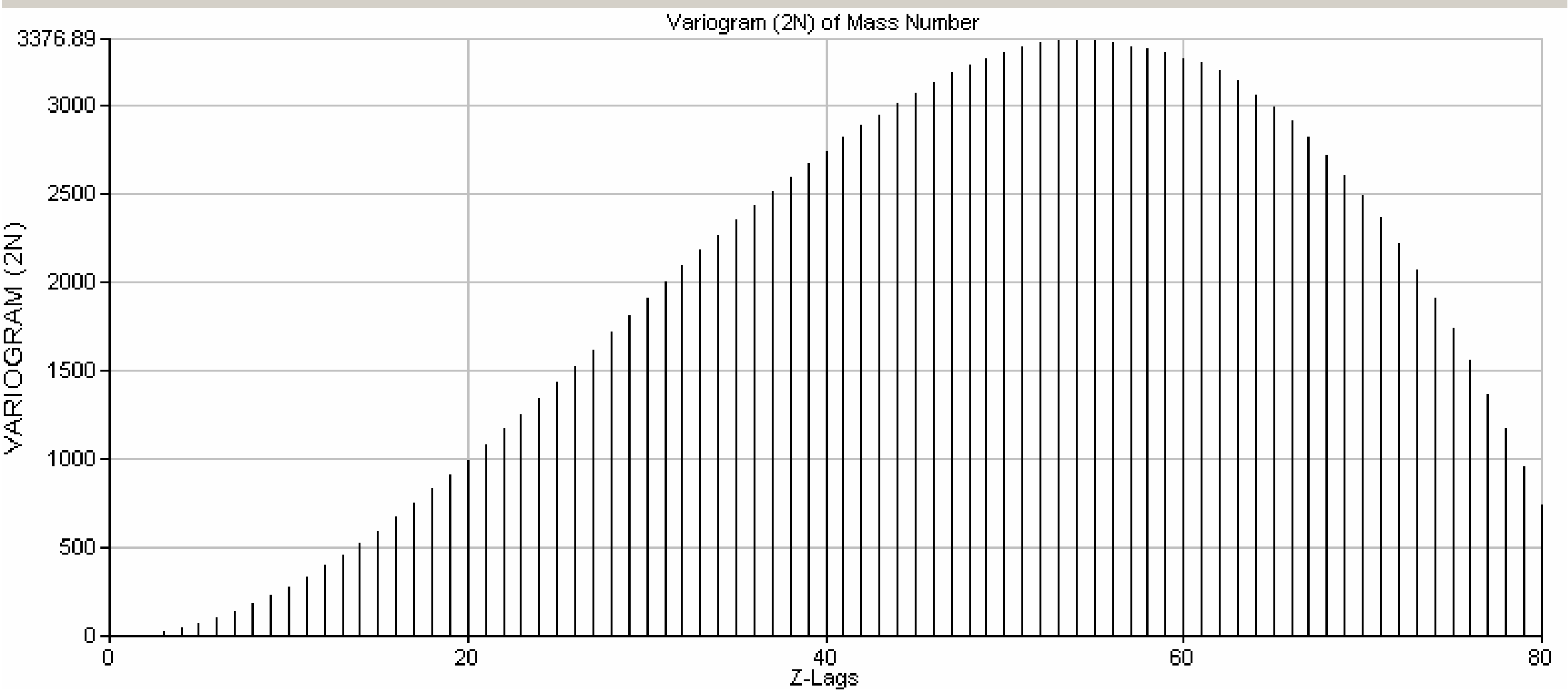

**Fig. 35a**

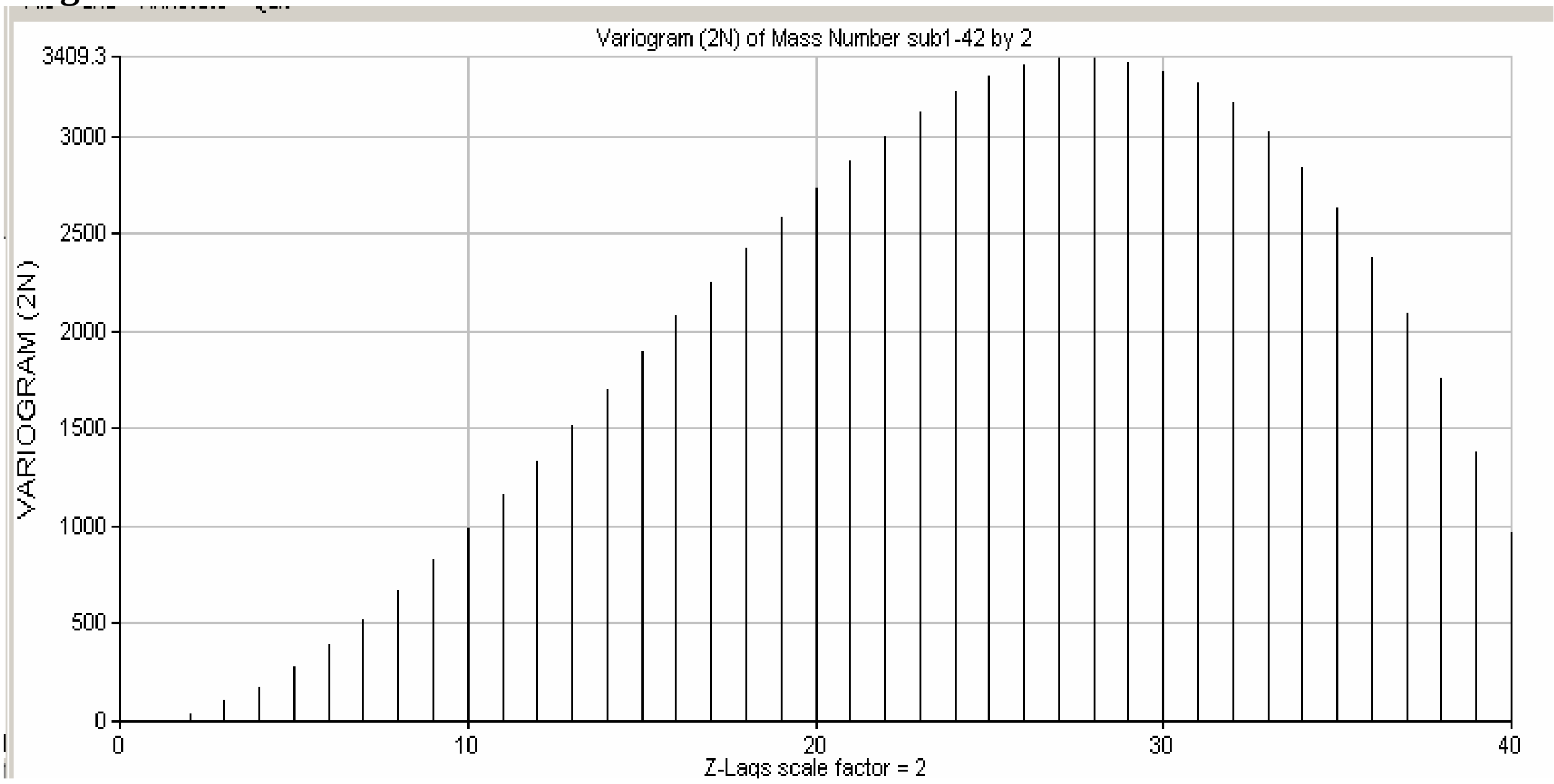

**Fig. 35b**

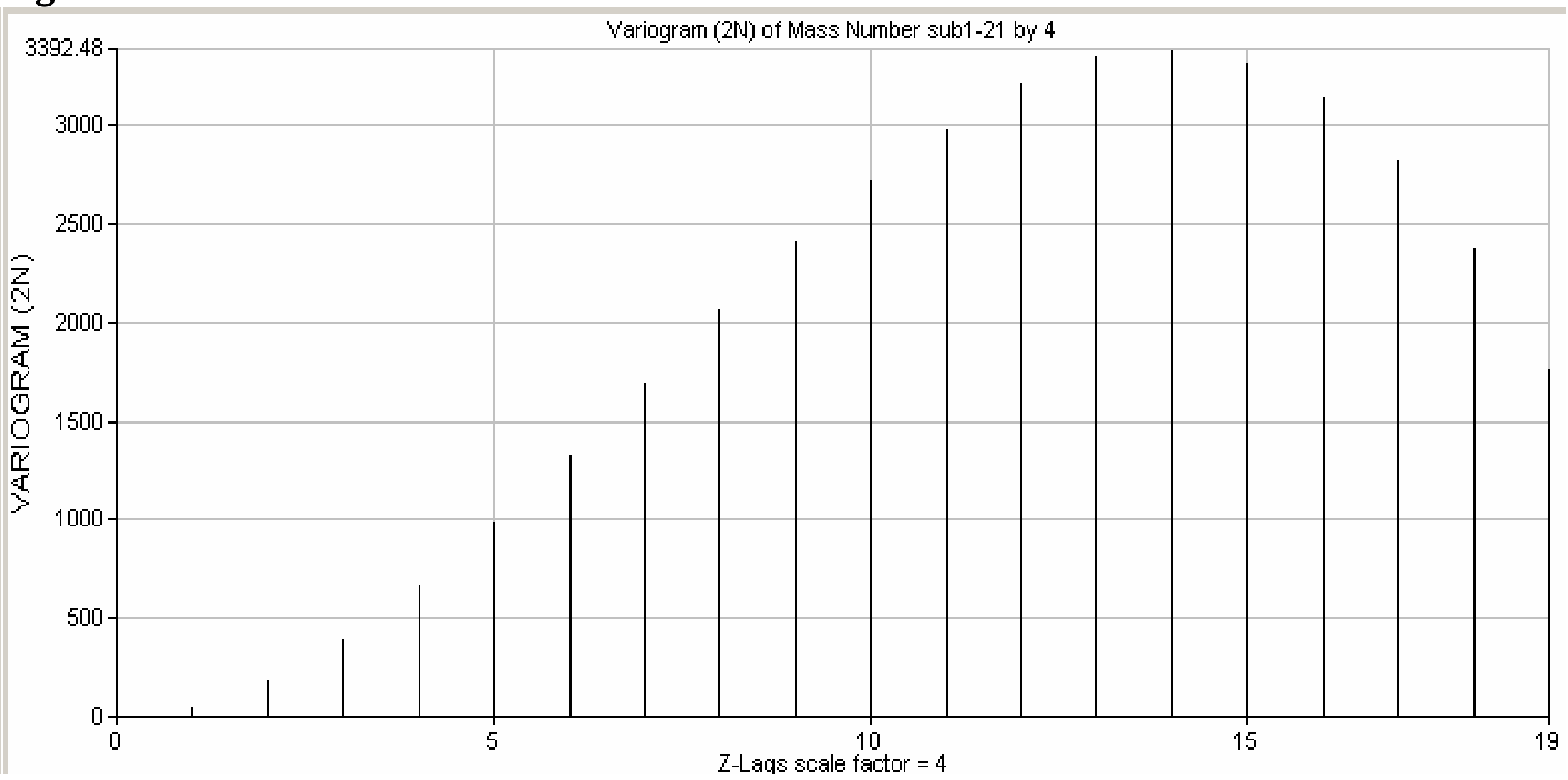

By the $\gamma_{2N}$ variogram we may now re-calculate the Generalized Fractal Dimension using a Ln-Ln scale. The results are given in the following scheme.

**Curve Fit Report**

Y Variable: C2. X Variable: C1.
Model Fit: C2=A+B*(C1) Simple Linear
**Parameter Estimates for All Groups**

| Groups | Count | Iter's | R2 | A | B |
|---|---|---|---|---|---|
| All | 49 | 4 | 0.99538 | 6.81261 | 1.70270 |

**Combined Plot Section**

**Fig.36**

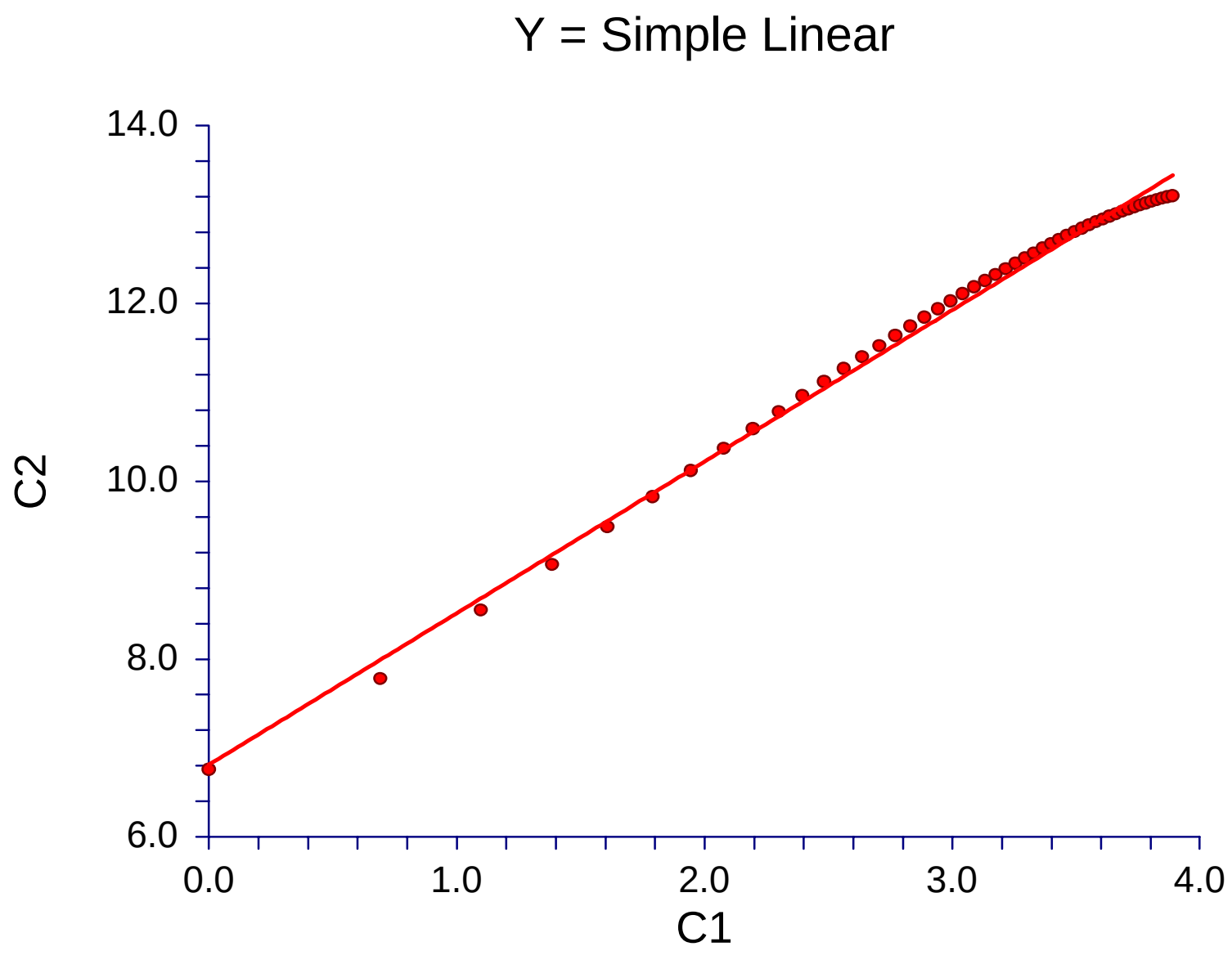

**Model Estimation Section**

| **Parameter Name** | **Parameter Estimate** | **Asymptotic Standard Error** | **Lower 95% C.L.** | **Upper 95% C.L.** |
|---|---|---|---|---|
| A | 6.81261 | 0.05209 | 6.70783 | 6.91739 |
| B | 1.70270 | 0.01692 | 1.66867 | 1.73674 |
| Iterations | 4 | Rows Read | 49 | |
| R-Squared | 0.995381 | Rows Used | 49 | |
| Random Seed | 11153 | Total Count | 49 | |

**Estimated Model**

(6.81260680697216)+(1.70270493930648)*(C1)

**Curve Fit Report**

Y Variable: C2.  X Variable: C1.

Model Fit: C2=A+B*(C1)  Simple Linear

**Plot Section**

**Fig.37**

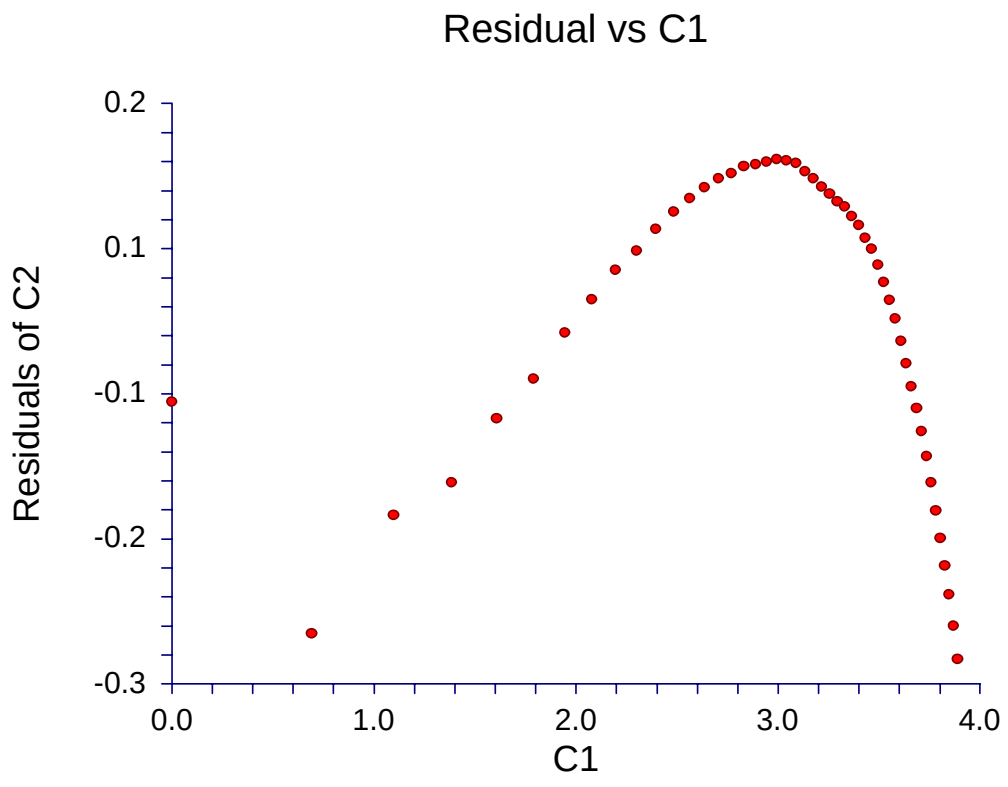

**Conclusion**: also in this case a non integer value of the Generalized Fractal Dimension is obtained. It results D=1.70270 with Hausdorff dimension $H_a = 0.85135$. Such values result in satisfactory accord with those previously obtained in the case of the standard variogram.

In conclusion, until here we have used the standard methodologies that generally one utilizes with the aim to ascertain the presence of non linear contributions in the investigated dynamics as well as to reconstruct phase space dynamics and to evaluate the possible presence of divergent features in the system, possibly of chaotic nature, and still the probable presence of a fractal regime in such dynamics. On the basis of the results that we have obtained, it seems very difficult to escape the conclusion that the process of increasing mass, regarding Atomic Weighs and Mass Number in atomic nuclei, concerns all the basic features of non linearity, divergence, possible chaoticity and fractality that we have only just indicated for systems with non linear dynamics. This is a conclusion that in some manner overthrows our traditional manner to approach nuclear matter. For this reason it requires still more detailed deepening. In the following sections we will support our conclusion by other detailed results.

## 6. Calculation of Hurst Exponent and Possible Presence of Fractional Brownian Behaviour In Atomic Weights and Mass Number of Atomic Nuclei

It is known that time series arise often from a random walk usually called Brownian motion. The Hurst exponent [12] in such cases is calculated to be 0.5.

This concept may be generalized introducing the Fractional Brownian Motion (fBM) which arises from integrating correlated –coloured noise.

The value of Hurst exponent helps us to identify the nature of the regime we have under examination. In detail, if the H exponent results greater than 0.5, we are in presence of persistence, that is to say, past trends also persist into the future. On the other hand, in presence of H exponent values less than 0.5 we conclude for anti persistence, indicating it in this case that past trends tend to reverse in the future.

In the present case the analysis is not performed having a time series but instead we consider the atomic number Z in $W_a(Z)$, the atomic weights, and in $A(Z)$, the Mass Number of atomic nuclei.

Our analysis gave the following results.

For atomic weights, $W_a(Z)$, we obtained the subsequent value:

Hurst exponent H = 0.9485604 ; SDH = 0.00625887 ; $r^2$ = 0.999645

Instead for Mass Number , $A(Z)$, we had the next value:

Hurst exponent H = 0.8953571 ; SDH = 0.0057648 ; $r^2$ = 0.999753.

Both the results obtained respectively for Atomic Weights and for Mass Number, enable us to conclude that:

1) we are in presence of a Fractional Brownian Regime in both the cases;
2) in both $W_a(Z)$ and $A(Z)$ the tendency is for the persistence that results more marked in $W_a(Z)$ respect to $A(Z)$;
3) in the case of the Atomic Weights, $W_a(Z)$, the value of Fractal Dimension results to be
   $D = 2 - H = 1.0514396$
   while in the case of Mass Number, $A(Z)$, the value of Fractal Dimension is
   $D = 2 - H = 1.1046429$ .

## 7. Recurrence Quantification Analysis – RQA

Further important information on the nature of the processes presiding over the mechanism of increasing mass in Atomic Weights and Mass Number of atomic nuclei may be obtained by using RQA, the Recurrence Quantification Analysis.

This is a kind of analysis that, as it is well known, was introduced by J.P Zbilut and C.L. Webber [13].

Such investigation offers a new opportunity to us. By it we may give a look to the process of increasing mass of atomic nuclei analyzing in detail the kind of dynamics that governs such mechanism. Therefore, the results of such investigation must be considered with particular attention owing to their relevance.

The features that we may investigate in detail are the following: first of all we may evaluate the level of recurrence, that is to say of “periodicity”, that such process exhibits. This is obtained by estimating the % Rec in RQA. Soon after we may also calculate the Determinism that is involved in such process. This is to say that we evaluate the level of predictability that it has. We estimate such features by %Det. in RQA. As third RQA variable we may also estimate the entropy and than the Max Line that is a measure linked proportionally to the inverse of the Lyapunov exponent. In brief, such measure enables us to evaluate still again the possible divergence involved in such mechanism.
Usually, when using RQA, one starts with an embedding procedure of the given time series and thus providing with a given reconstruction of the given time series in phase space. In our case such reconstruction was previously performed in previous sections and we obtained that we should use an embedding dimension $D = 2$ with a $Z - shift = 3$ in the case of Atomic Weights and a $Z - shift = 2$ in the case of Mass Numbers. However, in the present analysis our purpose is slightly different in the sense that we aim to preserve the embedded dimension $D = 2$ but we yearn for analyzing the behaviour of the basic RQA variables as %Rec., %Det., ENT., and Max Line shifting step by step the value of the atomic number $Z$ so to explore the mechanism as well as $Z$ increases step by step. In order to perform such kind of analysis a value of the distance $R$ should be correctly selected. Usually, the distance $R$ in RQA may be fixed rather empirically selecting a proper value so that %Rec. remain about 1%. However Zbilut and Webber [13] in their RQA software package introduced RQS that estimates recurrences at various distances and the cut off that one has at a particular distance respect to a flat behaviour. In this manner one selects the best optimized distance R to use in the analysis. We applied RQS software to select the proper distance and it was obtained that such value should be taken $R = 4$
We also ascertained that such selected value remained rather constant when increasing $Z$ step by step. In Fig. 38 we give some of the results that were obtained.

**Fig. 38**

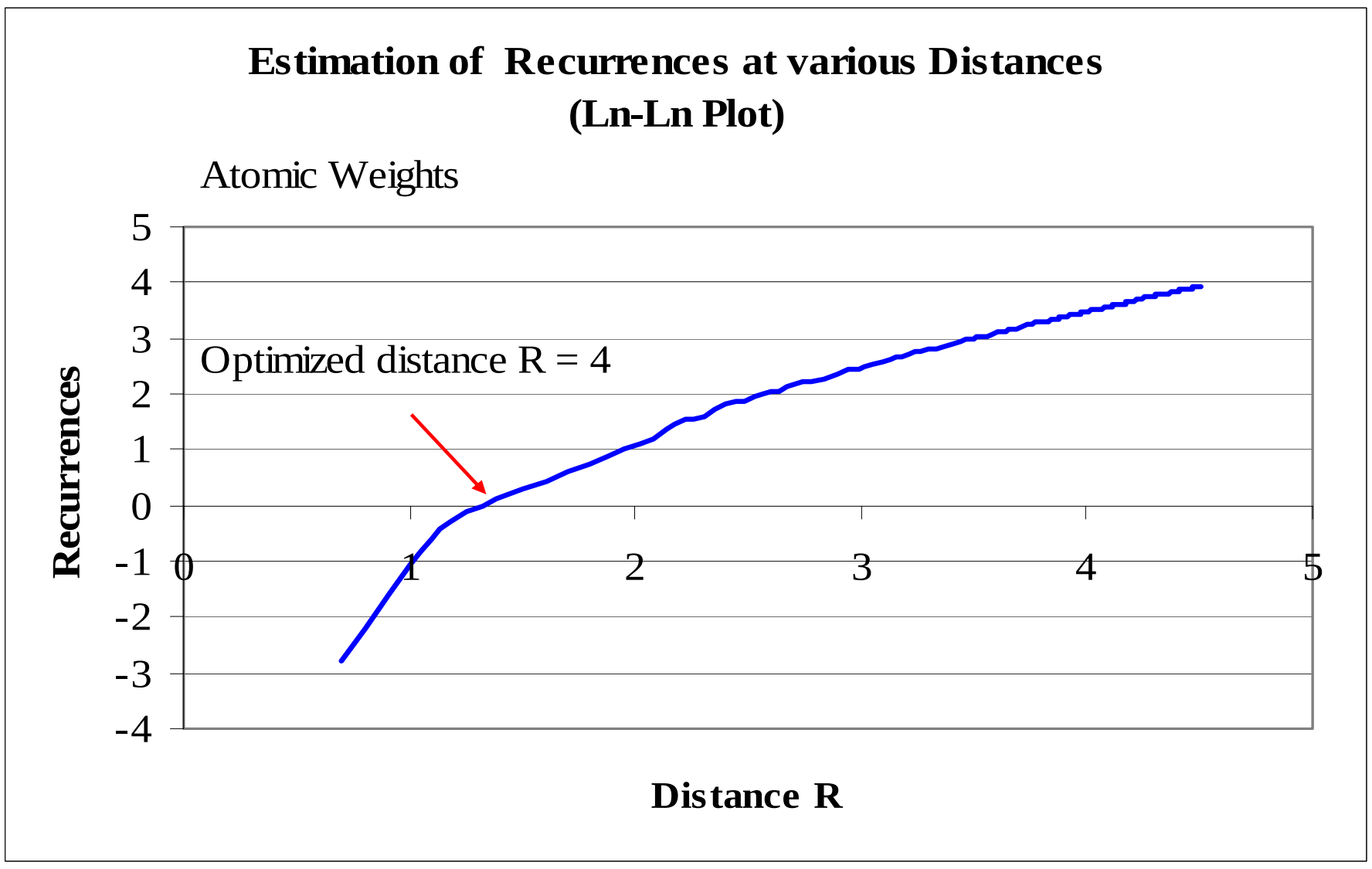

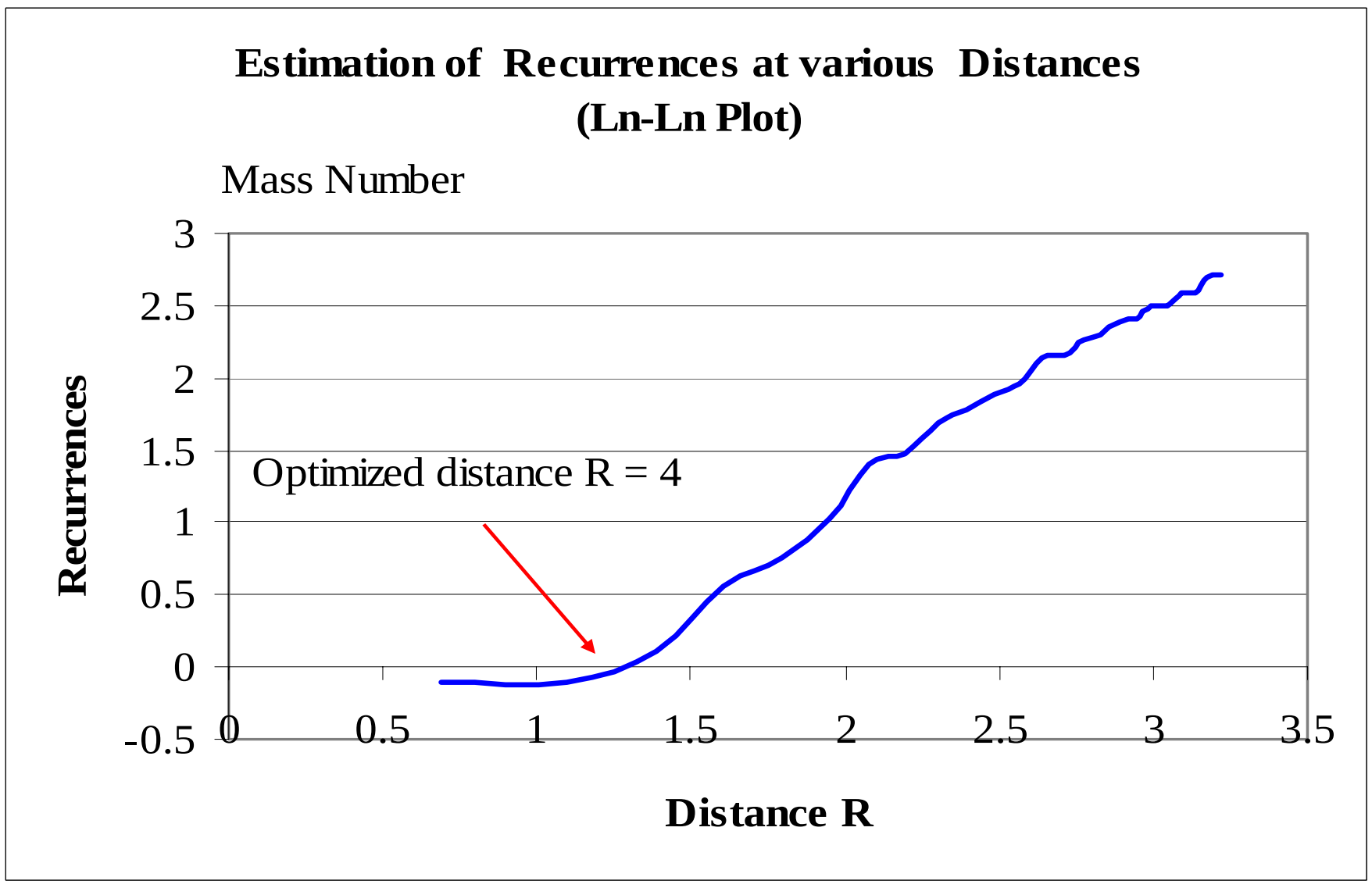

In conclusion we selected R=4 for the distance to use in RQA. The embedding dimension was chosen to be D=2 as it resulted by using FNN criterion and verifying this choice also for different $Z$ values. Finally, we decided to use the value L=3 for the Line Length.

We have obtained the following results.
Recurrence Quantification Analysis applied to Atomic Weights, $W_a(Z)$, for increasing values of $Z-shift$. The results obtained for %Rec., %Det., ENT., and MaxLine are reported in the following Table 1

**Table 1**

| Z-shift | %Rec. | %Det. | Entropy | Max-Line |
|---|---|---|---|---|
| 1 | 1.36 | 73.33 | 2.00 | 14 |
| 2 | 1.39 | 44.44 | 1.50 | 7 |
| 3 | 1.33 | 52.38 | 1.59 | 13 |
| 4 | 1.36 | 28.57 | 1.00 | 7 |
| 5 | 1.30 | 38.46 | 1.00 | 11 |
| 6 | 1.37 | 27.50 | 0.92 | 5 |
| 7 | 1.16 | 24.24 | 0.00 | 8 |
| 8 | 1.33 | 13.51 | 0.00 | 5 |
| 9 | 1.04 | 32.14 | 1.00 | 5 |
| 10 | 1.33 | 17.14 | 0.00 | 3 |
| 11 | 0.94 | 25.00 | 0.00 | 6 |
| 12 | 1.33 | 12.12 | 0.00 | 4 |
| 13 | 1.04 | 24.00 | 0.00 | 6 |
| 14 | 1.28 | 20.00 | 0.00 | 3 |
| 15 | 0.92 | 14.29 | 0.00 | 3 |
| 16 | 1.31 | 20.69 | 0.00 | 3 |
| 17 | 0.89 | 15.79 | 0.00 | 3 |
| 18 | 1.06 | 13.64 | 0.00 | 3 |
| 19 | 0.84 | 17.65 | 0.00 | 3 |
| 20 | 1.13 | 13.69 | 0.00 | 3 |
| 21 | 0.95 | 0.00 | - | - |

There are some results that deserve to be outlined .
%Rec. remains rather constant in correspondence of the different $Z-shift$ values with some fluctuations taking minima values mainly at $Z-shift$ = 11, 15, 19,21.
A graph is given in Fig.39.

**Fig. 39**

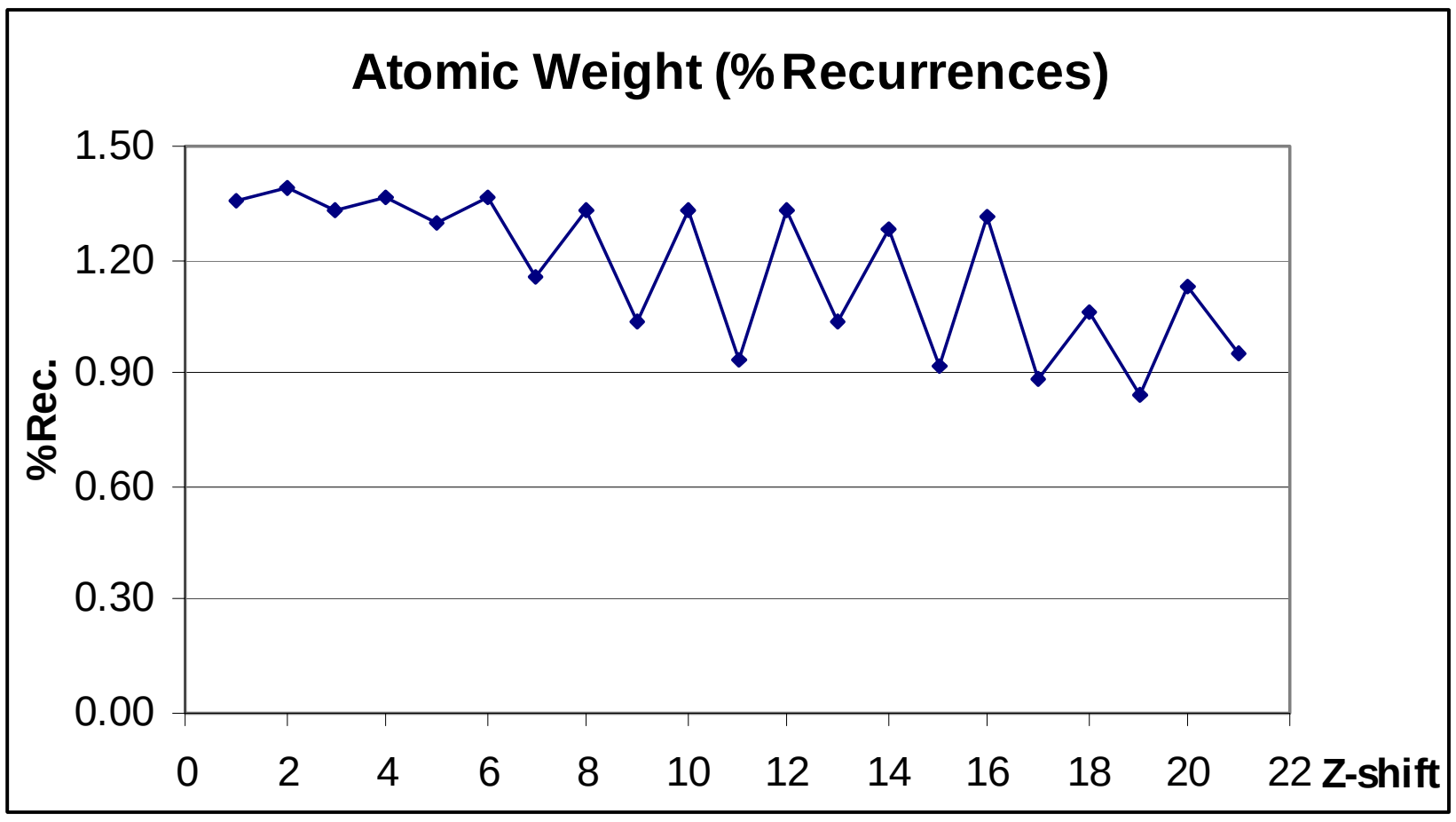

%Det. assumes rather low values also with a length Line L=3. It oscillates among maxima and minima for increasing values of $Z-shift$ as it is pictured in Fig.40 (a, b, c). Significantly, %Det. goes definitively to zero starting with $Z-shift$ =21.

**Fig. 40a**

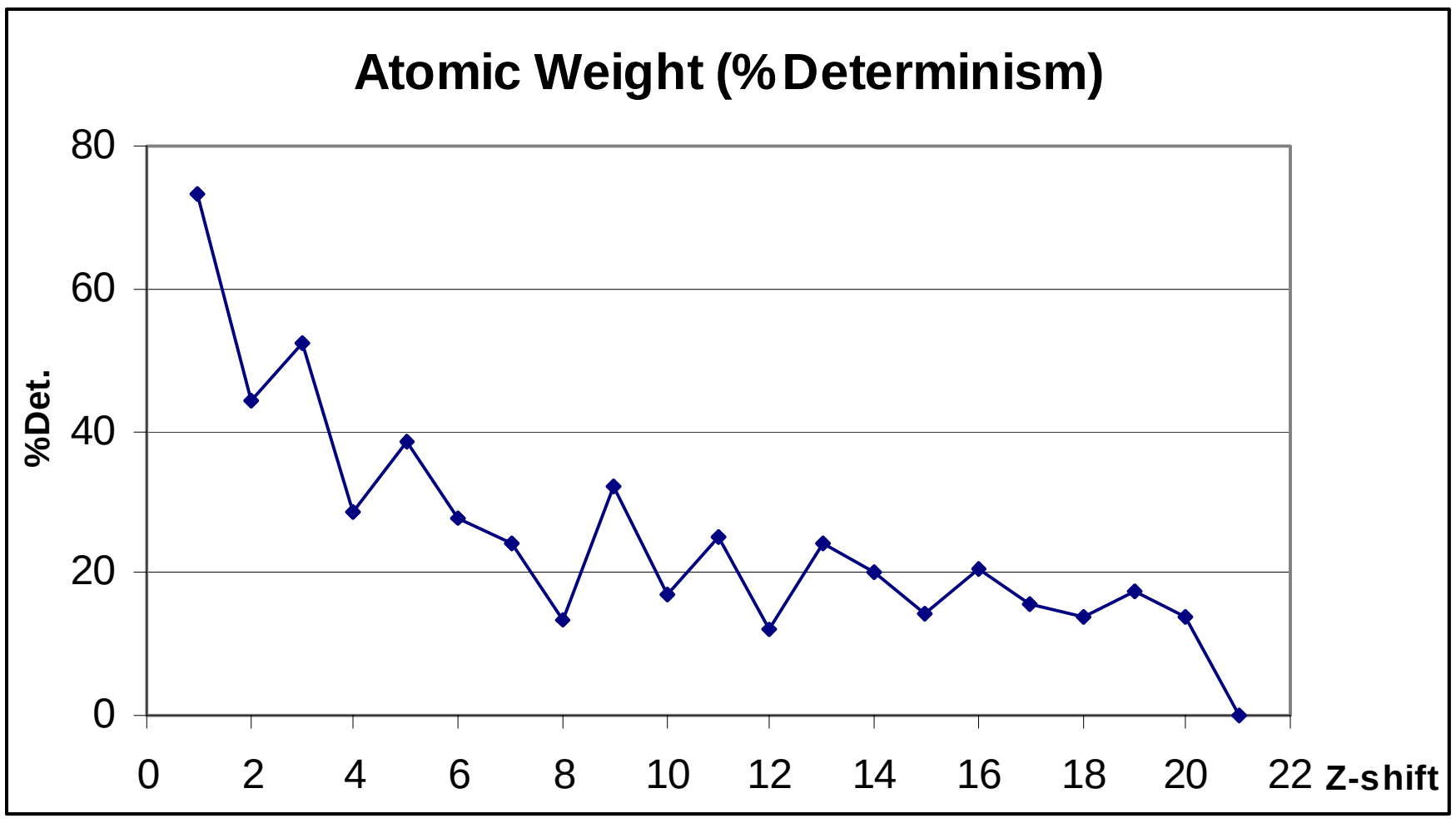

Rather interesting appear also the value we obtain for Entropy and Max Line as reported in the following figures.

**Fig. 40b**

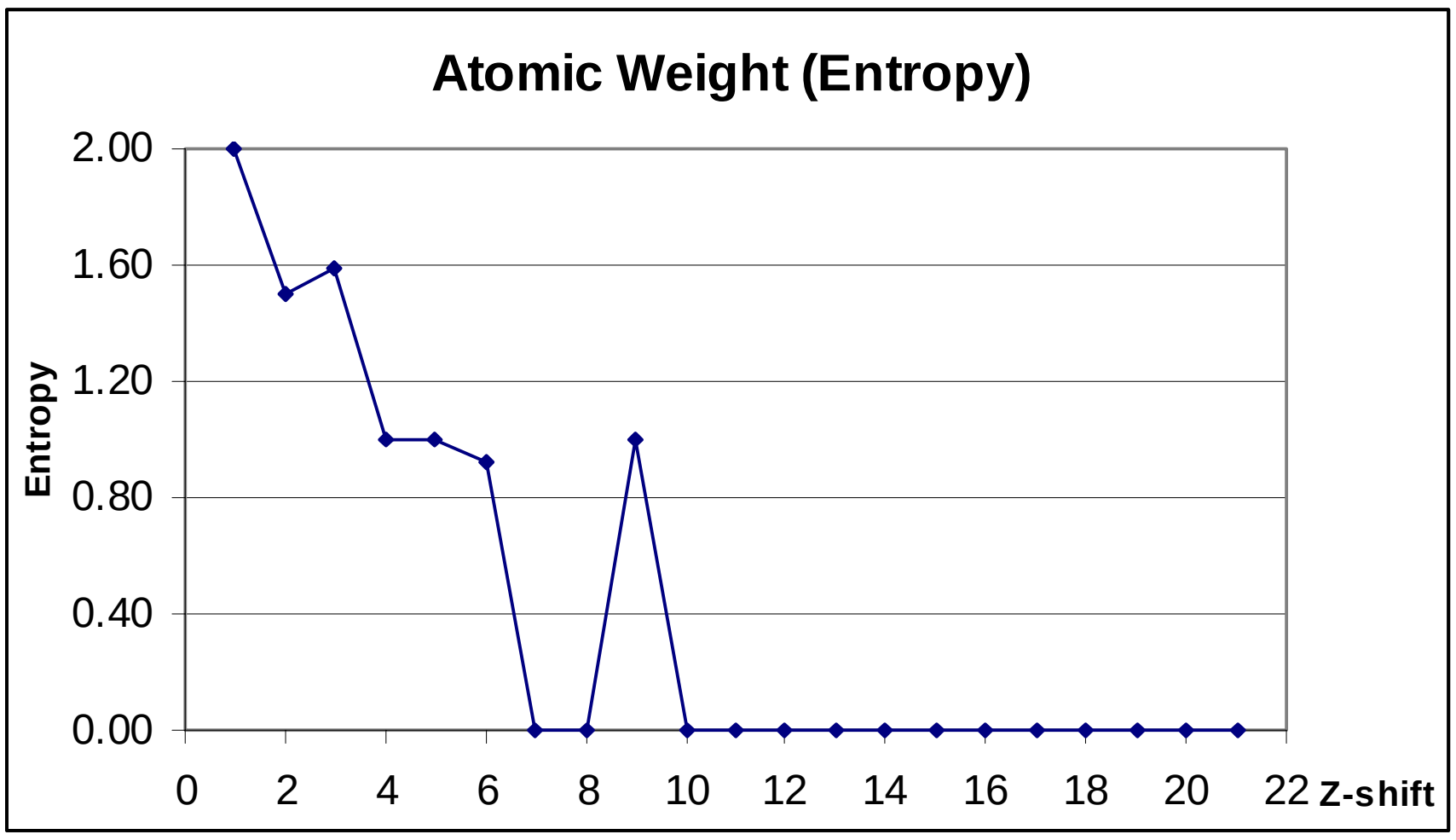

**Fig. 40c**

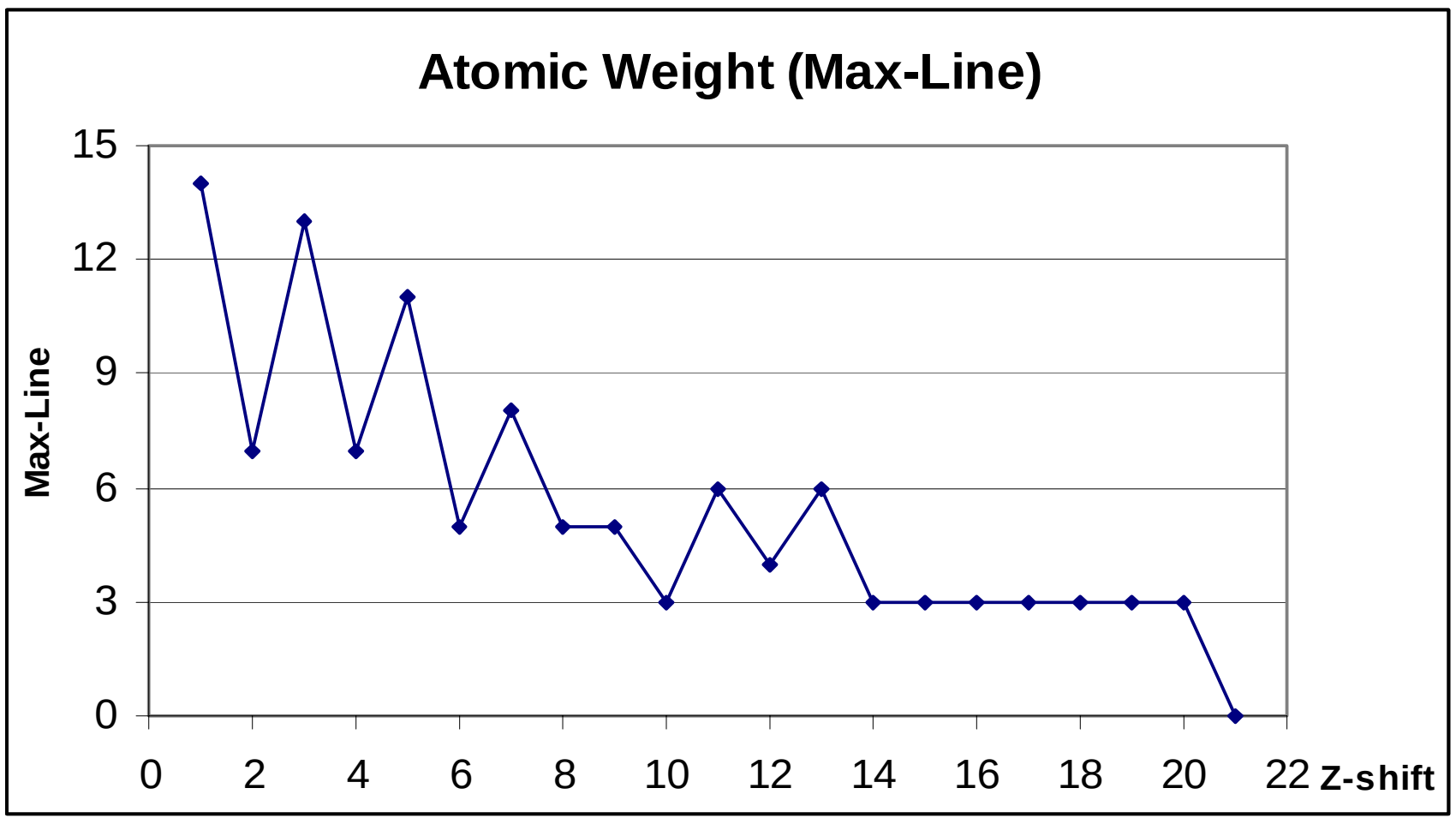

We may now pass to consider Recurrence Quantification Analysis in the case of Mass Numbers, $A(Z)$. The results for %Rec., %Det., ENT., and Max Line are given in Table2.

**Table 2**

| Z-shift | %Rec. | %Det. | Entropy | Max-Line |
|---|---|---|---|---|
| 1 | 1.20 | 77.50 | 1.37 | 14 |
| 2 | 1.33 | 30.23 | 0.92 | 7 |
| 3 | 1.23 | 41.03 | 1.00 | 13 |
| 4 | 1.43 | 36.36 | 0.81 | 7 |
| 5 | 1.23 | 37.84 | 1.00 | 11 |
| 6 | 1.30 | 21.05 | 1.00 | 5 |
| 7 | 1.09 | 38.71 | 1.00 | 8 |
| 8 | 1.44 | 20.00 | 1.00 | 5 |
| 9 | 1.19 | 31.25 | 1.00 | 6 |
| 10 | 1.41 | 8.11 | 0.00 | 3 |
| 11 | 1.25 | 18.75 | 0.00 | 6 |
| 12 | 1.33 | 21.21 | 1.00 | 4 |
| 13 | 1.28 | 41.94 | 1.59 | 6 |
| 14 | 1.58 | 27.03 | 0.92 | 4 |
| 15 | 1.14 | 46.15 | 1.59 | 6 |
| 16 | 1.45 | 9.38 | 0.00 | 3 |
| 17 | 1.31 | 21.43 | 0.00 | 3 |
| 18 | 1.59 | 21.21 | 1.00 | 4 |
| 19 | 1.19 | 25.00 | 0.00 | 3 |
| 20 | 0.92 | 16.67 | 0.00 | 3 |
| 21 | 0.79 | 0.00 | - | - |

%Rec. remains rather constant in correspondence of the different $Z-shift$ values with some fluctuations taking minima values mainly at $Z-shift = 7,9,11,13,15,..,19$.
A graph is given in Fig.41 (a, b, c, d)

**Fig.41a**

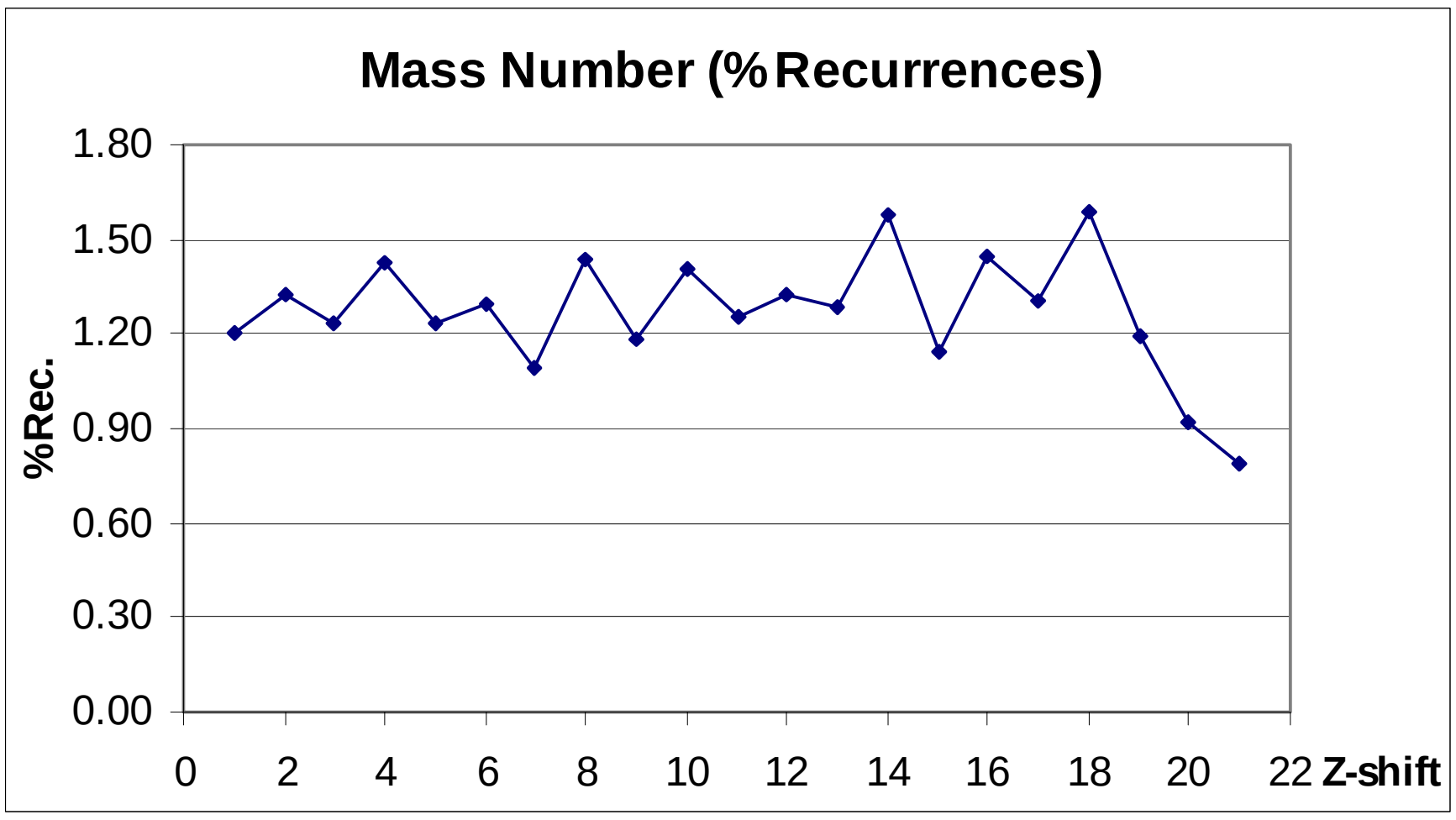

Fig. 41b

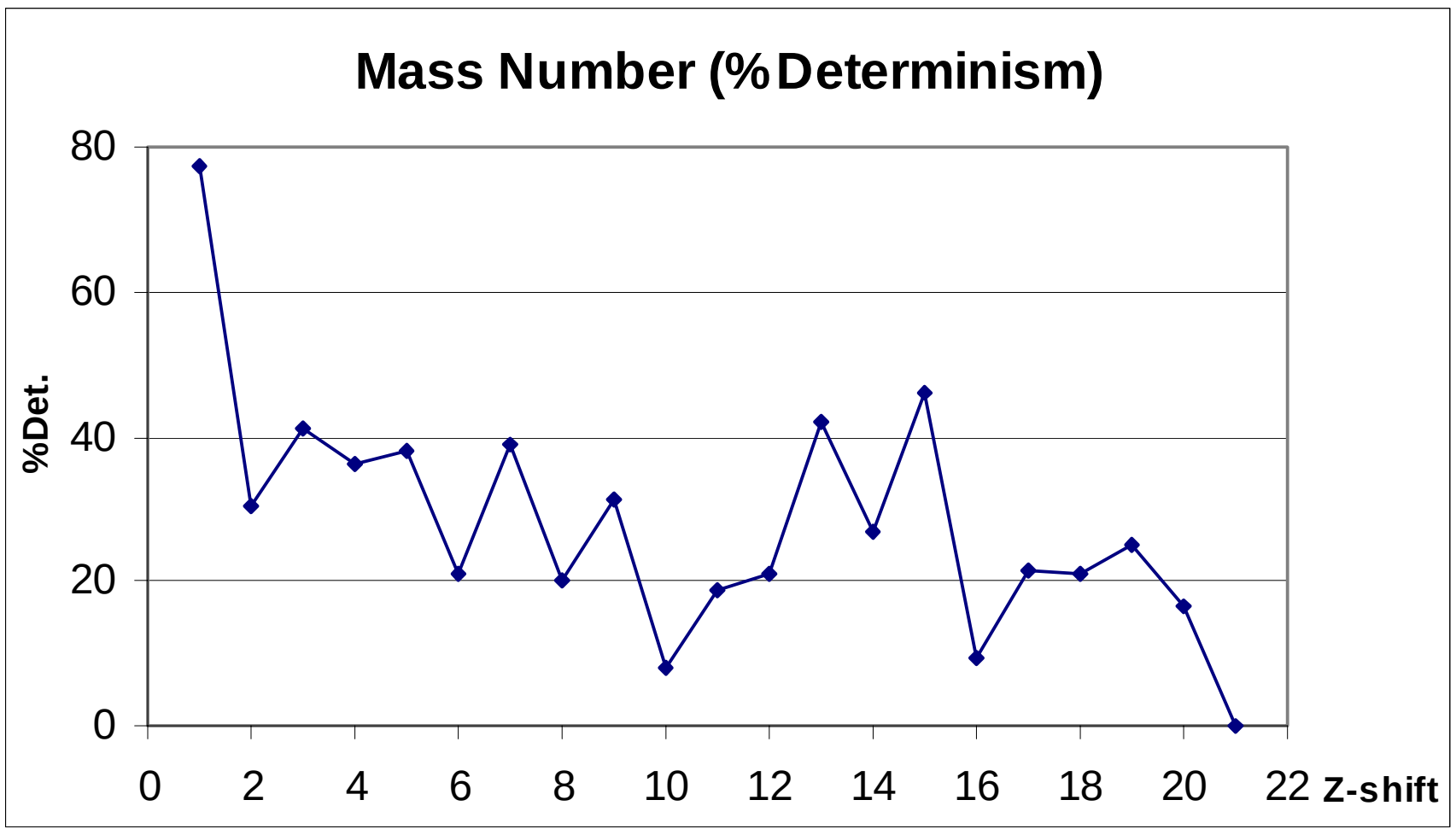

Fig. 41c

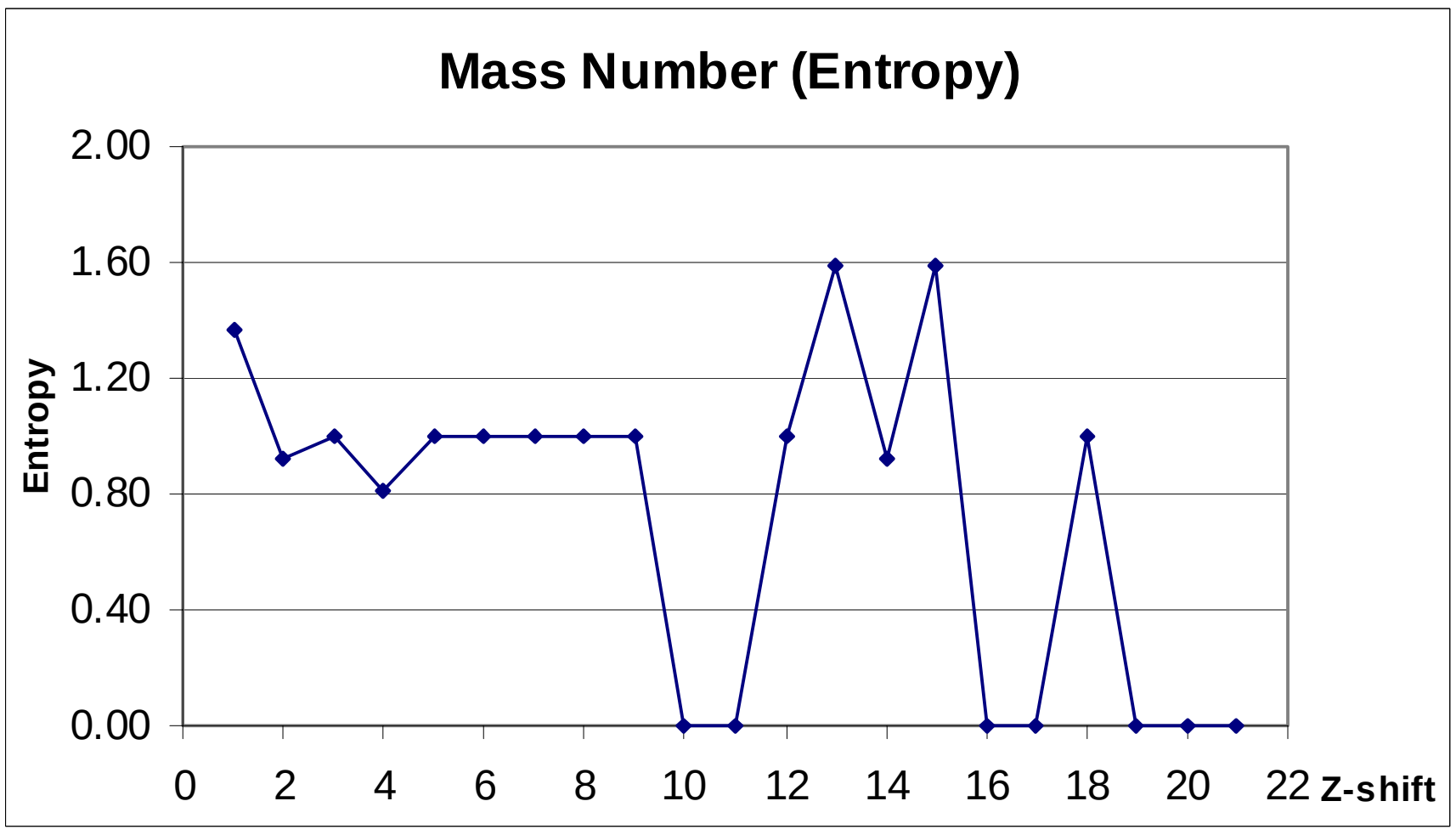

Fig. 41d

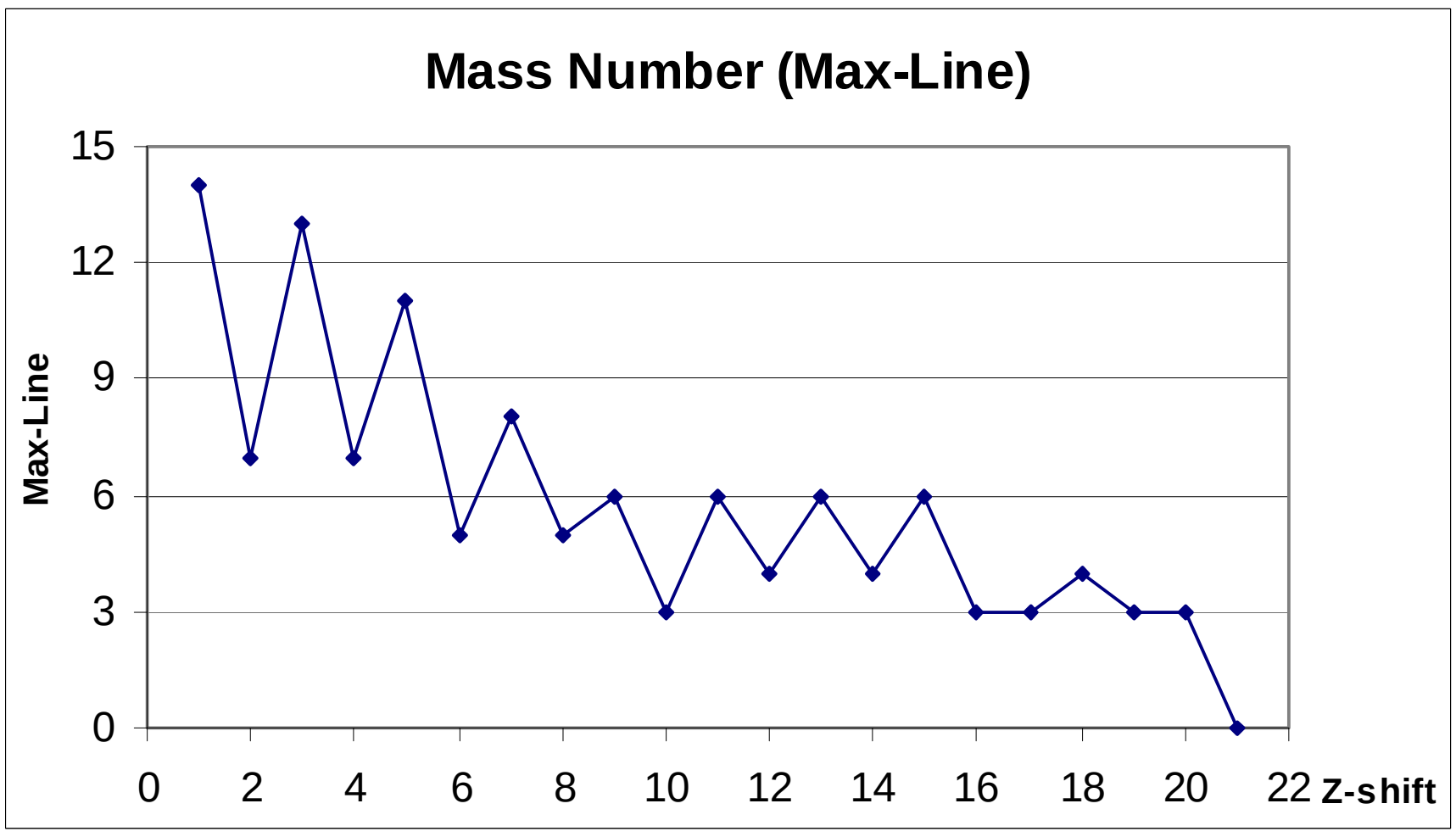

%Det. assumes rather low values also with a length Line L=3. It oscillates among maxima and minima for increasing values of $Z-shift$ as it is pictured in Fig.41b. Significantly, %Det. goes definitively to zero starting with $Z-shift$ =21.

In Fig.42 we have the comparison of %Det of Atomic Weights respect to %Det of Mass Numbers.

**Fig.42**

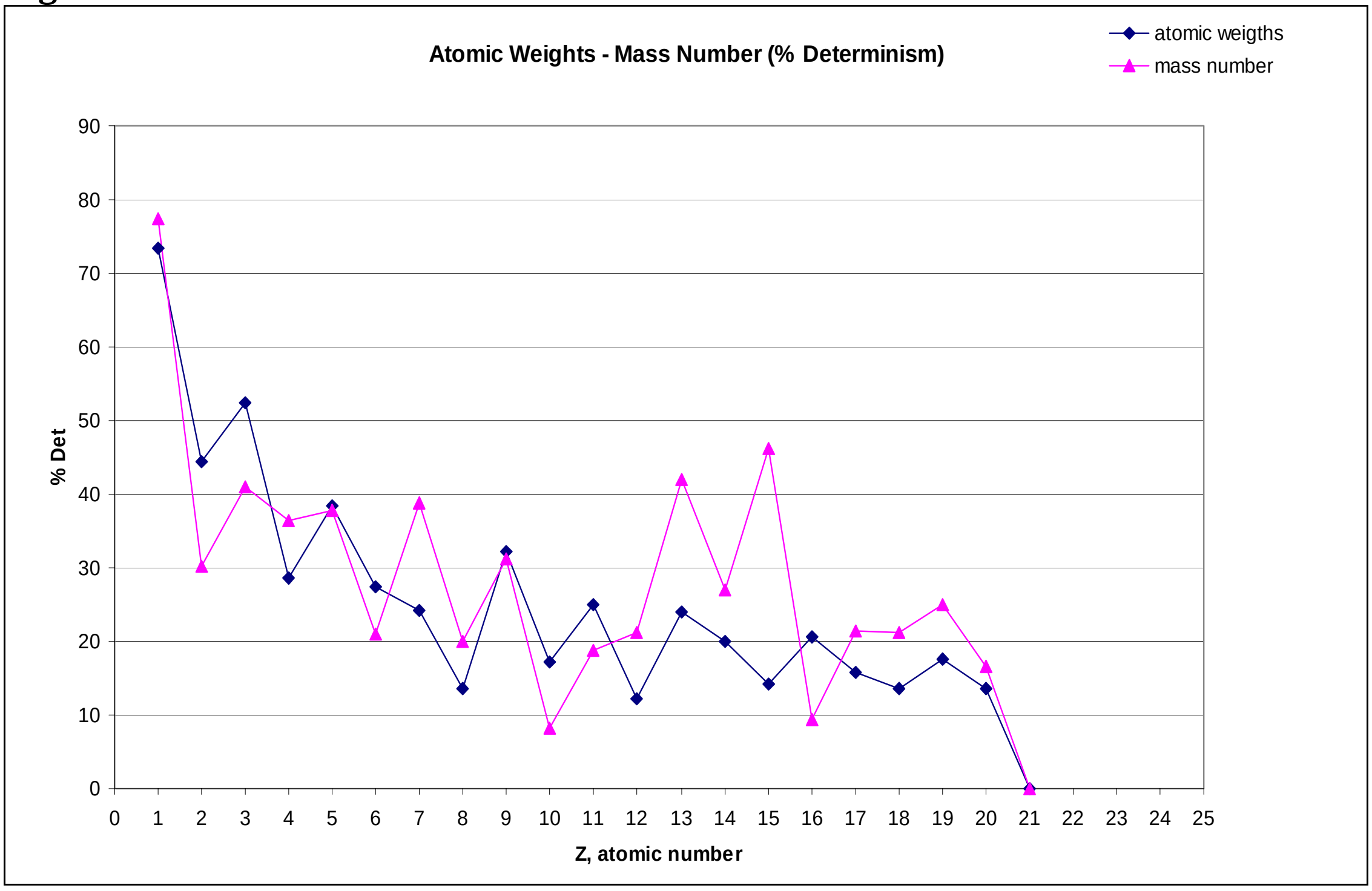

Looking at the results given in Tables 1 and 2 and linked figures, we deduce that for different values of $Z-shift$, the corresponding values of %Rec tend to show fluctuations. As it is well known, %Rec indicates in some manner presence of pseudoperiodicities. Therefore the rather small fluctuations of %Rec indicate that we are in presence of a mechanism of increasing mass that tends to preserve some kind of periodicity and self-resemblance with rather modest fluctuations The more interesting datum is given by %Det. In this case we have more marked oscillations showing that in the process of increasing mass of stable atomic nuclei we have phase of increasing stability as opposed to phases of decrease stability. Here the law is the mechanism of addition of nucleons that is realized at each step between the given nucleus and its subsequent as considered in our phase space representation. %Det oscillations indicate that the process of progressively addition of nucleons in nuclei happens on the basis of a complex non linear mechanism in which the determinism and thus the same predictability of subsequent Mass Number and /or Atomic Weights is very complex and so distant from a simple and linear regime of addition of matter that we have expected to hold for a very long time. Looking at the values of Entropy, expressed in bits, one finds that also in this case oscillations are dominant for increasing values of Z-shift. The same happens for MaxLine whose inverse gives estimation of the divergence of the system in consideration giving direct indication of a possible chaotic regime.
In conclusion, by using RQA we conclude that the mechanism of increasing mass in atomic nuclei is rather periodical and self-resemblance. We have obtained marked oscillations for the values of RQA variables. The important thing to remember here is that we are operating in a reconstructed phase space that takes into account o more an isolated nucleus as in the classical nuclear physics discussions, but each time pairs of nuclei in the embedded space with dimension D=2.The deriving behaviour of the mechanism of increasing mass of atomic nuclei evidences in this case all its complexity. We have now set of nuclei that evidence their oscillatory behaviour for % Rec, %Det, Entropy and Max Line.Such oscillatory behaviours of classes of nuclei result obviously connected to “periodicities” and mainly to classes of similarities that also stable nuclei seem to exhibit. The marked variations in the values of determinism indicate that the whole process results rather complex and it is regulated from phases of more stability and subsequent phases of increased instability.

In order to conclude such kind of research, and to confirm the new results that we have here indicated, we have performed the last kind analysis. In this last case we have in some manner overturned the scheme of the previous analysis in the sense that we have selected an embedding dimension D=1. The reader will remember that results by FFN gave same uncertainty in selecting the values D=1 or D=2. Our previous RQA was performed by using D=2 . In this final exploration we use D=1. In this condition of analysis a given value of delay and thus , in our case of $Z - shift$ , has no more sense . Each point in phase space is given by a value of $W_a(Z)$ or of $A(Z)$. To use RQA we have to select a distance , that is to say a Radius R. Using Euclidean distance , R will result to be the difference $\Delta A$ between two values of Mass Numbers in the case one utilizes $A(Z)$ for the analysis. In conclusion we have

$A_1 = Z_1 + N_1$ , $A_2 = Z_2 + N_2$ .

The distance , R, to use in RQA will result to be given

$\Delta A = \Delta Z + \Delta N$

We decided to use RQA considering L=3 as Line Length and R increasing step by step from 1 to209. In this manner we calculated %Rec, %Det, Entropy and Max Line, for increasing values of $Z = 1,2,3,.....$ . Note that in such kind of analysis we used also shuffled data in order to ascertain the validity of the obtained results. In addition , on the obtained data, we used also a Wald-Wolfowitz run test that we executed on %Det and on %Det / %Rec, and the probability that the results were obtained by chance, was found to be <0.001.

The results are now given in Tables 3, 4 and in Figures 43, 44, 45, 46, 47.

**Table 3**: Results of Recurrences quantification analysis of Mass Number with embedding D=1 and distance R ranging from 1 to 209

| R-distance | % Rec. | % Det. | %Det./%Rec. | R-distance | % Rec. | % Det. | %Det./%Rec. | R-distance | % Rec. | % Det. | %Det./%Rec. |
|---|---|---|---|---|---|---|---|---|---|---|---|
| 1 | 0.029 | 0.000 | 0.000 | 71 | 42.051 | 98.742 | 2.348 | 141 | 71.907 | 99.55 | 1.384 |
| 2 | 1.029 | 8.571 | 8.329 | 72 | 42.051 | 98.742 | 2.348 | 142 | 72.436 | 99.513 | 1.374 |
| 3 | 1.381 | 23.404 | 16.947 | 73 | 42.786 | 98.283 | 2.297 | 143 | 72.436 | 99.513 | 1.374 |
| 4 | 1.381 | 23.404 | 16.947 | 74 | 43.432 | 98.241 | 2.262 | 144 | 72.877 | 99.395 | 1.364 |
| 5 | 2.204 | 66.667 | 30.248 | 75 | 43.432 | 98.241 | 2.262 | 145 | 73.288 | 99.399 | 1.356 |
| 6 | 3.115 | 75.472 | 24.229 | 76 | 44.167 | 99.069 | 2.243 | 146 | 73.288 | 99.399 | 1.356 |
| 7 | 4.29 | 82.192 | 19.159 | 77 | 44.902 | 98.822 | 2.201 | 147 | 73.817 | 99.602 | 1.349 |
| 8 | 4.29 | 82.192 | 19.159 | 78 | 45.519 | 98.773 | 2.170 | 148 | 74.376 | 99.526 | 1.338 |
| 9 | 5.172 | 86.364 | 16.698 | 79 | 45.519 | 98.773 | 2.170 | 149 | 74.904 | 99.451 | 1.328 |
| 10 | 6.024 | 89.268 | 14.819 | 80 | 46.195 | 98.885 | 2.141 | 150 | 74.904 | 99.451 | 1.328 |
| 11 | 6.024 | 89.269 | 14.819 | 81 | 46.812 | 98.87 | 2.112 | 151 | 75.316 | 99.532 | 1.322 |
| 12 | 7.082 | 90.45 | 12.772 | 82 | 46.812 | 98.87 | 2.112 | 152 | 75.727 | 99.728 | 1.317 |
| 13 | 7.875 | 88.806 | 11.277 | 83 | 47.458 | 98.885 | 2.084 | 153 | 75.725 | 99.728 | 1.317 |
| 14 | 8.639 | 91.156 | 10.552 | 84 | 48.193 | 98.963 | 2.053 | 154 | 76.197 | 99.691 | 1.308 |
| 15 | 8.639 | 91.156 | 10.552 | 85 | 48.78 | 98.675 | 2.023 | 155 | 76.609 | 99.501 | 1.299 |
| 16 | 9.58 | 94.785 | 9.894 | 86 | 48.78 | 98.675 | 2.023 | 156 | 77.05 | 99.657 | 1.293 |
| 17 | 10.638 | 94.475 | 8.881 | 87 | 49.398 | 98.989 | 2.004 | 157 | 77.05 | 99.657 | 1.293 |
| 18 | 10.638 | 94.475 | 8.881 | 88 | 50.132 | 99.062 | 1.976 | 158 | 77.667 | 99.659 | 1.283 |
| 19 | 11.666 | 93.703 | 8.032 | 89 | 50.132 | 99.062 | 1.976 | 159 | 78.078 | 99.511 | 1.275 |
| 20 | 12.401 | 95.261 | 7.682 | 90 | 50.779 | 99.016 | 1.950 | 160 | 78.078 | 99.511 | 1.275 |
| 21 | 13.282 | 95.575 | 7.196 | 91 | 51.455 | 99.029 | 1.925 | 161 | 78.343 | 99.4 | 1.269 |
| 22 | 13.282 | 95.575 | 7.196 | 92 | 52.16 | 99.155 | 1.901 | 162 | 78.754 | 99.664 | 1.266 |
| 23 | 14.252 | 94.433 | 6.626 | 93 | 52.16 | 99.155 | 1.901 | 163 | 79.224 | 99.703 | 1.258 |
| 24 | 15.046 | 95.508 | 6.348 | 94 | 52.806 | 98.998 | 1.875 | 164 | 79.224 | 99.703 | 1.258 |
| 25 | 15.046 | 95.508 | 6.348 | 95 | 53.425 | 99.065 | 1.854 | 165 | 79.783 | 99.595 | 1.248 |
| 26 | 16.927 | 97.232 | 5.744 | 96 | 53.425 | 99.065 | 1.854 | 166 | 80.194 | 99.45 | 1.240 |
| 27 | 18.513 | 96.508 | 5.213 | 97 | 54.041 | 99.402 | 1.839 | 167 | 80.635 | 99.745 | 1.237 |
| 28 | 17.602 | 95.993 | 5.454 | 98 | 54.687 | 99.087 | 1.812 | 168 | 80.635 | 99.745 | 1.237 |
| 29 | 17.602 | 95.993 | 5.454 | 99 | 55.245 | 98.83 | 1.789 | 169 | 81.105 | 99.565 | 1.228 |
| 30 | 18.513 | 96.508 | 5.213 | 100 | 55.245 | 98.83 | 1.789 | 170 | 81.399 | 99.458 | 1.222 |
| 31 | 19.16 | 96.626 | 5.043 | 101 | 55.804 | 99.052 | 1.775 | 171 | 81.399 | 99.458 | 1.222 |
| 32 | 19.16 | 96.625 | 5.043 | 102 | 56.303 | 98.904 | 1.757 | 172 | 81.781 | 99.748 | 1.220 |
| 33 | 19.982 | 96.176 | 4.813 | 103 | 56.92 | 99.019 | 1.740 | 173 | 82.222 | 99.607 | 1.211 |
| 34 | 21.04 | 97.067 | 4.613 | 104 | 56.92 | 99.019 | 1.740 | 174 | 82.662 | 99.523 | 1.204 |
| 35 | 21.863 | 97.312 | 4.451 | 105 | 57.743 | 99.237 | 1.719 | 175 | 82.662 | 99.573 | 1.205 |
| 36 | 21.863 | 97.312 | 4.451 | 106 | 58.419 | 98.994 | 1.695 | 176 | 83.133 | 99.611 | 1.198 |
| 37 | 22.656 | 97.017 | 4.282 | 107 | 58.419 | 98.994 | 1.695 | 177 | 83.368 | 99.612 | 1.195 |
| 38 | 23.45 | 97.243 | 4.147 | 108 | 58.919 | 99.202 | 1.684 | 178 | 83.368 | 99.612 | 1.195 |
| 39 | 24.42 | 96.51 | 3.952 | 109 | 59.536 | 99.112 | 1.665 | 179 | 83.75 | 99.684 | 1.190 |
| 40 | 24.42 | 96.51 | 3.952 | 110 | 60.182 | 99.121 | 1.647 | 180 | 84.337 | 99.617 | 1.181 |
| 41 | 25.272 | 96.86 | 3.833 | 111 | 60.182 | 99.121 | 1.647 | 181 | 84.69 | 99.722 | 1.177 |
| 42 | 25.918 | 96.485 | 3.723 | 112 | 60.711 | 99.177 | 1.634 | 182 | 84.69 | 99.722 | 1.177 |
| 43 | 25.918 | 96.485 | 3.723 | 113 | 61.387 | 99.234 | 1.617 | 183 | 84.984 | 99.654 | 1.173 |
| 44 | 26.8 | 96.82 | 3.613 | 114 | 61.387 | 99.234 | 1.617 | 184 | 85.307 | 99.724 | 1.169 |
| 45 | 27.593 | 97.551 | 3.535 | 115 | 62.033 | 99.337 | 1.601 | 185 | 85.307 | 99.724 | 1.169 |
| 46 | 28.299 | 98.027 | 3.464 | 116 | 62.504 | 99.436 | 1.591 | 186 | 85.688 | 99.863 | 1.165 |
| 47 | 28.299 | 98.027 | 3.464 | 117 | 63.033 | 99.441 | 1.578 | 187 | 86.13 | 99.693 | 1.157 |
| 48 | 29.151 | 97.984 | 3.361 | 118 | 63.033 | 99.441 | 1.578 | 188 | 86.483 | 99.728 | 1.153 |
| 49 | 29.944 | 97.544 | 3.258 | 119 | 63.562 | 99.353 | 1.563 | 189 | 86.483 | 99.728 | 1.153 |
| 50 | 29.944 | 97.547 | 3.258 | 120 | 64.091 | 99.358 | 1.550 | 190 | 86.835 | 99.763 | 1.149 |
| 51 | 30.767 | 98.376 | 3.197 | 121 | 64.091 | 99.358 | 1.550 | 191 | 87.129 | 99.865 | 1.146 |
| 52 | 31.472 | 98.039 | 3.115 | 122 | 64.678 | 99.5 | 1.538 | 192 | 87.129 | 99.865 | 1.146 |
| 53 | 32.119 | 98.079 | 3.054 | 123 | 65.207 | 99.459 | 1.525 | 193 | 87.423 | 99.832 | 1.142 |
| 54 | 32.119 | 98.079 | 3.054 | 124 | 65.795 | 99.285 | 1.509 | 194 | 87.775 | 99.766 | 1.137 |
| 55 | 30.059 | 98.489 | 3.277 | 125 | 65.795 | 99.285 | 1.509 | 195 | 88.099 | 99.666 | 1.131 |
| 56 | 33.911 | 98.44 | 2.903 | 126 | 66.441 | 99.513 | 1.498 | 196 | 88.099 | 99.666 | 1.131 |
| 57 | 33.911 | 98.44 | 2.903 | 127 | 66.941 | 99.517 | 1.487 | 197 | 88.481 | 99.801 | 1.128 |
| 58 | 34.558 | 98.469 | 2.849 | 128 | 66.941 | 99.517 | 1.487 | 198 | 88.804 | 99.735 | 1.123 |
| 59 | 35.292 | 98.751 | 2.798 | 129 | 67.382 | 99.477 | 1.476 | 199 | 89.039 | 99.736 | 1.120 |
| 60 | 36.086 | 98.616 | 2.733 | 130 | 67.852 | 99.524 | 1.467 | 200 | 89.039 | 99.736 | 1.120 |
| 61 | 36.086 | 98.616 | 2.733 | 131 | 68.44 | 99.614 | 1.455 | 201 | 89.362 | 99.704 | 1.116 |
| 62 | 36.938 | 98.329 | 2.662 | 132 | 68.44 | 99.614 | 1.455 | 202 | 89.686 | 99.803 | 1.113 |
| 63 | 37.702 | 98.051 | 2.601 | 133 | 69.486 | 99.49 | 1.432 | 203 | 89.686 | 99.803 | 1.113 |
| 64 | 37.702 | 98.051 | 2.601 | 134 | 69.556 | 99.493 | 1.430 | 204 | 90.009 | 99.739 | 1.108 |
| 65 | 38.29 | 98.388 | 2.570 | 135 | 69.909 | 99.496 | 1.423 | 205 | 90.332 | 99.707 | 1.104 |
| 66 | 39.024 | 98.494 | 2.524 | 136 | 69.906 | 99.496 | 1.423 | 206 | 90.538 | 99.838 | 1.103 |
| 67 | 39.788 | 98.523 | 2.476 | 137 | 70.291 | 99.373 | 1.414 | 207 | 90.538 | 99.838 | 1.103 |
| 68 | 39.788 | 98.523 | 2.476 | 138 | 70.79 | 99.377 | 1.404 | 208 | 90.831 | 99.871 | 1.100 |
| 69 | 40.406 | 98.545 | 2.439 | 139 | 70.79 | 99.377 | 1.404 | 209 | 91.155 | 99.742 | 1.094 |
| 70 | 41.17 | 98.787 | 2.399 | 140 | 71.378 | 99.506 | 1.394 | | | | |

**Fig. 43**

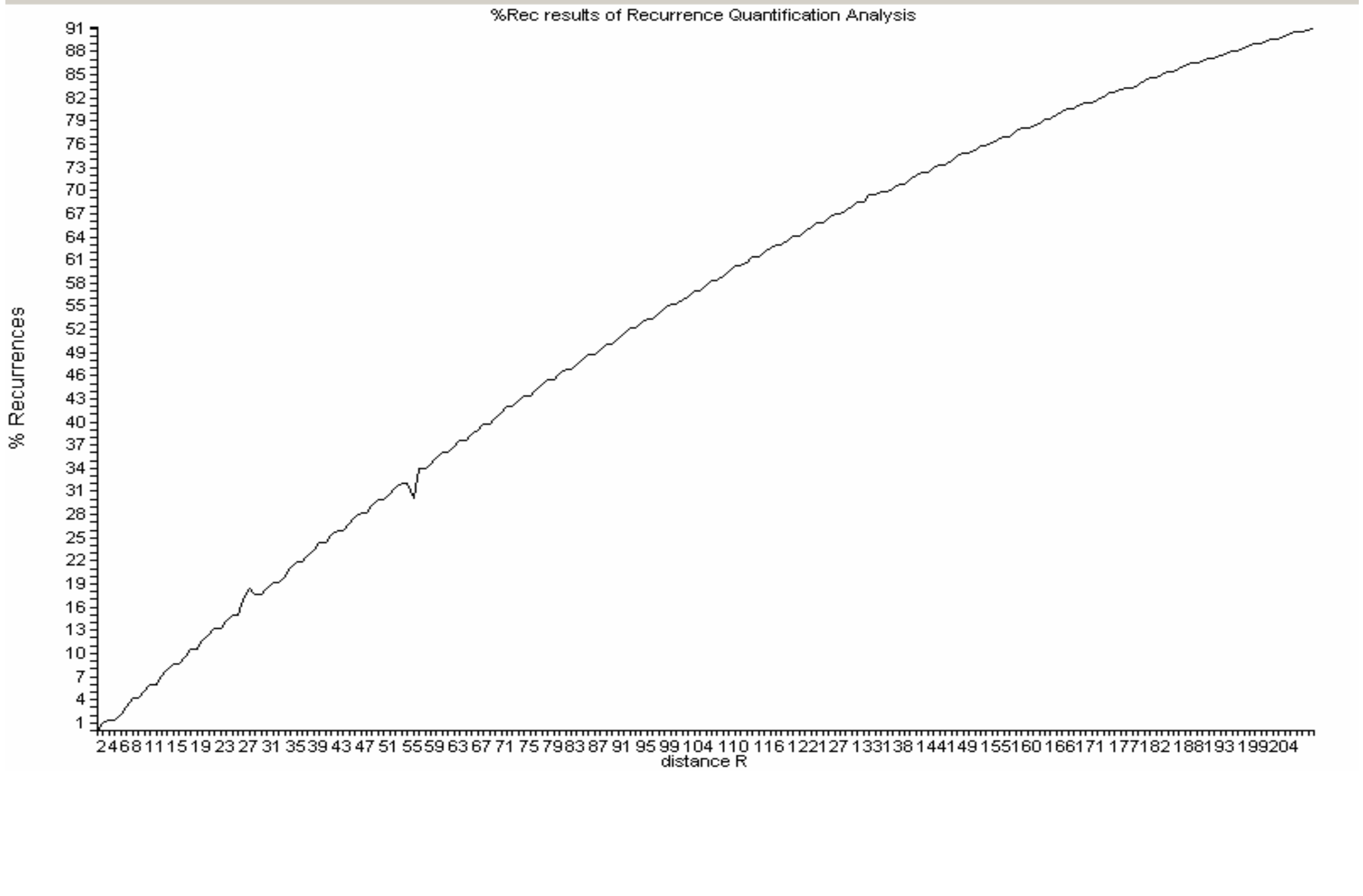

Fig. 44

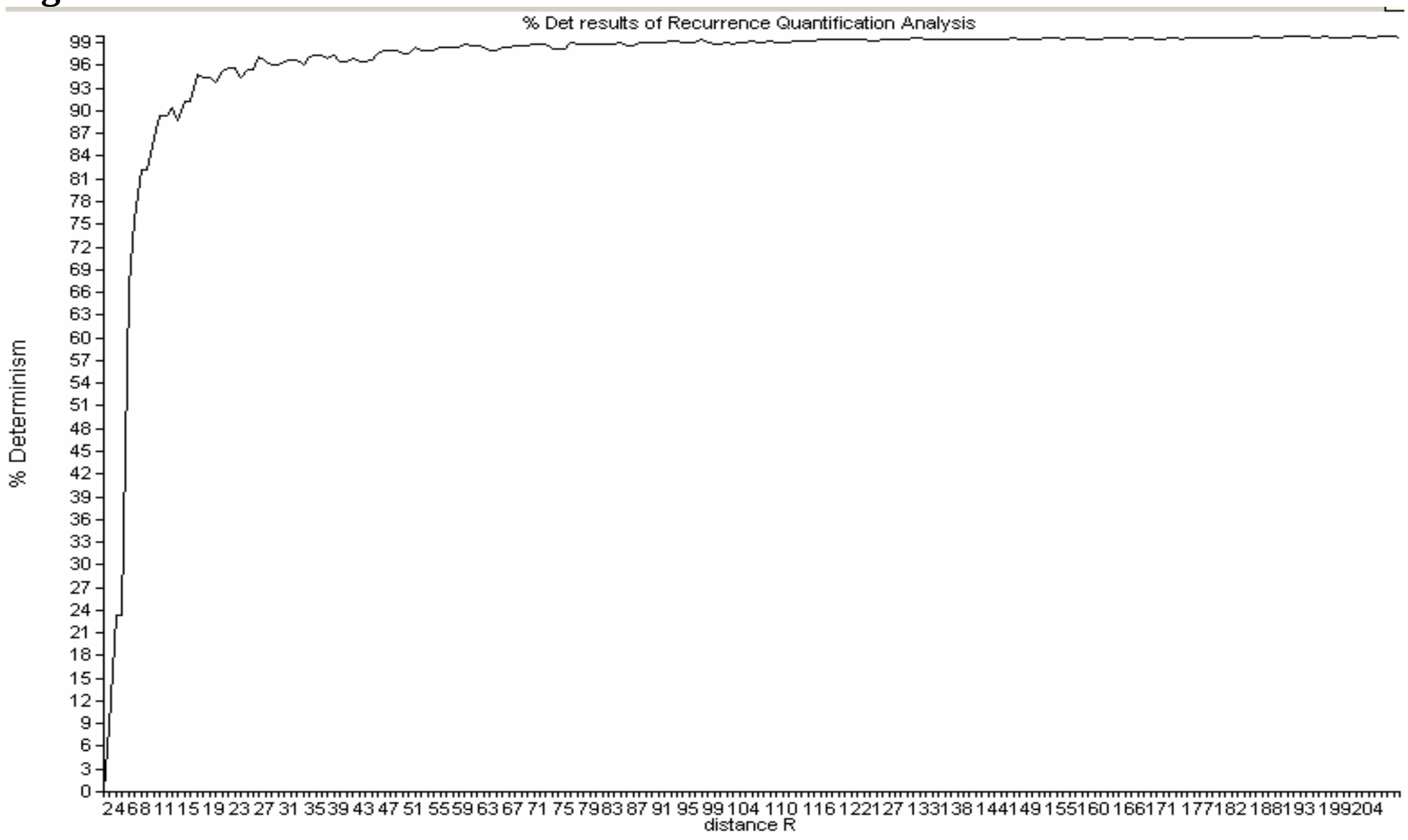
% Det results of Recurrence Quantification Analysis
% Determinism
distance R

Fig. 45

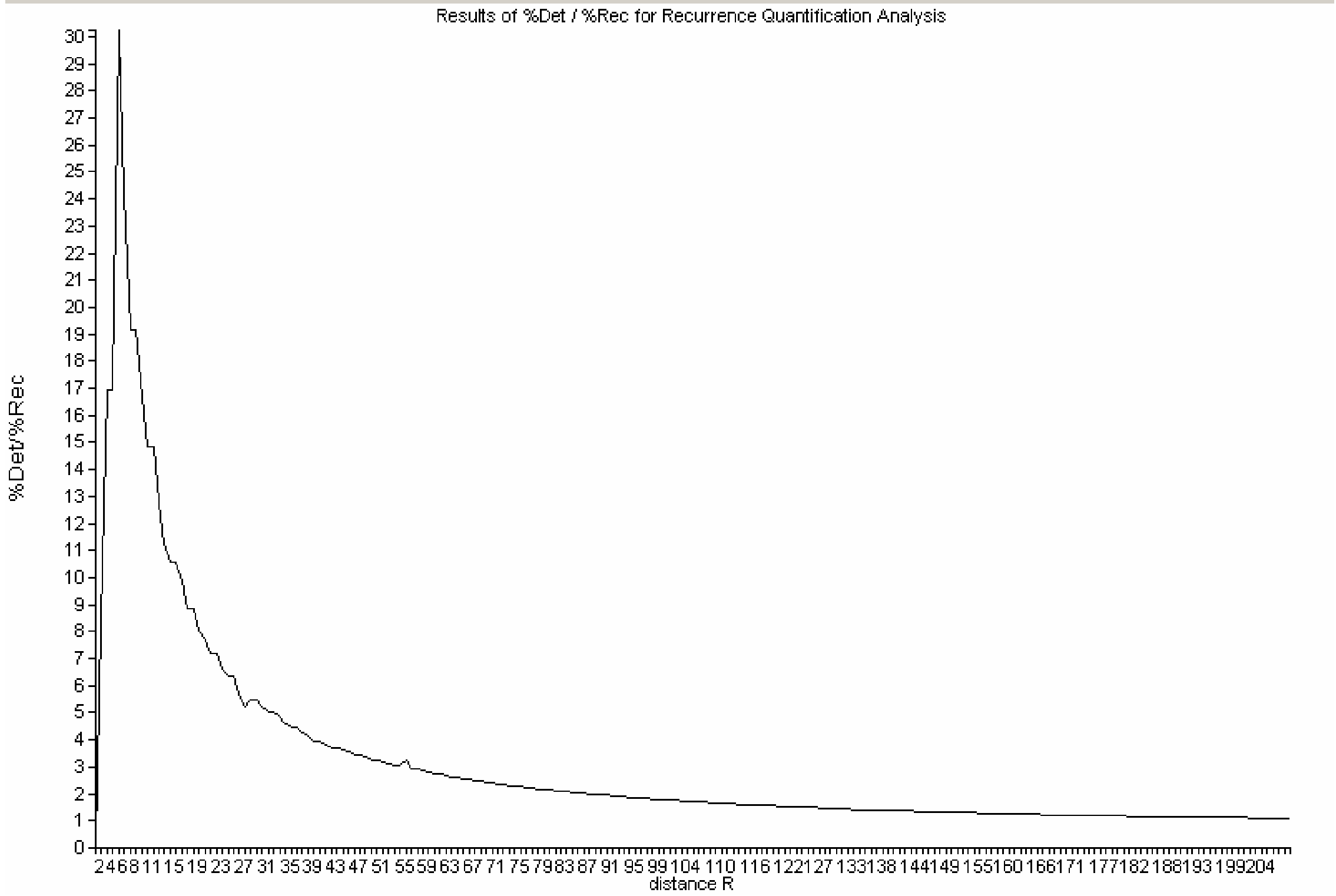
Results of %Det / %Rec for Recurrence Quantification Analysis
%Det/%Rec
distance R

**Fig. 46**

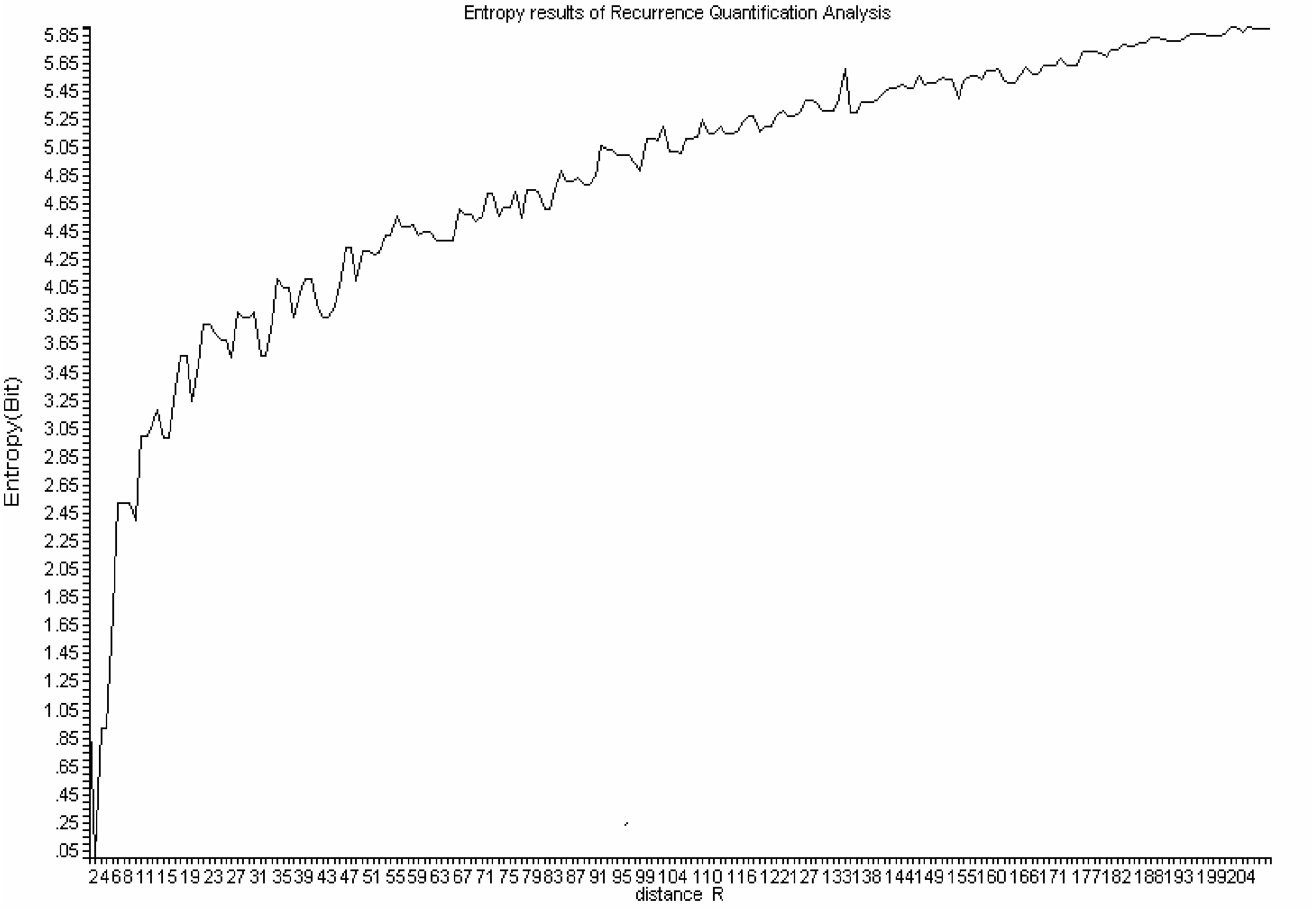

**Fig. 47**

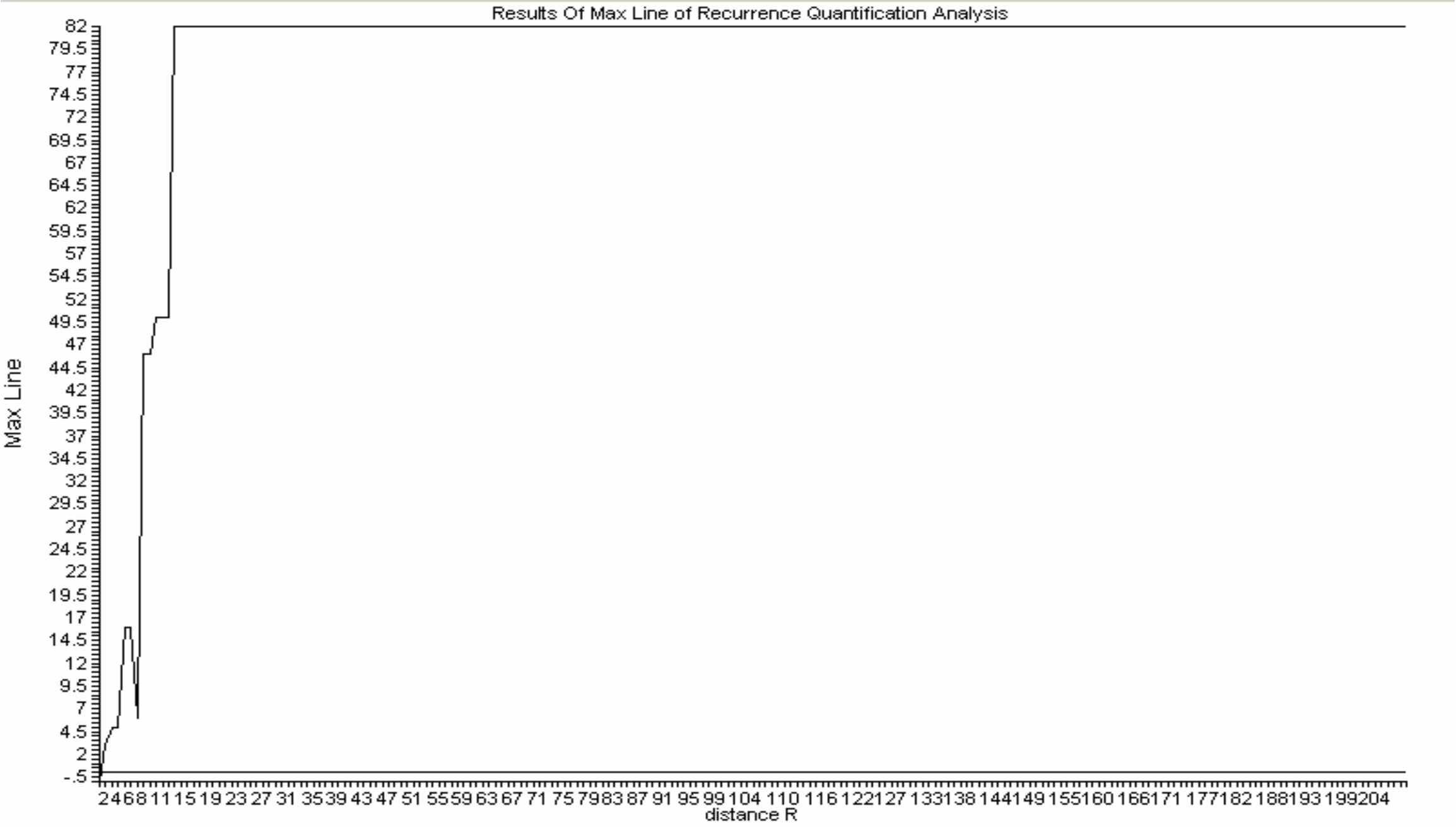

The obtained results may be considered of valuable interest since they indicate possible new properties for Mass Number of atomic nuclei.

At increasing values of Radius R, % Rec and % Det increase, as it is trivially expected in some general case, but the interesting new thing is that, after some regular increasing values of %Rec and %Det, occurring every two or three step, soon after the values of RQA variables reach values of stability that so remain for two steps in the increasing values of R. In other terms, in presence of increasing R, we have corresponding increasing values of % Rec, %Det, Entropy, followed by a phase in which, still for increasing R, the values of RQA variables remain instead constant.
This is certainly a new mechanism of increasing mass of atomic nuclei that deserves to be carefully explained.

**Table 4**

| d=2 | Z | N | Z | N | Δ Z | Δ N |
|---|---|---|---|---|---|---|
| | 3 | 4 | 4 | 5 | 1 | 1 |
| | 4 | 5 | 5 | 6 | 1 | 1 |
| | 6 | 6 | 7 | 7 | 1 | 1 |
| | 7 | 7 | 8 | 8 | 1 | 1 |
| | 26 | 30 | 28 | 30 | 2 | 0 |
| | 38 | 50 | 40 | 50 | 2 | 0 |
| | 52 | 78 | 54 | 78 | 2 | 0 |
| | 56 | 82 | 58 | 82 | 2 | 0 |
| | 57 | 82 | 59 | 82 | 2 | 0 |
| | 66 | 98 | 68 | 98 | 2 | 0 |
| | 77 | 116 | 78 | 117 | 1 | 1 |
| | 78 | 117 | 79 | 118 | 1 | 1 |

| d=3 | Z | N | Z | N | Δ Z | Δ N |
|---|---|---|---|---|---|---|
| | 1 | 0 | 2 | 2 | 1 | 2 |
| | 2 | 2 | 3 | 4 | 1 | 2 |
| | 4 | 5 | 6 | 6 | 2 | 1 |
| | 5 | 6 | 7 | 7 | 2 | 1 |
| | 8 | 8 | 9 | 10 | 1 | 2 |
| | 10 | 10 | 11 | 12 | 1 | 2 |
| | 12 | 12 | 13 | 14 | 1 | 2 |
| | 14 | 14 | 15 | 16 | 1 | 2 |
| | 16 | 16 | 17 | 18 | 1 | 2 |
| | 21 | 24 | 22 | 26 | 1 | 2 |
| | 22 | 26 | 23 | 28 | 1 | 2 |
| | 24 | 28 | 25 | 30 | 1 | 2 |
| | 25 | 30 | 28 | 30 | 3 | 0 |
| | 26 | 30 | 27 | 32 | 1 | 2 |
| | 37 | 48 | 38 | 50 | 1 | 2 |
| | 40 | 50 | 41 | 52 | 1 | 2 |
| | 43 | 56 | 44 | 58 | 1 | 2 |
| | 45 | 58 | 46 | 60 | 1 | 2 |
| | 52 | 78 | 55 | 78 | 3 | 0 |
| | 56 | 82 | 59 | 82 | 3 | 0 |
| | 59 | 82 | 60 | 84 | 1 | 2 |
| | 68 | 98 | 69 | 100 | 1 | 2 |
| | 73 | 108 | 74 | 110 | 1 | 2 |
| | 74 | 110 | 75 | 112 | 1 | 2 |
| | 76 | 116 | 78 | 117 | 2 | 1 |
| | 80 | 122 | 81 | 124 | 1 | 2 |
| | 81 | 124 | 82 | 126 | 1 | 2 |

| d=4 | Z | N | Z | N | Δ Z | Δ N |
|---|---|---|---|---|---|---|
| | 3 | 4 | 5 | 6 | 2 | 2 |
| | 6 | 6 | 8 | 8 | 2 | 2 |
| | 8 | 8 | 10 | 10 | 2 | 2 |
| | 9 | 10 | 11 | 12 | 2 | 2 |
| | 10 | 10 | 12 | 12 | 2 | 2 |
| | 11 | 12 | 13 | 14 | 2 | 2 |
| | 12 | 12 | 14 | 14 | 2 | 2 |
| | 13 | 14 | 15 | 16 | 2 | 2 |
| | 14 | 14 | 16 | 16 | 2 | 2 |
| | 15 | 16 | 17 | 18 | 2 | 2 |
| | 17 | 18 | 19 | 20 | 2 | 2 |
| | 22 | 26 | 24 | 28 | 2 | 2 |
| | 23 | 28 | 25 | 30 | 2 | 2 |
| | 24 | 28 | 26 | 30 | 2 | 2 |
| | 25 | 30 | 27 | 32 | 2 | 2 |
| | 27 | 32 | 29 | 34 | 2 | 2 |
| | 33 | 42 | 35 | 44 | 2 | 2 |
| | 34 | 46 | 36 | 48 | 2 | 2 |
| | 36 | 48 | 38 | 50 | 2 | 2 |
| | 37 | 48 | 39 | 50 | 2 | 2 |
| | 39 | 50 | 41 | 52 | 2 | 2 |
| | 42 | 56 | 44 | 58 | 2 | 2 |
| | 43 | 56 | 45 | 58 | 2 | 2 |
| | 44 | 58 | 46 | 60 | 2 | 2 |
| | 45 | 58 | 47 | 60 | 2 | 2 |
| | 58 | 82 | 60 | 84 | 2 | 2 |
| | 59 | 82 | 61 | 84 | 2 | 2 |
| | 67 | 98 | 69 | 100 | 2 | 2 |
| | 72 | 108 | 74 | 110 | 2 | 2 |
| | 77 | 116 | 79 | 118 | 2 | 2 |
| | 81 | 124 | 83 | 126 | 2 | 2 |

| d=6 | Z | N | Z | N | Δ Z | Δ N |
|---|---|---|---|---|---|---|
| | 1 | 0 | 3 | 4 | 2 | 4 |
| | 7 | 7 | 10 | 10 | 3 | 3 |
| | 19 | 20 | 21 | 24 | 2 | 4 |
| | 21 | 24 | 23 | 28 | 2 | 4 |
| | 24 | 28 | 28 | 30 | 4 | 2 |
| | 28 | 30 | 30 | 34 | 2 | 4 |
| | 29 | 34 | 31 | 38 | 2 | 4 |
| | 31 | 38 | 33 | 42 | 2 | 4 |
| | 32 | 42 | 34 | 46 | 2 | 4 |
| | 35 | 44 | 37 | 48 | 2 | 4 |
| | 36 | 48 | 40 | 50 | 4 | 2 |
| | 41 | 52 | 43 | 56 | 2 | 4 |
| | 48 | 66 | 50 | 70 | 2 | 4 |
| | 49 | 66 | 51 | 70 | 2 | 4 |
| | 51 | 70 | 53 | 74 | 2 | 4 |
| | 53 | 74 | 55 | 78 | 2 | 4 |
| | 54 | 78 | 56 | 82 | 2 | 4 |
| | 55 | 78 | 57 | 82 | 2 | 4 |
| | 56 | 82 | 60 | 84 | 4 | 2 |
| | 57 | 82 | 61 | 84 | 4 | 2 |
| | 62 | 90 | 64 | 94 | 2 | 4 |
| | 63 | 90 | 65 | 94 | 2 | 4 |
| | 64 | 94 | 66 | 98 | 2 | 4 |
| | 65 | 94 | 67 | 98 | 2 | 4 |
| | 69 | 100 | 71 | 104 | 2 | 4 |
| | 70 | 104 | 72 | 108 | 2 | 4 |
| | 71 | 104 | 73 | 108 | 2 | 4 |
| | 73 | 108 | 75 | 112 | 2 | 4 |
| | 75 | 112 | 77 | 116 | 2 | 4 |
| | 80 | 122 | 82 | 126 | 2 | 4 |

| d=7 | Z | N | Z | N | Δ Z | Δ N |
|---|---|---|---|---|---|---|
| | 2 | 2 | 5 | 6 | 3 | 4 |
| | 3 | 4 | 7 | 7 | 4 | 3 |
| | 4 | 5 | 8 | 8 | 4 | 3 |
| | 6 | 6 | 9 | 10 | 3 | 4 |
| | 8 | 8 | 11 | 12 | 3 | 4 |
| | 10 | 10 | 13 | 14 | 3 | 4 |
| | 12 | 12 | 15 | 16 | 3 | 4 |
| | 14 | 14 | 17 | 18 | 3 | 4 |
| | 16 | 16 | 19 | 20 | 3 | 4 |
| | 21 | 24 | 24 | 28 | 3 | 4 |
| | 22 | 26 | 25 | 30 | 3 | 4 |
| | 23 | 28 | 28 | 30 | 5 | 2 |
| | 24 | 28 | 27 | 32 | 3 | 4 |
| | 26 | 30 | 29 | 34 | 3 | 4 |
| | 43 | 56 | 46 | 60 | 3 | 4 |
| | 47 | 60 | 48 | 66 | 1 | 6 |
| | 48 | 66 | 51 | 70 | 3 | 4 |
| | 50 | 70 | 53 | 74 | 3 | 4 |
| | 54 | 78 | 57 | 82 | 3 | 4 |
| | 55 | 78 | 58 | 82 | 3 | 4 |
| | 56 | 82 | 61 | 84 | 5 | 2 |
| | 61 | 84 | 62 | 90 | 1 | 6 |
| | 62 | 90 | 65 | 94 | 3 | 4 |
| | 64 | 94 | 67 | 98 | 3 | 4 |
| | 65 | 94 | 68 | 98 | 3 | 4 |
| | 70 | 104 | 73 | 108 | 3 | 4 |
| | 72 | 108 | 75 | 112 | 3 | 4 |
| | 78 | 117 | 80 | 122 | 2 | 5 |
| | 80 | 122 | 83 | 126 | 3 | 4 |

| d=8 | Z | N | Z | N | Δ Z | Δ N |
|---|---|---|---|---|---|---|
| | 1 | 0 | 4 | 5 | 3 | 5 |
| | 2 | 2 | 6 | 6 | 4 | 4 |
| | 5 | 6 | 9 | 10 | 4 | 4 |
| | 6 | 6 | 10 | 10 | 4 | 4 |
| | 8 | 8 | 12 | 12 | 4 | 4 |
| | 9 | 10 | 13 | 14 | 4 | 4 |
| | 10 | 10 | 14 | 14 | 4 | 4 |
| | 11 | 12 | 15 | 16 | 4 | 4 |
| | 12 | 12 | 16 | 16 | 4 | 4 |
| | 13 | 14 | 17 | 18 | 4 | 4 |
| | 15 | 16 | 19 | 20 | 4 | 4 |
| | 16 | 16 | 18 | 22 | 2 | 6 |
| | 16 | 16 | 20 | 20 | 4 | 4 |
| | 18 | 22 | 22 | 26 | 4 | 4 |
| | 20 | 20 | 22 | 26 | 2 | 6 |
| | 22 | 26 | 26 | 30 | 4 | 4 |
| | 23 | 28 | 27 | 32 | 4 | 4 |
| | 25 | 30 | 29 | 34 | 4 | 4 |
| | 26 | 30 | 30 | 34 | 4 | 4 |
| | 34 | 46 | 38 | 50 | 4 | 4 |
| | 37 | 48 | 41 | 52 | 4 | 4 |
| | 40 | 50 | 42 | 56 | 2 | 6 |
| | 42 | 56 | 46 | 60 | 4 | 4 |
| | 43 | 56 | 47 | 60 | 4 | 4 |
| | 46 | 60 | 48 | 66 | 2 | 6 |
| | 47 | 60 | 49 | 66 | 2 | 6 |
| | 52 | 78 | 56 | 82 | 4 | 4 |
| | 54 | 78 | 58 | 82 | 4 | 4 |
| | 55 | 78 | 59 | 82 | 4 | 4 |
| | 60 | 84 | 62 | 90 | 2 | 6 |
| | 61 | 84 | 63 | 90 | 2 | 6 |
| | 64 | 94 | 68 | 98 | 4 | 4 |
| | 68 | 98 | 70 | 104 | 2 | 6 |
| | 74 | 110 | 76 | 116 | 2 | 6 |
| | 75 | 112 | 78 | 117 | 3 | 5 |
| | 79 | 118 | 81 | 124 | 2 | 6 |

| d=9 | Z | N | Z | N | ΔZ | ΔN |
|---|---|---|---|---|---|---|
| | 3 | 4 | 8 | 8 | 5 | 4 |
| | 5 | 6 | 10 | 10 | 5 | 4 |
| | 7 | 7 | 11 | 12 | 4 | 5 |
| | 9 | 10 | 14 | 14 | 5 | 4 |
| | 11 | 12 | 16 | 16 | 5 | 4 |
| | 15 | 16 | 18 | 22 | 3 | 6 |
| | 15 | 16 | 20 | 20 | 5 | 4 |
| | 19 | 20 | 22 | 26 | 3 | 6 |
| | 25 | 30 | 30 | 34 | 5 | 4 |
| | 33 | 42 | 36 | 48 | 3 | 6 |
| | 34 | 46 | 39 | 50 | 5 | 4 |
| | 35 | 44 | 38 | 50 | 3 | 6 |
| | 36 | 48 | 41 | 52 | 5 | 4 |
| | 39 | 50 | 42 | 56 | 3 | 6 |
| | 40 | 50 | 43 | 56 | 3 | 6 |
| | 41 | 52 | 44 | 58 | 3 | 6 |
| | 42 | 56 | 47 | 60 | 5 | 4 |
| | 46 | 60 | 49 | 66 | 3 | 6 |
| | 51 | 70 | 52 | 78 | 1 | 8 |
| | 52 | 78 | 57 | 82 | 5 | 4 |
| | 54 | 78 | 59 | 82 | 5 | 4 |
| | 60 | 84 | 63 | 90 | 3 | 6 |
| | 67 | 98 | 70 | 104 | 3 | 6 |
| | 68 | 98 | 71 | 104 | 3 | 6 |
| | 71 | 104 | 74 | 110 | 3 | 6 |
| | 74 | 110 | 77 | 116 | 3 | 6 |
| | 77 | 116 | 80 | 122 | 3 | 6 |

| d=10 | Z | N | Z | N | ΔZ | ΔN |
|---|---|---|---|---|---|---|
| | 1 | 0 | 5 | 6 | 4 | 6 |
| | 2 | 2 | 7 | 7 | 5 | 5 |
| | 4 | 5 | 9 | 10 | 5 | 5 |
| | 7 | 7 | 12 | 12 | 5 | 5 |
| | 17 | 18 | 21 | 24 | 4 | 6 |
| | 21 | 24 | 25 | 30 | 4 | 6 |
| | 22 | 26 | 28 | 30 | 6 | 4 |
| | 27 | 32 | 31 | 38 | 4 | 6 |
| | 30 | 34 | 32 | 42 | 2 | 8 |
| | 31 | 38 | 35 | 44 | 4 | 6 |
| | 32 | 42 | 36 | 48 | 4 | 6 |
| | 33 | 42 | 37 | 48 | 4 | 6 |
| | 34 | 46 | 40 | 50 | 6 | 4 |
| | 35 | 44 | 39 | 50 | 4 | 6 |
| | 38 | 50 | 42 | 56 | 4 | 6 |
| | 39 | 50 | 43 | 56 | 4 | 6 |
| | 41 | 52 | 45 | 58 | 4 | 6 |
| | 50 | 70 | 52 | 78 | 2 | 8 |
| | 52 | 78 | 58 | 82 | 6 | 4 |
| | 65 | 94 | 69 | 100 | 4 | 6 |
| | 66 | 98 | 70 | 104 | 4 | 6 |
| | 67 | 98 | 71 | 104 | 4 | 6 |
| | 70 | 104 | 74 | 110 | 4 | 6 |
| | 75 | 112 | 79 | 118 | 4 | 6 |
| | 76 | 116 | 80 | 122 | 4 | 6 |
| | 78 | 117 | 81 | 124 | 3 | 7 |

| d=11 | Z | N | Z | N | ΔZ | ΔN |
|---|---|---|---|---|---|---|
| | 1 | 0 | 6 | 6 | 5 | 6 |
| | 4 | 5 | 10 | 10 | 6 | 5 |
| | 6 | 6 | 11 | 12 | 5 | 6 |
| | 8 | 8 | 13 | 14 | 5 | 6 |
| | 10 | 10 | 15 | 16 | 5 | 6 |
| | 12 | 12 | 17 | 18 | 5 | 6 |
| | 14 | 14 | 19 | 20 | 5 | 6 |
| | 18 | 22 | 23 | 28 | 5 | 6 |
| | 20 | 20 | 23 | 28 | 3 | 8 |
| | 21 | 24 | 26 | 30 | 5 | 6 |
| | 22 | 26 | 27 | 32 | 5 | 6 |
| | 24 | 28 | 29 | 34 | 5 | 6 |
| | 28 | 30 | 31 | 38 | 3 | 8 |
| | 29 | 34 | 32 | 42 | 3 | 8 |
| | 30 | 34 | 33 | 42 | 3 | 8 |
| | 31 | 38 | 34 | 46 | 3 | 8 |
| | 32 | 42 | 37 | 48 | 5 | 6 |
| | 35 | 44 | 40 | 50 | 5 | 6 |
| | 38 | 50 | 43 | 56 | 5 | 6 |
| | 45 | 58 | 48 | 66 | 3 | 8 |
| | 51 | 70 | 54 | 78 | 3 | 8 |
| | 52 | 78 | 59 | 82 | 7 | 4 |
| | 53 | 74 | 56 | 82 | 3 | 8 |
| | 55 | 78 | 60 | 84 | 5 | 6 |
| | 59 | 82 | 62 | 90 | 3 | 8 |
| | 63 | 90 | 66 | 98 | 3 | 8 |
| | 64 | 94 | 69 | 100 | 5 | 6 |
| | 66 | 98 | 71 | 104 | 5 | 6 |
| | 69 | 100 | 72 | 108 | 3 | 8 |
| | 73 | 108 | 76 | 116 | 3 | 8 |
| | 74 | 110 | 78 | 117 | 4 | 7 |
| | 79 | 118 | 82 | 126 | 3 | 8 |

| d=13 | Z | N | Z | N | ΔZ | ΔN |
|---|---|---|---|---|---|---|
| | 1 | 0 | 7 | 7 | 6 | 7 |
| | 3 | 4 | 10 | 10 | 7 | 6 |
| | 5 | 6 | 12 | 12 | 7 | 6 |
| | 7 | 7 | 13 | 14 | 6 | 7 |
| | 9 | 10 | 16 | 16 | 7 | 6 |
| | 13 | 14 | 18 | 22 | 5 | 8 |
| | 13 | 14 | 20 | 20 | 7 | 6 |
| | 16 | 16 | 21 | 24 | 5 | 8 |
| | 17 | 18 | 22 | 26 | 5 | 8 |
| | 19 | 20 | 24 | 28 | 5 | 8 |
| | 21 | 24 | 28 | 30 | 7 | 6 |
| | 23 | 28 | 30 | 34 | 7 | 6 |
| | 26 | 30 | 31 | 38 | 5 | 8 |
| | 33 | 42 | 38 | 50 | 5 | 8 |
| | 34 | 46 | 41 | 52 | 7 | 6 |
| | 37 | 48 | 42 | 56 | 5 | 8 |
| | 39 | 50 | 44 | 58 | 5 | 8 |
| | 40 | 50 | 45 | 58 | 5 | 8 |
| | 41 | 52 | 46 | 60 | 5 | 8 |
| | 44 | 58 | 49 | 66 | 5 | 8 |
| | 47 | 60 | 50 | 70 | 3 | 10 |
| | 48 | 66 | 53 | 74 | 5 | 8 |
| | 50 | 70 | 55 | 78 | 5 | 8 |
| | 53 | 74 | 58 | 82 | 5 | 8 |
| | 54 | 78 | 61 | 84 | 7 | 6 |
| | 57 | 82 | 62 | 90 | 5 | 8 |
| | 58 | 82 | 63 | 90 | 5 | 8 |
| | 61 | 84 | 64 | 94 | 3 | 10 |
| | 62 | 90 | 67 | 98 | 5 | 8 |
| | 63 | 90 | 68 | 98 | 5 | 8 |
| | 70 | 104 | 75 | 112 | 5 | 8 |
| | 72 | 108 | 77 | 116 | 5 | 8 |
| | 74 | 110 | 79 | 118 | 5 | 8 |
| | 76 | 116 | 81 | 124 | 5 | 8 |
| | 78 | 117 | 82 | 126 | 4 | 9 |

| d=14 | Z | N | Z | N | ΔZ | ΔN |
|---|---|---|---|---|---|---|
| | 4 | 5 | 11 | 12 | 7 | 7 |
| | 7 | 7 | 14 | 14 | 7 | 7 |
| | 15 | 16 | 21 | 24 | 6 | 8 |
| | 21 | 24 | 27 | 32 | 6 | 8 |
| | 25 | 30 | 31 | 38 | 6 | 8 |
| | 32 | 42 | 38 | 50 | 6 | 8 |
| | 33 | 42 | 39 | 50 | 6 | 8 |
| | 35 | 44 | 41 | 52 | 6 | 8 |
| | 36 | 48 | 42 | 56 | 6 | 8 |
| | 37 | 48 | 43 | 56 | 6 | 8 |
| | 38 | 50 | 44 | 58 | 6 | 8 |
| | 39 | 50 | 45 | 58 | 6 | 8 |
| | 41 | 52 | 47 | 60 | 6 | 8 |
| | 46 | 60 | 50 | 70 | 4 | 10 |
| | 47 | 60 | 51 | 70 | 4 | 10 |
| | 52 | 78 | 60 | 84 | 8 | 6 |
| | 53 | 74 | 59 | 82 | 6 | 8 |
| | 56 | 82 | 62 | 90 | 6 | 8 |
| | 57 | 82 | 63 | 90 | 6 | 8 |
| | 60 | 84 | 64 | 94 | 4 | 10 |
| | 61 | 84 | 65 | 94 | 4 | 10 |
| | 62 | 90 | 68 | 98 | 6 | 8 |
| | 68 | 98 | 72 | 108 | 4 | 10 |
| | 73 | 108 | 78 | 117 | 5 | 9 |
| | 78 | 117 | 83 | 126 | 5 | 9 |

| d=15 | Z | N | Z | N | ΔZ | ΔN |
|---|---|---|---|---|---|---|
| | 1 | 0 | 8 | 8 | 7 | 8 |
| | 2 | 2 | 9 | 10 | 7 | 8 |
| | 4 | 5 | 12 | 12 | 8 | 7 |
| | 6 | 6 | 13 | 14 | 7 | 8 |
| | 8 | 8 | 15 | 16 | 7 | 8 |
| | 10 | 10 | 17 | 18 | 7 | 8 |
| | 12 | 12 | 19 | 20 | 7 | 8 |
| | 18 | 22 | 25 | 30 | 7 | 8 |
| | 20 | 20 | 25 | 30 | 5 | 10 |
| | 22 | 26 | 29 | 34 | 7 | 8 |
| | 27 | 32 | 32 | 42 | 5 | 10 |
| | 30 | 34 | 35 | 44 | 5 | 10 |
| | 31 | 38 | 36 | 48 | 5 | 10 |
| | 32 | 42 | 39 | 50 | 7 | 8 |
| | 33 | 42 | 40 | 50 | 7 | 8 |
| | 36 | 48 | 43 | 56 | 7 | 8 |
| | 38 | 50 | 45 | 58 | 7 | 8 |
| | 43 | 56 | 48 | 66 | 5 | 10 |
| | 46 | 60 | 51 | 70 | 5 | 10 |
| | 49 | 66 | 52 | 78 | 3 | 12 |
| | 52 | 78 | 61 | 84 | 9 | 6 |
| | 56 | 82 | 63 | 90 | 7 | 8 |
| | 60 | 84 | 65 | 94 | 5 | 10 |
| | 65 | 94 | 70 | 104 | 5 | 10 |
| | 67 | 98 | 72 | 108 | 5 | 10 |
| | 68 | 98 | 73 | 108 | 5 | 10 |
| | 69 | 100 | 74 | 110 | 5 | 10 |
| | 72 | 108 | 78 | 117 | 6 | 9 |
| | 75 | 112 | 80 | 122 | 5 | 10 |
| | 77 | 116 | 82 | 126 | 5 | 10 |

In Table 4 we give the scheme of increasing R corresponding to $\Delta A$ and the corresponding variations in the number of nucleons as they are induced step by step. Obviously this table 4 cannot be complete. However, the exposition of the process, also limited to few cases of interest, will contribute to elucidate the mechanism under consideration. In brief for $\Delta A = 2$ we have oscillation in the values of RQA variables but they soon after return to be stable for $\Delta A = 3$ and $\Delta A = 4$. After we pass to $\Delta A = 6$ where again RQA variables are unstable but they return to be stable for $\Delta A = 7$ and $\Delta A = 8$. The next step is $\Delta A = 9$ with instability, followed from stable values for $\Delta A = 10$ and $\Delta A = 11$. We may continue with $\Delta A = 13$ that is unstable but followed from stable $\Delta A = 14$ and $\Delta A = 15$. The same thing happens for $\Delta A = 16$ *or* 20 *or* 23 *or* 27 *or* 30 *or* 34 *or* 38........*or* 80 *or* ........*or* 119 *or*............*or* 180 *or*........... . To each given unstable $\Delta A$ value , will correspond two subsequent stable values that respectively will be given at $\Delta A = 17$ and $\Delta A = 18$; at $\Delta A = 21$ and $\Delta A = 22$, ……….. at $\Delta A = 120$ and $\Delta A = 121$, ….. at $\Delta A = 181$ and $\Delta A = 182$.

Instabilities are present every three or four increasing values of $\Delta A$. Systematically, each of them is followed by stable values at the two subsequent increments of $\Delta A$.

In conclusion the law seems as it follows: for each pair of nuclei , fixed the value of $\Delta A$ with unstable value of the RQA variables, the addition of one nucleon by two subsequent steps stabilizes the values of such variables. Obviously, for each selected value of $\Delta A$ we have a class of pair of nuclei as indicated as example in Table 4.

In conclusion, the use of RQA variables has cleared that we are in presence of new features for atomic nuclei that deserve to be properly explained. We intend to say that the next step of the present research should be now to link the different results that have been obtained with concrete evidences expressible in terms of basic concepts of nuclear physics. If on one hand some of such new findings are just evident by itself on the other hand we cannot ignore that in this paper we have moved more on the line of the notions as they are contained in the methods that we have used. More concretely : referring as example to the basic results that we have obtained by using RQA, and, in particular, to the last results as given by using embedding dimension D=1 and reported in Tables 3 and 4 and in Figures 43-47, we cannot ignore that we have to consider now pairs of nuclei with given $\Delta A$ and thus to identify pairs of subsequent stable nuclei and, following this way, to find some new regularities in Z, N and to give new classifications of nuclei to different groups using such regularities. In short, the results that we have obtained should reveal new regularities about ground states of nuclei not found so easily by other methods. Consequently, this new approach might be very useful and important. The aim is to pursue such research work in our future investigations.

## 8. Conclusions

In the present paper we have introduced a preliminary but complete analysis of Atomic Weights and Mass Number using the methods of non linear analysis.

We have obtained some results that appear to be of some interest in understanding the basic foundations of nuclear matter. As methodology, we have applied the tests of autocorrelation function and of Mutual Information. We have also provided to a reconstruction of the experimental data in phase space giving results on Lyapunov spectrum and Correlation Dimension. We have performed an analysis to establish the presence of a power law in data on Atomic Weights and Mass Number and such kind of analysis has been completed by using the technique of the variogram. The results seem to confirm the presence of a fractal regime in the process of increasing mass of atomic nuclei. The estimation of Husrt exponent has enabled us to indicate that we would be in presence of a fractional Brownian regime with long range correlations.

To summarize: Some preliminary results have been obtained. The mechanism of increasing mass in atomic nuclei reveals itself to be a nonlinear mechanism marked by a non integer value of Correlation dimension in phase space reconstruction. The presence of positive Lyapunov exponents indicate that the system of mass increasing is divergent and thus possibly chaotic. By using an identified Power Law and the variogram technique we may conclude that we are in presence of a fractal regime, a fractional Brownian regime.

The most relevant results have been obtained by using RQA. The process under our investigation results to be not fully deterministic when considering an embedding dimension D=2. We are in presence of self-resemblance and pseudo periodicities that show small fluctuations at increasing value of $Z-shift$ while instead Determinism shows consistent variations at increasing values of such parameter. Also Entropy and Max Line reveal the same tendency. Therefore, in the same framework of stable nuclei we have phase of increasing stability or increasing instability, depending on the mechanism of composition of the considered atomic nuclei and on the differences that they exhibit in the values of their Atomic Weights and of Mass Number. A final important result is obtained by using RQA in phase space reconstruction using embedding dimension D=1 and increasing Radius R corresponding to net differences in Mass number of the considered atomic nuclei. In this case, in phase space reconstruction, RQA involves pairs of nuclei in our analysis. New properties are identified at the increasing values of $\Delta A$. In particular, determinism oscillates but at some regular distances it also shows definite constant values as well as the other RQA variables . This confirms that we are in presence of a mechanism of increasing mass of atomic nuclei in which phases of stability result subsequent to phases of instability possibly marked from conditions of order-disorder like transitions. We have to consider pairs of nuclei with fixed $\Delta A$ and to identify pairs of subsequent stable nuclei that indicate new regularities in Z, N that we need to indicate in detail . We have to classify nuclei pertaining to different groups using these new regularities. This approach might be of valuable interest and it will constitute the object of our future work.
In this framework, the next step of the present investigation will be also to analyze data corresponding to values of binding energies for atomic nuclei. Possibly the complex of such results will give the possibility to indicate new perspectives in the elaboration of more accurate nuclear models of nuclear matter.

**Acknowledgement**
Many thanks are due to M. Pitkanen for his continuous and stimulating interest, suggestions and encouragement through this work.

Software NDT by J. Reiss and VRA by E. Kononov were also used for general non linear analysis.